\PassOptionsToPackage{unicode}{hyperref}
\PassOptionsToPackage{hyphens}{url}
\PassOptionsToPackage{dvipsnames,svgnames,x11names}{xcolor}
\documentclass[12pt]{article}

\usepackage{amsmath,amssymb,verbatim,setspace}
\usepackage{iftex}
\ifPDFTeX
  \usepackage[T1]{fontenc}
  \usepackage[utf8]{inputenc}
  \usepackage{textcomp} 
\else 
  \usepackage{unicode-math}
  \defaultfontfeatures{Scale=MatchLowercase}
  \defaultfontfeatures[\rmfamily]{Ligatures=TeX,Scale=1}
\fi
\usepackage{lmodern}
\ifPDFTeX\else  
\fi
\IfFileExists{upquote.sty}{\usepackage{upquote}}{}
\IfFileExists{microtype.sty}{
  \usepackage[]{microtype}
  \UseMicrotypeSet[protrusion]{basicmath} 
}{}
\makeatletter
\@ifundefined{KOMAClassName}{
  \IfFileExists{parskip.sty}{%
    \usepackage{parskip}
  }{
    \setlength{\parindent}{0pt}
    \setlength{\parskip}{6pt plus 2pt minus 1pt}}
}{
  \KOMAoptions{parskip=half}}
\makeatother
\usepackage{xcolor}

\usepackage{bm}

\usepackage{algorithm}
\usepackage{algpseudocode}
\usepackage{multirow} 
\usepackage{amsthm}
\usepackage{placeins}
\newtheoremstyle{nonitalic}  
  {3pt}                      
  {3pt}                      
  {\normalfont}              
  {}                         
  {\color{blue}\itshape}     
  {.}                        
  {.5em}                     
  {\thmname{#1}~\thmnumber{#2}\thmnote{~(\textit{#3})}}                         
\theoremstyle{nonitalic}
\newtheorem{customtheorem}{Theorem}
\newtheorem{customlemma}{Lemma}

\makeatletter
\ifx\paragraph\undefined\else
  \let\oldparagraph\paragraph
  \renewcommand{\paragraph}{
    \@ifstar
      \xxxParagraphStar
      \xxxParagraphNoStar
  }
  \newcommand{\xxxParagraphStar}[1]{\oldparagraph*{#1}\mbox{}}
  \newcommand{\xxxParagraphNoStar}[1]{\oldparagraph{#1}\mbox{}}
\fi
\ifx\subparagraph\undefined\else
  \let\oldsubparagraph\subparagraph
  \renewcommand{\subparagraph}{
    \@ifstar
      \xxxSubParagraphStar
      \xxxSubParagraphNoStar
  }
  \newcommand{\xxxSubParagraphStar}[1]{\oldsubparagraph*{#1}\mbox{}}
  \newcommand{\xxxSubParagraphNoStar}[1]{\oldsubparagraph{#1}\mbox{}}
\fi
\makeatother

\usepackage{longtable,booktabs,array}
\usepackage{calc} 
\usepackage{etoolbox}
\makeatletter
\patchcmd\longtable{\par}{\if@noskipsec\mbox{}\fi\par}{}{}
\makeatother
\IfFileExists{footnotehyper.sty}{\usepackage{footnotehyper}}{\usepackage{footnote}}
\makesavenoteenv{longtable}
\usepackage{graphicx}
\makeatletter
\def\maxwidth{\ifdim\Gin@nat@width>\linewidth\linewidth\else\Gin@nat@width\fi}
\def\maxheight{\ifdim\Gin@nat@height>\textheight\textheight\else\Gin@nat@height\fi}
\makeatother
\setkeys{Gin}{width=\maxwidth,height=\maxheight,keepaspectratio}
\makeatletter
\def\fps@figure{htbp}
\makeatother

\makeatletter
\@ifpackageloaded{caption}{}{\usepackage{caption}}
\AtBeginDocument{%
\ifdefined\contentsname
  \renewcommand*\contentsname{Table of contents}
\else
  \newcommand\contentsname{Table of contents}
\fi
\ifdefined\listfigurename
  \renewcommand*\listfigurename{List of Figures}
\else
  \newcommand\listfigurename{List of Figures}
\fi
\ifdefined\listtablename
  \renewcommand*\listtablename{List of Tables}
\else
  \newcommand\listtablename{List of Tables}
\fi
\ifdefined\figurename
  \renewcommand*\figurename{Figure}
\else
  \newcommand\figurename{Figure}
\fi
\ifdefined\tablename
  \renewcommand*\tablename{Table}
\else
  \newcommand\tablename{Table}
\fi
}
\@ifpackageloaded{float}{}{\usepackage{float}}
\floatstyle{ruled}
\@ifundefined{c@chapter}{\newfloat{codelisting}{h}{lop}}{\newfloat{codelisting}{h}{lop}[chapter]}
\floatname{codelisting}{Listing}

\makeatother
\makeatletter
\@ifpackageloaded{caption}{}{\usepackage{caption}}
\@ifpackageloaded{subcaption}{}{\usepackage{subcaption}}
\makeatother

\ifLuaTeX
  \usepackage{selnolig}  
\fi
\usepackage[]{natbib}
\setcitestyle{semicolon}
\usepackage{bookmark}

\IfFileExists{xurl.sty}{\usepackage{xurl}}{} 
\hypersetup{
  pdftitle={Information Borrowing from Partially Compatible Trajectories for Estimation of Dynamic Treatment Regimes},
  pdfauthor={Chloe Si; David A. Stephens; Erica E.M. Moodie},
  pdfkeywords={adaptive treatment strategies, dynamic information borrowing, inverse probability weighting, variance reduction, partial adherence, precision medicine},
  colorlinks=true,
  linkcolor={blue},
  filecolor={Maroon},
  citecolor={Blue},
  urlcolor={Blue},
  pdfcreator={LaTeX via pandoc}}

\newcommand{\anon}{1}

\begin{document}

\def\spacingset#1{\renewcommand{\baselinestretch}%
{#1}\small\normalsize} \spacingset{1}


\if1\anon
{
  \title{\bf Information Borrowing from Partially Compatible Trajectories for Estimation of Dynamic Treatment Regimes
  }
  \author{
    Chloe Si \thanks{McGill University, Montreal, QC H3A 0B9, Canada. Email: chuqiao.si@mail.mcgill.ca}\\
    Department of Mathematics and Statistics\\
    David A. Stephens \thanks{
    McGill University, Montreal, QC H3A 0B9, Canada. Email: david.stephens@mcgill.ca}\\
    Department of Mathematics and Statistics\\
    Erica E.M. Moodie \thanks{McGill University, Montreal, QC H3A 1A3, Canada. Email: erica.moodie@mcgill.ca}\\
    Department of Epidemiology, Biostatistics, and Occupational Health
}
  \date{} 

  \maketitle
} \fi

\if0\anon
{
  \bigskip
  \bigskip
  \bigskip
  \begin{center}
    {\LARGE\bf Information Borrowing from Partially Compatible Trajectories for Estimation of Dynamic Treatment Regimes}
\end{center}
  \medskip
} \fi

\bigskip
\begin{abstract}
Dynamic Treatment Regimes (DTRs) provide a systematic framework for optimizing sequential decision-making in chronic disease management, where therapies must adapt to patients' evolving clinical profiles. Inverse probability weighting (IPW) is a cornerstone methodology for estimating regime values from observational data due to its intuitive formulation and established theoretical properties, yet standard IPW estimators face significant limitations, including variance instability and data inefficiency. A fundamental but underexplored source of inefficiency lies in the strict alignment requirement between observed and target treatment trajectories, which fails to account for partial compatibility and discards substantial information from individuals with only minimal deviations from the regime. We propose two novel methodologies that relax the strict inclusion rule through flexible compatibility mechanisms. Both methods provide computationally tractable alternatives that can be easily integrated into existing IPW workflows, offering more efficient approaches to DTR estimation. Theoretical analysis demonstrates that both estimators preserve consistency while achieving superior finite-sample efficiency compared to standard IPW, and comprehensive simulation studies confirm improved stability. We illustrate the practical utility of our methods through an application to HIV treatment data from the AIDS Clinical Trials Group Study 175 (ACTG175). 
\end{abstract}

\noindent%
\vfill

\newpage
\spacingset{1.8} 

\section{Introduction}

Dynamic Treatment Regimes (DTRs) provide a principled framework for sequential decision-making in chronic disease management, where treatment strategies must adapt to patients’ evolving clinical profiles and responses \citep{murphy2003optimal, robins2004optimal}. These regimes are particularly important in settings such as HIV/AIDS management, oncology, and mental health treatment, where multi-stage interventions must be tailored to patient characteristics 
to optimize long-term outcomes \citep{cain2010start, wang2012evaluation}. 
The central objective is to identify the optimal treatment regime, defined as the sequence of decision rules that maximizes the expected outcome, referred to as the \textit{value} of the regime.

A broad class of statistical methods has been developed for regime estimation. Indirect approaches such as Q-learning \citep{watkins1992q} and A-learning \citep{robins2004optimal, blatt2004learning} rely on dynamic programming principles, whereas direct approaches estimate regime values explicitly and optimize over candidate regimes. Among these, marginal structural models (MSMs) \citep{robins2000marginal, orellana2010dynamic} and the associated inverse probability weighting (IPW) or augmented IPW (AIPW) estimators \citep{robins1994estimation, rotnitzky1998semiparametric} form the foundational tools. Recent work has incorporated flexible modeling strategies, including Bayesian nonparametrics \citep{xu2016bayesian} and machine learning techniques \citep{murray2018bayesian}.

Despite these innovations, IPW remains the primary approach for its intuitive formulation, established theoretical properties, and computational tractability. However, standard IPW estimators for discrete treatments face well-documented drawbacks that can severely limit their utility. The main concern is variance issue, particularly when estimated treatment probabilities approach zero, leading to extreme weights and unstable estimates. 
Numerous refinements have been proposed to address this instability, including weight stabilization \citep{hernan2000marginal}, truncation and trimming procedures \citep{cole2008constructing, lee2011weight}, and covariate balancing approaches \citep{hainmueller2012entropy,imai2014covariate, zhao2017entropy, li2018balancing}.

However, comparatively less attention has been paid to a different but equally fundamental source of inefficiency: the binary rule inherent in standard IPW, under which only individuals whose observed treatment path exactly matches the target regime contribute to estimation. This inclusion criterion discards all non-compatible trajectories, even those who deviate minimally with the regime, resulting in large variance and data inefficiency, especially in multi-stage settings. A growing body of work seeks to overcome the discrepancies between assigned or prescribed treatment and treatment actually received, including introducing partial compliance strata, categorical adherence measures and explicit modeling of the adherence process through g-estimation approaches \citep{artman2024marginal, zhu2025accounting, spicker2025optimal}. Related ideas have also appeared in the DTR learning literature; for example, \citet{ye2024stage} relaxes the strict inclusion requirement by weighting trajectories according to the number of stages where the observed treatment agrees with the regime, within a framework that evaluates cumulative rewards across many (e.g., five or more) stages. Together, these developments point to a growing consensus that strict exclusion of non-compliers or non-compatible trajectories are neither necessary nor optimal for causal effect estimation or for learning dynamic treatment strategies in realistic clinical settings.

A complementary perspective emerges when we consider continuous treatment settings, where analogous limitations have 
successfully been addressed by smooth weighting approaches for causal effect estimation. Established methods employ kernel-based or density-based weights that borrow strength from individuals receiving treatment doses close to the target \citep{imai2004causal, hirano2004, kennedy2017non}. These ideas naturally motivate rethinking compatibility in discrete DTR settings, particularly when discrete treatment categories are determined by underlying continuous biological processes and near-boundary individuals are clinically interchangeable. 

Inspired by these considerations, we propose two novel estimators that relax the strict inclusion rule through flexible compatibility mechanisms. The Generalized Compatibility Weighting (GCW) estimator replaces the compatibility indicator with a probability-based compatibility score, serving as a proxy for how closely an individual aligns with the target regime based on their covariates. 
By weighting observations proportional to their probability of receiving the regime-recommended treatment, GCW partially incorporates individuals who deviate from the assigned treatment but remain highly compatible with it. The Bootstrap Compatibility Windowing (BCW) estimator introduces covariate-based tolerance windows around treatment thresholds, recognizing that individuals near decision boundaries may be informative about regime performance, drawing conceptual inspiration from kernel methods \citep{hirano2004} and bootstrap-based bandwidth selection \citep{wand1994kernel, hall1995blocking}. We establish that both estimators preserve desirable asymptotic properties, reduce finite-sample variance, are offer straightforward computation that can be integrated into existing IPW workflows with minimal modifications. 

The remainder of this article proceeds as follows. Section~\ref{sec:background} establishes the formal framework for DTRs and reviews standard IPW methodology, highlighting key limitations. Section~\ref{sec:method} introduces our proposed estimators, including algorithmic implementations and theoretical properties. Section~\ref{sec:simulation} presents comprehensive simulation studies evaluating finite-sample performance. 
Section~\ref{sec:application} demonstrates an application to HIV treatment using the AIDS Clinical Trials Group Study 175 (ACTG175). Section~\ref{sec:discussion} concludes. 

\section{Background}
\label{sec:background}
\subsection{Dynamic Treatment Regimes: Notation and Framework}

We adopt standard notation where uppercase letters denote random variables and lowercase letters represent their realizations. Bold is used for vectors and matrices. Consider a $T$-stage dynamic treatment regime with discrete decision points indexed by $t \in \{1, 2, \ldots, T\}$. At each stage $t$, let $\bm{X}_{t,i}$ denote the vector of treatment-free covariates observed for individual $i$, and let $A_{t,i}$ represent the treatment assigned. At the conclusion of stage $T$, we observe outcome $Y_i$. We use overbar notation to denote cumulative histories: $\bar{\bm{X}}_t = (\bm{X}_1, \ldots, \bm{X}_t)$ and $\bar{\bm{A}}_t = (A_1, \ldots, A_t)$ represent the covariate and treatment histories up to stage $t$, respectively. For simplicity, $\bar{\bm X}$ and $\bar{\bm A}$ refer to the full histories through stage $T$. For a given treatment sequence $\bm{a} = (a_1, a_2, \ldots, a_T)$,  denote by $Y^{\bm{a}}$ the potential outcome that would be observed if the individual had received treatment history $\bm{a}$. Throughout, we assume the treatment space $\mathcal{A}_t$ is discrete (e.g., binary or categorical) and the outcome space is continuous.

A dynamic treatment regime $d = (d_1, d_2, \ldots, d_T)$ is formally defined as a sequence of decision rules that map patient history to treatment recommendations. Each component $d_t$ is a measurable function $d_t: (\bar{\bm{X}}_t, \bar{\bm{A}}_{t-1}) \to \mathcal{A}_t$ that assigns treatment $a_t \in \mathcal{A}_t$ based on the observed covariate history and previous treatment decisions. We focus specifically on the class of threshold-based regimes, where each decision rule $d_t$ at stage $t$ is defined by a set of threshold parameters $\bm{\psi}_t = (\psi_{t,1}, \ldots, \psi_{t,p_t})$ corresponding to the $p_t$-dimensional covariate vector $\bm{X}_t = (X_{t,1}, \ldots, X_{t, p_t})$. Treatment assignment at stage $t$ follows threshold-based rules of the form ``treat if ${X}_{t,j} \lesseqgtr \psi_{t,j}$ for $j = 1, \ldots, p_t$'' \citep{orellana2010dynamic, cain2010start}. These rules can be extended to include \textit{and/or} statements if multiple covariates are considered in treatment allocation decisions at a given stage. The complete regime is characterized by the collection of threshold parameters across all stages $\Psi = (\bm{\psi}_1, \ldots, \bm{\psi}_T)$.

Let $\mathcal{D}_{\Psi}$ denote the class of all threshold-based DTRs indexed by parameter vector $\Psi$. The optimal DTR $d^{\text{opt}}$ is defined as the regime that maximizes expected outcome:
$$
d^{\text{opt}} = \arg\max_{d \in \mathcal{D}_{\Psi}} \mathbb{E}[Y^d],
$$
where $Y^d$ is the potential outcome under regime $d$. Estimating $d^{\text{opt}}$ from observational data requires addressing confounding and selection bias through appropriate causal strategies.

To enable causal inference from observational data, we impose the following standard assumptions: (A1) consistency, (A2) SUTVA, (A3) sequential ignorability, and (A4) positivity, along with regularity conditions (A5) i.i.d.\ sampling and (A6) bounded outcomes. The full statements of assumptions (A1) to (A6) are provided in the supplementary materials~A.

\subsection{Inverse Probability Weighting: Formulation}
The \textit{value} of a given regime $d$ is defined as $V(d) = \mathbb{E}[Y^d]$. Under assumptions (A1)-(A4), this quantity can be estimated from observational data through inverse probability weighting.

Let $d_{t,i} := d_t(\bar{\bm{X}}_{t,i}, \bar{\bm A}_{t-1,i})$ denote the treatment recommended to individual $i$ at stage $t$. The stage-$t$ weight is:
$$w_{t,i}^{\text{I}} = \frac{\mathbb{I}(A_{t,i}=d_{t,i})}{P(A_{t,i} \mid \bar{\bm X}_{t,i}, \bar{\bm A}_{t-1,i})},$$
and $\bar w_{t,i}^{\mathrm{I}}=\prod_{s=1}^t w_{s,i}^{\mathrm{I}}$ is the cumulative weight. The IPW estimator of the regime value is:
$$
\widehat{V}_{\text{IPW}} = \frac{1}{n} \sum_{i=1}^n \left\{\prod_{t=1}^T w_{t,i}^{\text{I}} \right\} Y_i = \frac{1}{n} \sum_{i=1}^n \bar{w}_{T,i}^{\text{I}} Y_i.
$$
In practice, we estimate the probability of treatment $P(A_{t,i} \mid \bar{\bm X}_{t,i},\bar{\bm A}_{t-1,i})$ using propensity score (PS) models. With correct PS specification and standard causal assumptions, the IPW estimator is unbiased for the value of the regime $d$. 
To improve finite-sample stability, we consider a normalized variant that remains consistent for $V(d)$ as $n \to \infty$: $$\widehat{V}_{\text{nIPW}}(d) = \frac{\sum_{i=1}^n \bar{w}_{T,i}^{\text{I}} Y_i}{\sum_{i=1}^n \bar{w}_{T,i}^{\text{I}}}.$$

\subsection{Augmented IPW: A Doubly Robust Extension}
The augmented IPW (AIPW) estimator improves efficiency by incorporating regression of structural mean models, which we shall refer to somewhat informally as \textit{outcome regression models}. Let $Q_{T+1}^d=Y$, ${w}_{0,i}^{\text{I}} = 1$ and define the sequence
\[
Q_t(\bar{\bm x}_t, \bar{\bm a}_t)
= 
\mathbb{E}\!\left[
  Q_{t+1}^d(\bar{\bm X}_{t+1}, \bar{\bm A}_t)
  \,\middle|\,
  \bar{\bm X}_t = \bar{\bm x}_t,\,
  \bar{\bm A}_t = \bar{\bm a}_t
\right] \quad t = T, T-1, \dots, 1.
\]
After fitting $\widehat Q_t(\cdot,\cdot)$ from observed pairs
$(\bar{\bm X}_t,\bar{\bm A}_t)$, we evaluate $Q_t$ under regime $d_t$ via:
\[
Q_t^d:= Q_t^d(\bar{\bm x}_t, \bar{\bm a}_{t-1})
= Q_t(\bar{\bm x}_t, \bar{\bm a}_{t-1}, d_t).
\]
The AIPW estimator is:
$$
\widehat{V}_{\text{AIPW}}(d) = \frac{1}{n}\sum_{i=1}^n \left\{\bar{w}_{T,i}^{\text{I}} Y_i + \sum_{t=1}^T (\bar{w}_{t-1,i}^{\text{I}} -\bar{w}_{t,i}^{\text{I}}) Q_t^d\right\} = \frac{1}{n} \sum_{i=1}^n  \left\{Q_1^d + \sum_{t=1}^T \bar{w}_{t,i}^{\text{I}} (Q_{t+1}^d - Q_t^d)\right\},
$$
with its normalized form obtained by scaling weights 
by $\sum_{i=1}^n\bar w_{t,i}^{\mathrm{I}}$. Under mild conditions, AIPW achieves \textit{double robustness}: it is consistent for $V(d)$ if either the propensity score or the outcome model is correctly specified \citep{robins1994estimation, bang2005doubly}.

\section{Methodology}
\label{sec:method}

In this section, we introduce two approaches that preserve the general structure of IPW while relaxing the strict inclusion rule. Section~\ref{sec:method1} proposes a probability-based modification that replaces the binary indicator with a generalized compatibility function. Section~\ref{sec:method2} presents a data-adaptive alternative that defines compatibility through covariate proximity. We describe parameter selection and establish theoretical properties for both approaches.

\subsection{The Generalized Compatibility Weighting Estimator}
\label{sec:method1}
The standard IPW estimator employs a binary inclusion rule through the indicator function $\mathbb{I}(A_{t,i} = d_{t,i})$ which equals 1 if individual $i$'s stage $t$ observed treatment matches the regime recommendation and 0 otherwise. Our goal is to relax this hard inclusion criterion to improve data efficiency and reduce estimation variation.

\subsubsection{Generalized Compatibility Function}
To allow partial contributions from non-compatible trajectories, we define the stage-$t$ compatibility score:
$$m_{t,i} := m_{t,i}(\bar{\bm{X}}_{t,i}, \bar{\bm A}_{t-1,i}, A_t) = \mathbb{I}(A_{t,i} = d_t) + \gamma_{t,i} \cdot \mathbb{I}(A_{t,i} \neq d_t),$$
where $\gamma_{t,i} \in (0, 1)$ represents the relative contribution weight assigned to individual $i$ at stage $t$ when they deviate from the prescribed regime. 

Following \citet{hernan2006estimating}, we stabilize these weights by defining
$$D_{t,i} := D_{t,i}(\bar{\bm{X}}_{t,i},\bar{\bm{A}}_{t-1,i}) = P(A_{t,i} = d_{t,i} \mid \bar{\bm{X}}_{t,i}, \bar{\bm{A}}_{t-1,i}) + \gamma_{t,i} \cdot P(A_{t,i} \neq d_{t,i} \mid \bar{\bm{X}}_{t,i}, \bar{\bm{A}}_{t-1,i}),$$
and constructing the generalized compatibility weight as
$$w_{t,i}^{\text{G}} = \dfrac{m_{t,i}}{D_{t,i}}, \quad \bar{w}_{t, i}^{\text{G}} = \prod_{l=1}^t w_{l,i}^{\text{G}}.$$
\begin{customlemma}[Stabilization Property]\label{lem:GCWweight}
The GCW weight satisfies $\mathbb{E} [w_{t}^{\text{G}}] = 1$ and $\mathbb{E}[\bar{w}_{t}^{\text{G}}] = 1$ for all $t=1,\, \ldots, \, T$. A proof is provided in the supplementary materials~B.
\end{customlemma}

\subsubsection{The GCW Estimator and Bias Characterization} 
The GCW estimator is constructed as:
$$\widehat{V}_{\text{GCW}} = \dfrac{1}{n} \sum_{i=1}^n \left\{\prod_{t=1}^T w_{t,i}^{\text{G}} \right\}  Y_i = \dfrac{1}{n} \sum_{i=1}^n \bar{w}_{T,i}^{\text{G}}  Y_i.$$
Unlike the standard IPW estimator, GCW introduces a controlled bias that can be explicitly characterized and bounded. This bias arises from the inclusion of non-compatible observations and represents the fundamental trade-off inherent in our approach. The following lemma provides a complete decomposition of this bias.

\begin{customlemma}[Bias Decomposition]\label{lem:biasdecomp}
Under assumptions (A1)$-$(A5) and correct specification of the propensity score model, the bias of the GCW estimator decomposes as:
\[
\mathbb{E}[\widehat{V}_\text{GCW}(d)] - \mathbb{E}[Y^d] = \mathbb{E}\left[
\big(1-\theta(\bar{\bm X})\big)\,
\big(\mathbb{E}[Y^{\neg d}\mid \bar{\bm X}] - \mathbb{E}[Y^{d}\mid \bar{\bm X}]\big)
\right],
\]
where $\theta(\bar{\bm{X}}) = \prod_{t=1}^T \left(p_t / D_t\right) $ with $p_t = P(A_t = d_t \mid \bar{\bm X}_t, \bar{\bm a}_{t-1})$ and $\mathbb{E}\left[Y^{\neg d} \mid \bar{\bm{X}} \right]$ represents a weighted expectation of non-compatible outcomes at covariate level $\bar{\bm{X}}$. 

This decomposition shows that: (i) the bias is proportional to $(1-\theta(\bar{\bm{X}}))$ and can be controlled through the choice of $\gamma_{t}$, and (ii) when $\theta(\bar{\bm{X}})=1$ (i.e, $\gamma_{t}=0,\, \forall t$), the original unbiased IPW estimator is recovered. A proof is provided in the supplementary materials~B.
\end{customlemma}

\subsubsection{Parameter Selection and Asymptotic Properties}

The bias decomposition in Lemma~\ref{lem:biasdecomp} suggests constraining bias via
\[
1-\theta(\bar{\bm X}) = 1-\prod_{t=1}^T \frac{p_t}{D_t} \;\leq\; \epsilon_n,
\qquad \epsilon_n \to 0.
\]
A sufficient condition is
\begin{equation}
\label{eq:eq1}
    1-\frac{p_t}{D_t} \leq \epsilon_t = \epsilon_n / T \iff \gamma_t \leq \frac{\epsilon_t p_t}{(1-\epsilon_t)(1-p_t)}.
\end{equation}
We recommend choosing the bias-control parameter $\epsilon_n=c\,n^{-k}$ with constants $c>0$ and $k>0$. This choice ensures the bias vanishes asymptotically while providing meaningful variance reduction in finite samples. The constants $(c,k)$ can be tuned based on the desired balance bias-variance trade-off, as we demonstrate in our simulation studies.

\begin{customtheorem}[Consistency of GCW]\label{thm:GCWasymp}
Under (A1)$-$(A6), correct specification of the propensity model, and the bias-control parameter $\epsilon_n=c\,n^{-k}\to 0$ with $\gamma_t$ satisfying equation~\ref{eq:eq1}, both the unnormalized and normalized GCW estimators are consistent for $\mathbb{E}[Y^d]$:
\[
\widehat V_{\mathrm{GCW}}(d)
=\frac{1}{n}\sum_{i=1}^n \bar{w}_{T,i}^{\text{G}} Y_i \xrightarrow{p} \mathbb{E}[Y^d],
\qquad
\widehat V_{\mathrm{nGCW}}(d)
=\frac{\sum_{i=1}^n \bar{w}_{T,i}^{\text{G}} Y_i}{\sum_{i=1}^n \bar{w}_{T,i}^{\text{G}}} \xrightarrow{p} \mathbb{E}[Y^d].
\]
A proof is provided in the supplementary materials~B.
\end{customtheorem}

\subsubsection{Variance Reduction Properties}
A key advantage of the GCW estimator is its ability to achieve variance reduction while maintaining asymptotic unbiasedness. The following Theorem quantifies this improvement.
\begin{customtheorem}[Finite-Sample Variance Reduction]\label{thm:GCWvar}
Under assumptions (A1)$-$(A6), correct specification of the propensity score model, and $\epsilon_n = c n^{-k} \to 0$ with $\gamma_t$ satisfying equation~\ref{eq:eq1}, the difference between variances of nIPW and nGCW satisfies:
\[
\operatorname{Var}(\widehat{V}_{\text{nIPW}}) - \operatorname{Var}(\widehat{V}_{\text{nGCW}}) = O(n^{-k-1}) > 0,
\]
for finite $n$ and sufficiently small $\epsilon_n$, with the difference vanishing as $n \to \infty$.
\end{customtheorem}
A proof is given in supplementary materials~B. The rate $O(n^{-k-1})$ indicates that the variance benefit becomes more pronounced for smaller values of $k$, though this comes at the cost of slower bias convergence. This trade-off can be optimized based on sample size, compatibility rates, and the desired balance between bias and variance in specific applications.
\paragraph*{MSE quantification}
We can bound the bias by
\[
\operatorname{Bias} = \mathbb{E}\left[
\big(1-\theta(\bar{\bm X})\big)\,
\big(\mathbb{E}[Y^{\neg d}\mid \bar{\bm X}] - \mathbb{E}[Y^{d}\mid \bar{\bm X}]\big)
\right] \leq \epsilon_n \mathbb{E}[\Delta(\bar{\bm X})] = O(\epsilon_n) = O(n^{-k}).
\]
Combining this with the variance result gives
\[
\operatorname{MSE}(\widehat{V}_{\text{nIPW}}) - \operatorname{MSE}(\widehat{V}_{\text{nGCW}}) = \operatorname{Var}(\widehat{V}_{\text{nIPW}}) - \operatorname{Var}(\widehat{V}_{\text{nGCW}}) - \operatorname{Bias}^2 = O(n^{-k-1}) - O(n^{-2k}).
\]
When $k>1$, the variance reduction dominates the squared bias term, yielding asymptotic MSE improvement. In finite samples, choosing $k \leq 1$ generally provides a favourable balance, enabling meaningful variance reduction while keeping bias small.

\subsubsection{Augmented Generalized  Compatibility Weighting}
In analogy to AIPW, we define the augmented estimator by replacing IPW weights with GCW weights:
$$
\widehat{V}_{\text{AGCW}}(d)  = \frac{1}{n} \sum_{i=1}^n  \left\{Q_1^d + \sum_{t=1}^T \bar{w}_{t,i}^{\text{G}} (Q_{t+1}^d - Q_t^d)\right\},
$$
and its normalized version:
$$
\widehat{V}_{\text{nAGCW}}(d)  = \frac{1}{n}\sum_{i=1}^n Q_1^d + \sum_{t=1}^T \frac{\sum_{i=1}^n \bar{w}_{t,i}^{\text{G}}(Q_{t+1}^d - Q_t^d)}{\sum_{i=1}^n \bar{w}^{\text{G}}_{t,i}}.
$$
The AGCW estimators inherit the analogous theoretical guarantees as in GCW: AGCW remains consistent under standard double robustness conditions and offers additional variance gains relative to AIPW and GCW. The following results formalize these properties.
\begin{customtheorem}[Double Robustness of AGCW]\label{thm:AGCWdblrob}
Under assumptions (A1)$-$(A6), correct specification of either the propensity model or the outcome model, and $\epsilon_n=c\,n^{-k}\to 0$ with $\gamma_t$ satisfying equation~\ref{eq:eq1}, both the unnormalized and normalized AGCW estimators are consistent for $\mathbb{E}[Y^d]$:
\[
\widehat{V}_{\text{AGCW}}(d)  \xrightarrow{p} \mathbb{E}[Y^d], \quad 
\widehat{V}_{\text{nAGCW}}(d)  \xrightarrow{p} \mathbb{E}[Y^d].
\]
A proof is given in the supplementary materials~B.
\end{customtheorem}

\begin{customtheorem}[Variance Reduction of AGCW]\label{thm:AGCWvarred}
Under the conditions of Theorem~\ref{thm:AGCWdblrob}, the nAGCW estimator has asymptotic variance no larger than that of nAIPW for sufficiently small $\epsilon_n$, and no larger than that of nGCW when the outcome model is correctly specified. A proof is provided in the supplementary materials~B.
\end{customtheorem}

\subsubsection{Computational Algorithm}
The results established in the preceding sections of 3.1 demonstrate that generalized compatibility weighting provides a principled approach to relaxing the strict inclusion rule while maintaining consistency and improving finite-sample efficiency. Algorithm 1 presents the implementation of the normalized GCW and AGCW estimators, which preserves the basic structure of standard IPW algorithms and adds minimal complexity.

\begin{spacing}{1.25}
\begin{algorithm}[t]
\caption{Normalized (Augmented) Generalized  Compatibility Weighting Estimators}
\begin{algorithmic}[1]
\State \textbf{Input:} 
Data $\{(\bar{\bm X}_i, \bar{\bm A}_i, Y_i)\}_{i=1}^n$, regime $d$, and $(c,k)$ for $\epsilon_n=c n^{-k}$.
\State Compute stage-wise bias levels $\epsilon_t=\epsilon_n/T$.
\State Estimate stage-wise propensity $P_t(a_t\mid\bar{\bm X}_t,\bar{\bm A}_{t-1})$ for $t=1,\dots,T$.
\State Fit outcome regression models to obtain $Q_t^d$ and $Q_{t+1}^d$ for $t=1,\dots,T$.
\For{$i = 1$ to $n$}
    \State Initialize cumulative weight: $\bar{w}^{\text G}_{0,i} = 1$.
    \For{$t = 1$ to $T$}
        \State Compute compatibility probability: $p_{t,i} = P(A_{t,i} = d_{t,i} \mid \bar{\bm{X}}_{t,i}, \bar{\bm{A}}_{t-1,i}).$

        \State Choose stage-wise $\gamma_{t,i}$ satisfying the bias constraint:$\gamma_{t,i} = \frac{\epsilon_t p_{t,i}}{(1-\epsilon_t)(1 - p_{t,i})}.$

        \State Compute compatibility score:
        $m_{t,i}= \mathbb{I}(A_{t,i}=d_{t,i})+ \gamma_{t,i}\,\mathbb{I}(A_{t,i}\neq d_{t,i}).$

        \State Compute stabilizing denominator:
        $D_{t,i} = p_{t,i} + \gamma_{t,i}(1 - p_{t,i}).$

        \State Construct stage-wise GCW weight:
        $w_{t,i}^{\text G} = \frac{m_{t,i}}{D_{t,i}}.$

        \State Update cumulative weight:
        $\bar{w}^{\text G}_{t,i}= \bar{w}^{\text G}_{t-1,i} \cdot w_{t,i}^{\text G}.$
    \EndFor
\EndFor
\State \textbf{Compute nGCW:}
\[
\widehat V_{\mathrm{nGCW}}(d)=
\frac{\sum_{i=1}^n \bar w^{\mathrm G}_{T,i} Y_i}
     {\sum_{i=1}^n \bar w^{\mathrm G}_{T,i}}.
\]

\State \textbf{Compute nAGCW:}
\[
\widehat V_{\mathrm{nAGCW}}(d)
=
\frac{1}{n}\sum_{i=1}^n Q_{1,i}^d
+ \sum_{t=1}^T
  \frac{\sum_{i=1}^n \bar w^{\mathrm G}_{t,i}(Q_{t+1,i}^d-Q_{t,i}^d)}
       {\sum_{i=1}^n \bar w^{\mathrm G}_{t,i}}.
\]

\State \textbf{Output:} $\widehat V_{\mathrm{nGCW}}(d)$, $\widehat V_{\mathrm{nAGCW}}(d)$.
\end{algorithmic}
\end{algorithm}
\end{spacing}

\subsection{The Bootstrap Compatibility Windowing Estimator}
\label{sec:method2}

For threshold-based regimes, individuals with covariates close to the decision boundary may be clinically similar to compatible individuals and can therefore provide useful information even when their observed treatment differs from the regime recommendation. This motivates a covariate-based relaxation of the inclusion criterion, which we refer to as Bootstrap Compatibility Windowing (BCW).

\subsubsection{The Windowed Regime Formulation}
Consider a threshold-based regime $d$ characterized by stage-specific thresholds $(\psi_1, \ldots, \psi_T)$, where treatment is administered at stage $t$ if the covariate satisfies $X_{t,i} \leq \psi_t$. For simplicity, we assume a single covariate controls treatment at each stage, though the framework extends naturally to multi-covariate settings. Under the standard IPW setup, an individual is considered compatible at stage $t$ if:
$$\left(X_{t,i} \leq \psi_t \; \land \; A_{t,i} = 1 \right) \; \vee \; \left(X_{t,i} > \psi_t \; \land \; A_{t,i} = 0\right).$$

To incorporate individuals near the decision boundary, we introduce directional tolerance parameters $\delta = (\delta_1, \ldots, \delta_T)$ that define a relaxed regime $d^{(\delta)}$. At stage $t$, we specify lower and upper tolerances $\delta_t^{l}$ and $\delta_t^{u}$ that expand the compatibility rule to:
$$\left(X_{t,i} \leq \psi_t + \delta_t^u \; \land \; A_{t,i} = 1 \right) \; \vee \; \left(X_{t,i} > \psi_t - \delta_t^l \; \land \; A_{t,i} = 0\right).$$
This formulation recognizes that individuals with covariates slightly above or below the threshold as approximately compatible with the regime.

\paragraph*{Directional Window Constraints.} The tolerance parameters $\delta_t^l, \delta_t^u$ must respect both the natural bounds of the covariate space and the regime structure. For a threshold $\psi_t \in [\underline{x}_t, \overline{x}_t]$, we impose the following constraints:
$$\delta_t^l \leq \min(\delta_t^{\text{max}}, \; \psi_t - \underline{x}_t), \quad \delta_t^u \leq \min(\delta_t^{\text{max}}, \; \overline{x}_t - \psi_t ),$$
such that the windowed regime remains well-defined and clinically meaningful.

\paragraph*{Grid Range Selection.} For a threshold $\psi_t \in [\underline{x}_t, \overline{x}_t]$, we suggest choosing the maximum window half-width $$\delta_t^{\text{max}} = \dfrac{\overline{x}_t - \underline{x}_t}{q} \quad \text{with} \quad q \in [20, 50].$$ This choice restricts the largest candidate windows to roughly 2\% $-$ 5\% of the covariate range, which keeps the window sufficiently local to justify treating near-boundary individuals as clinically comparable, while still yielding enough observations within each window for stable bootstrap estimation.

\subsubsection{Adaptive Window Size Selection via Bootstrap}
The optimal window size $\delta$ is selected via a data-adaptive procedure that empirically balances bias and variance using bootstrap resampling.
\paragraph*{Grid Construction.} For each stage $t$, we define a finite grid $\mathcal{G}_t$ as:
$$\mathcal{G}_t = \{0, s_t, 2s_t, \ldots, M_ts_t\}, \quad s_t \in (0, \delta_t^{\text{max}}], \quad  M_ts_t \leq \delta_t^{\text{max}}.$$
The grid step $s_t$ should reflect the measurement resolution for discrete covariates or a small fraction of $\delta_t^{\text{max}}$ for continuous ones.

For each candidate window $\delta \in \mathcal{G}_1 \times \cdots \times \mathcal{G}_T$, we compute the BCW estimate for the $\delta$-altered regime $d^{(\delta)}$ across $B$ bootstrap replicates as $\widehat{V}_{\text{BCW}}^{(b)}(d) = n^{-1}\sum_{i=1}^n\{\prod_{t=1}^T w_{t,i}^{(b)}\}Y_i^{(b)}$, where $w_{t,i}^{(b)} = \mathbb{I}(A_{t,i}^{(b)} = d^{(\delta),(b)}_{t,i})\, P(A_{t,i}^{(b)} \mid \bar{\bm X}_{t,i}^{(b)}, \bar{\bm A}_{t-1,i}^{(b)})^{-1}$. The optimal window minimizes the penalized loss $\mathcal{L}(\delta) = \widehat{\mathrm{Var}}(\delta) + \lambda_{\text{bias}}\widehat{\mathrm{Bias}}^2(\delta)$, where $\widehat{\mathrm{Var}}(\delta) = (B-1)^{-1}\sum_{b=1}^B\{\widehat{V}_{\text{BCW}}^{(b)}(d) - \mathbb{E}_B[\widehat{V}_{\text{BCW}}(d)]\}^2$, $\mathbb{E}_B[\widehat{V}_{\text{BCW}}(d)] = B^{-1}\sum_{b=1}^B \widehat{V}_{\text{BCW}}^{(b)}(d)$, and $\widehat{\mathrm{Bias}}(\delta) = \mathbb{E}_B[\widehat{V}_{\text{BCW}}(d)] - \widehat{V}_{\text{IPW}}(d)$. The penalty parameter $\lambda_{\text{bias}}$ controls the bias–variance trade-off and $\lambda_{\text{bias}}=1$ yields the usual mean-squared error loss.

The optimal window is selected as 
$\delta_{\text{opt}} = \arg \min_\delta \mathcal{L}(\delta)$
and the BCW estimator is:
$$\widehat{V}_{\text{BCW}}(d) = \dfrac{1}{n} \sum_{i=1}^n \left\{\prod_{t=1}^T w_{t,i}^{\text{B}}\right\} Y_i, \text{ where } w_{t,i}^{\text{B}} = \frac{\mathbb{I}\left( A_{t,i} = d^{(\delta_{\text{opt}})}_{t,i} \right)}{P\left(A_{t,i} \mid \bar{\bm X}_{t,i}, \bar{\bm A}_{t-1, i}\right)}.$$

\subsubsection{Asymptotic Properties}
The consistency of the BCW estimator relies on the key property that the data-selected window shrinks to zero as the sample size increases.
We make the following assumption: 
\begin{itemize}
    \item[(A7)]  Adaptive Window Convergence: The bootstrap optimization procedure selects a window such that: $\| \delta_{\text{opt}} \|_{\infty} \to 0$ as $n \to \infty.$
\end{itemize}
This assumption is reasonable and analogous to bandwidth conditions in kernel estimation \citep{wand1994kernel}. As the sample size increases, the standard IPW estimator becomes more stable, reducing the need for relaxed inclusion. 

\begin{customtheorem}[Consistency of BCW]\label{thm:method2asymp}
Under assumptions (A1)$-$(A7) and correct specification of the propensity score model, both the unnormalized and normalized BCW estimators are consistent for $\mathbb{E}[Y^d]$:
\[
\widehat V_{\mathrm{BCW}}(d)
=\frac{1}{n}\sum_{i=1}^n \bar{w}_{T,i}^{\text{B}} Y_i \xrightarrow{p} \mathbb{E}[Y^d],
\qquad
\widehat V_{\mathrm{nBCW}}(d)
=\frac{\sum_{i=1}^n \bar{w}_{T,i}^{\text{B}} Y_i}{\sum_{i=1}^n \bar{w}_{T,i}^{\text{B}}} \xrightarrow{p} \mathbb{E}[Y^d].
\]
A proof is provided in the supplementary materials~B.
\end{customtheorem}

\paragraph*{Variance Dominance Property.}
When the bias penalty is set to zero, that is, $\lambda_{\text{bias}} = 0$, the BCW estimator is guaranteed to perform no worse than standard IPW in terms of variance, since the loss function reduces to $\mathcal{L}(\delta) = \widehat{\text{Var}}(\delta)$. With $\delta_t^{\max}$ kept small, the resulting bias is limited even in this purely variance-driven criterion.

\subsubsection{Augmented Bootstrap Compatibility Windowing}
Parallel to AIPW and AGCW, we define the augmented BCW estimator as
$$
\widehat{V}_{\text{ABCW}}(d)  = \frac{1}{n} \sum_{i=1}^n  \left\{Q_1^d + \sum_{t=1}^T \bar{w}_{t,i}^{\text{B}} (Q_{t+1}^d - Q_t^d)\right\},
$$
with the normalized version
$$
\widehat{V}_{\text{nABCW}}(d)  = \frac{1}{n}\sum_{i=1}^n Q_1^d + \sum_{t=1}^T \frac{\sum_{i=1}^n \bar{w}_{t,i}^{\text{B}}(Q_{t+1}^d - Q_t^d)}{\sum_{i=1}^n \bar{w}^{\text{B}}_{t,i}}.
$$

\begin{customtheorem}[Double Robustness of ABCW]\label{thm:ABCWdblrob}
Under (A1)$-$(A7), correct specification of either the propensity model or the outcome model, both the unnormalized and normalized ABCW estimators are consistent for $\mathbb{E}[Y^d]$:
\[
\widehat{V}_{\text{ABCW}}(d)  \xrightarrow{p} \mathbb{E}[Y^d], \quad 
\widehat{V}_{\text{nABCW}}(d)  \xrightarrow{p} \mathbb{E}[Y^d].
\]
Assumption (A7) implies that the bootstrap-selected window satisfies 
$\|\delta_{\mathrm{opt}}\|_\infty \to 0$, which in turn yields
$w_{t,i}^{\text{B}} \to w_{t,i}^{\text{I}}$ in probability for all $t$ and $i$.
Hence both $\widehat{V}_{\text{ABCW}}(d)$ and $\widehat{V}_{\text{nABCW}}(d)$
are asymptotically equivalent to the corresponding AIPW estimators.
Since (n)AIPW is doubly robust, the same doubly robust consistency property
holds for (n)ABCW.
\end{customtheorem}

\FloatBarrier
\subsubsection{Computational Implementation} 
We provide a practical algorithm for computing nBCW and nABCW in Algorithm 2.
\begin{spacing}{1.25} 
\begin{algorithm}[t]
\caption{Normalized (Augmented) Bootstrap Compatibility Windowing Estimators}
\begin{algorithmic}[1]

\State \textbf{Input:} 
Data $\{(\bar{\bm{X}}_{i}, \bar{\bm{A}}_{i}, Y_i)\}_{i=1}^n$,
target regime $d$, window threshold $\delta_{\text{max}}$, grid steps $\{s_t\}$, number of bootstraps $B$, bias penalty $\lambda_{\mathrm{bias}}$.

\State Estimate stage-specific propensity score: $P_t(a_t \mid \bar{\bm{X}}_{t}, \bar{\bm{A}}_{t-1} ),\quad t=1,\ldots,T.$

\State Fit the outcome regression models to obtain 
$Q_t^d$ and $Q_{t+1}^d$ for $t=1,\ldots,T$.

\State Construct grids $\mathcal{G}_t = \{0, s_t, 2s_t, \ldots, \delta_t^{\text{max}}\}$ and define the full grid  $\mathcal{G} = \mathcal{G}_1 \times \cdots \times \mathcal{G}_T$.

\State Generate bootstrap datasets 
$\mathcal{D}^{(b)}=\{(\bar{\bm{X}}^{(b)}_{i},\bar{\bm{A}}^{(b)}_{i},Y_i^{(b)})\}_{i=1}^n$, 
for $b=1,\ldots,B$.

\For{each $\delta\in\mathcal{G}$}
    \State Define the windowed regime $d^{(\delta)}$.
    \For{$b=1$ to $B$}
        \State Compute bootstrap estimate $\widehat V^{(b)}_{\mathrm{nBCW}}(d)$ or $\widehat V^{(b)}_{\mathrm{nABCW}}(d)$ depending on whether augmentation is used.
    \EndFor

   \State Compute bootstrap mean $\mathbb{E}_B[\widehat V_{\mathrm{n(A)BCW}}(d)]$, variance $\widehat{\mathrm{Var}}(\delta)$, and bias $\widehat{\mathrm{Bias}}(\delta)$.

    \State Compute penalized loss:  
    $\mathcal{L}(\delta)=\widehat{\mathrm{Var}}(\delta)+\lambda_{\mathrm{bias}}\widehat{\mathrm{Bias}}^2(\delta)$.

\EndFor

\State Choose optimal window:
$
\delta_{\mathrm{opt}}=\arg\min_{\delta\in\mathcal{G}} \mathcal{L}(\delta).
$

\State \textbf{Compute nBCW with $\bar{w}_{T,i}^{B}$ calculated from $\delta_{\mathrm{opt}}$:}
\[
\widehat{V}_{\mathrm{nBCW}}(d)
= \frac{\sum_{i=1}^n \bar{w}_{T,i}^{B} Y_i}
       {\sum_{i=1}^n \bar{w}_{T,i}^{B}}.
\]

\State \textbf{Compute nABCW with $\bar{w}_{T,i}^{B}$ calculated from $\delta_{\mathrm{opt}}$:}
\[
\widehat V_{\mathrm{nABCW}}(d)
= \frac{1}{n}\sum_{i=1}^n Q_{1,i}^d
+ \sum_{t=1}^T
\frac{
\sum_{i=1}^n \bar{w}_{t,i}^{B} (Q_{t+1,i}^d - Q_{t,i}^d)
}{
\sum_{i=1}^n \bar{w}_{t,i}^{B}
}.
\]

\State \textbf{Output:} 
$\widehat{V}_{\mathrm{nBCW}}(d)$, $\widehat{V}_{\mathrm{nABCW}}(d)$.

\end{algorithmic}
\end{algorithm}
\end{spacing}

\section{Simulation Study}
\label{sec:simulation}
We conduct simulation studies to evaluate the finite-sample performance of the proposed methods, including both the normalized GCW and BCW estimators (nGCW, nBCW) and their augmented extensions (nAGCW, nABCW), in comparison with the standard nIPW and nAIPW approaches. Our evaluation focuses on several key aspects: (i) demonstrate that the proposed methods achieve reduced variance in regime value estimation, including agreement between analytically derived variances and Monte Carlo empirical variances (ii) assess bias control and coverage properties of the proposed estimators across a range of sample sizes, and (iii) evaluate the ability to correctly identify the optimal DTR, which is the central measure for practical utility in clinical settings.

\subsection{Simulation 1}
\label{sim:sim1}
We consider a two-stage DTR representing HIV management decisions over a 12-month period. We adopt a similar setup to \citet{moodie2007demystifying}. The data-generating process is structured as follows: baseline CD4: $X_1 \sim \mathcal{N}(450, 150)$; stage I treatment: $A_1 \sim Ber(p1)$, with $p_1 = \text{expit}(2-0.006X_1)$; 6-month CD4: $X_2 \sim \mathcal{N}(1.25X_1, 60)$; stage II treatment: $A_2 \sim Ber(p_2)$, with $p_2 = \text{expit} (0.8-0.004X_2)$; 1-year CD4: $Y \sim \mathcal{N}(400+1.6X_1, 60) - \mu_1 - \mu_2$ where $\mu_1 = A_1 (X_1 - 350)$, $\mu_2 = A_2 (2X_2 - 900)$. 

We study threshold-based regimes of the form: ``treat if $X_i \leq \psi_i$'' for $i = 1,2$ with $\psi_1 \in [150, 500]$ and $\psi_2 \in [200, 600]$. A grid of candidate thresholds $\mathcal{G}_{(\psi_1,\psi_2)}$ is created by taking the cross product of values in increments of 5 units for both $\psi_1$ and $\psi_2$. The true optimal regime is $(\psi_1,\psi_2)=(350,450)$ with a value of 1212.22.

To compare the performance of the estimators, we generate datasets of size $n=$200, 500, 1000, each replicated $R=500$ times. For each Monte Carlo replicate, regime value estimates are computed across $\mathcal{G}_{(\psi_1,\psi_2)}$. Coverage is evaluated using $B=500$ bootstrap replications. For the nGCW estimator, we fix $k=0.5$ and $c=1$ as a benchmark tuning choice, corresponding to a moderate level of borrowing while maintaining stability across regimes. Sensitivity to alternative values of $c$ is examined in supplementary materials~C.
For nBCW, we adopt the MSE loss with penalty $\lambda_{\text{bias}}=1$ and directional windows $\delta=\{0,2,4,6,8,10\}$ for both covariates.

Table~\ref{tab:Sim1_t1} summarizes the performance of the estimated optimal regime. For $\psi_1$ and $\psi_2$, we report (i) the mean and standard deviation (SD), (ii) the median and interquartile range (IQR), (iii) the root mean square error (RMSE) at the estimated optimum, and (iv) the coverage of the 95\% confidence intervals (CI). Across all sample sizes, the estimates $(\hat{\psi}_1,\hat{\psi}_2)$ remain close to the true optima, with variability steadily decreasing as $n$ increases. Both adaptive weighting methods generally produce less dispersed threshold estimates than nIPW, with the improvement being more pronounced when the sample size is small. In particular, nGCW achieves the smallest RMSE for both thresholds across almost all sample sizes, indicating improved accuracy in locating the optimal decision boundaries. Coverage remains close to the nominal level in most settings.



\FloatBarrier
\begin{table}[htbp]
\centering
\caption{Summary of estimated thresholds, RMSE, and coverage for nIPW, nGCW and nBCW based on 500 replications.}
\label{tab:Sim1_t1}
\setlength{\tabcolsep}{3pt}
{\fontsize{10pt}{10pt}\selectfont
\renewcommand{\arraystretch}{1.5}
\begin{tabular}{@{}ccccccccccc@{}}
\toprule
\multirow{2}{*}{$n$} & \multirow{2}{*}{Method} &
\multicolumn{2}{c}{Mean (SD)} &
\multicolumn{2}{c}{Median (IQR)} &
\multicolumn{2}{c}{RMSE} &
\multicolumn{2}{c}{Coverage} \\
\cmidrule(lr){3-4} \cmidrule(lr){5-6} \cmidrule(lr){7-8} \cmidrule(lr){9-10}
& & $\psi_1$ & $\psi_2$ & $\psi_1$ & $\psi_2$ & $\psi_1$ & $\psi_2$ & $\psi_1$ & $\psi_2$ \\
\midrule
200 & nIPW & 351.59 (73.00) & 485.70 (57.05) & 355.00 (105.00) & 485.00 (75.00) & 59.51 & 54.44 & 0.964 & 0.894 \\
    & nGCW & 351.40 (66.68) & 483.15 (53.61) & 355.00 (90.00)  & 485.00 (70.00) & 53.06 & 51.15 & 0.970 & 0.884 \\
    & nBCW & 349.84 (68.37) & 483.02 (55.08) & 355.00 (90.00)  & 480.00 (75.00) & 54.34 & 52.14 & 0.970 & 0.900 \\
\midrule\midrule
500 & nIPW & 347.46 (56.63) & 472.71 (42.03) & 350.00 (80.00) & 475.00 (55.00) & 45.08 & 38.23 & 0.980 & 0.906 \\
    & nGCW & 346.12 (53.78) & 471.40 (40.69) & 350.00 (70.00) & 470.00 (50.00) & 42.38 & 36.70 & 0.982 & 0.910 \\
    & nBCW & 346.26 (54.31) & 470.12 (41.71) & 350.00 (70.00) & 470.00 (60.00) & 42.20 & 37.16 & 0.980 & 0.908 \\
\midrule\midrule
1000 & nIPW & 346.25 (43.42) & 465.39 (33.19) & 350.00 (55.00) & 465.00 (45.00) & 34.15 & 29.31 & 0.988 & 0.954 \\
     & nGCW & 346.75 (41.66) & 464.45 (32.46) & 350.00 (55.00) & 465.00 (45.00) & 32.93 & 28.61 & 0.988 & 0.956 \\
     & nBCW & 347.30 (42.61) & 464.79 (33.13) & 350.00 (60.00) & 465.00 (45.00) & 33.70 & 29.07 & 0.990 & 0.958 \\
\bottomrule
\end{tabular}
}
\end{table}

Table~\ref{tab:Sim1_t2} reports grid-level performance of the estimators across the candidate regime set $\mathcal{G}_{(\psi_1,\psi_2)}$, including averaged variance, effective sample sizes (ESS) and the Spearman rank correlation. The ESS is calculated as $\mathrm{ESS}=(\sum_i w_i)^2/(\sum_i w_i^2)$ with larger values indicating more stable and less extreme weights \citep{mccaffrey2013tutorial}. The correlation evaluates how well the estimated value surface preserves the ordering of candidate regimes. Specifically, for each replication, we rank all regimes according to their estimated values and compute the Spearman correlation between this ranking and the ranking induced by the true regime values. A larger correlation indicates a closer ranking to the truth and 1.0 is perfect match. We also report the proportion of grid points where nGCW and nBCW achieve variance reduction relative to nIPW, together with regime-level performance metrics evaluated on an independent sample of size $n=10{,}000$, including the proportion of individuals optimally treated (POT) and the expected outcome under the estimated optimal regime.

Across all sample sizes, nGCW consistently achieves variance reduction relative to nIPW across the entire grid while also producing larger ESS, indicating improved weight stability. The nBCW estimator yields milder variance reduction but still improves ESS relative to nIPW in most settings. The magnitude of these improvements is particularly pronounced at smaller sample sizes. For example, when $n=200$, nGCW reduces the average variance by approximately 20\% and increases the ESS by about 15\% relative to nIPW. Both adaptive weighting methods also slightly improve the ranking of candidate regimes, as reflected by higher Spearman correlations with the true regime values. These improvements translate into better regime-level performance: both nGCW and nBCW produce higher agreement with the true optimal treatment assignments and higher expected outcomes compared with nIPW. The gains are most noticeable at smaller sample sizes.

\FloatBarrier
\begin{table}[htbp]
\centering
\caption{Performance of nIPW, nGCW and nBCW over all regimes. Measurements are reported as mean (standard deviation) across 500 Monte Carlo replications.}
\label{tab:Sim1_t2}
\resizebox{\textwidth}{!}{
{\fontsize{10pt}{10pt}\selectfont
\renewcommand{\arraystretch}{1.5}
\begin{tabular}{cccccccc}
\toprule
$n$ & Method & Var & ESS & Corr & \% Var $\downarrow$ (grid) & POT & Value \\
\midrule
200  & nIPW & 504.14 (237.94) & 76.80 (17.49) & 0.728 (0.102) & ---    & 0.743 (0.189) & 1118.76 (33.65) \\
     & nGCW & 399.00 (178.52) & 87.97 (19.85) & 0.756 (0.094) & 100.00 & 0.768 (0.174) & 1122.54 (29.27) \\
     & nBCW & 484.72 (221.00) & 77.47 (17.28) & 0.739 (0.099) & 89.31  & 0.763 (0.180) & 1121.61 (30.19) \\
\midrule\midrule
500  & nIPW & 190.06 (87.84) & 192.69 (44.28) & 0.865 (0.056) & ---    & 0.829 (0.125) & 1131.90 (17.35) \\
     & nGCW & 164.84 (73.77) & 210.24 (48.18) & 0.877 (0.052) & 100.00 & 0.840 (0.121) & 1132.95 (16.45) \\
     & nBCW & 185.44 (83.94) & 193.66 (43.85) & 0.870 (0.054) & 79.10  & 0.841 (0.116) & 1133.15 (15.63) \\
\midrule\midrule
1000 & nIPW & 104.18 (42.60) & 384.85 (89.01) & 0.922 (0.035) & ---    & 0.878 (0.084) & 1137.24 (8.45) \\
     & nGCW & 94.12 (37.19)  & 409.62 (94.60) & 0.927 (0.033) & 100.00 & 0.883 (0.078) & 1137.72 (7.23) \\
     & nBCW & 102.30 (40.96) & 386.40 (88.37) & 0.925 (0.034) & 65.78  & 0.879 (0.082) & 1137.38 (8.30) \\
\bottomrule
\end{tabular}}}
\end{table}

We also evaluate the normalized augmented estimators nAGCW, nABCW, and nAIPW under the same simulation configuration; we further conduct sensitivity analyses for all estimators under different hyperparameter choices. All augmented estimators exhibit substantially reduced variance and improved coverage. In addition, we compare the analytical variance of nGCW and nAGCW with the Monte Carlo (MC) variance. As the sample size increases, analytical and MC variances converge, and augmentation further sharpens this convergence. These results confirm that the analytical variance is a reliable and computationally efficient alternative to MC estimation, particularly when repeated sampling is computationally expensive. All results can be found in the supplementary materials~C. 

Overall, Simulation 1 demonstrates that both non-augmented and augmented adaptive weighting estimators provide meaningful variance reduction and improved performance characteristics relative to IPW, while maintaining accurate identification of the optimal treatment regimes. These findings indicate that adaptive weighting methods offer a practical and effective approach for improving the stability, accuracy, and overall reliability of estimated dynamic treatment regimes without compromising clinical interpretability.

\subsection{Simulation 2}

We next consider a simulation setting designed to closely reflect our application. We focus on a one-stage DTR for HIV management in which treatment is jointly determined by baseline CD4 and weight. The data-generating process is specified as follows: baseline CD4: $X_1 \sim \mathcal{N}(450, 150)$; baseline weight: $X_2 \sim \mathcal{N}(50, 20)$; stage I treatment: $A_1 \sim Ber(p_1)$ with $p_1 = \text{expit}(-4.5 + 0.005X_1 + 0.02X_2)$; 1-year CD4: $Y \sim \mathcal{N}\big( X_1 + 2X_2 + \tau(X_1) A_1, \; 20 \big)$ with $\tau(X_1) = 100 \exp\!\left\{ - \big( (X_1 - 350)/75 \big)^2 \right\} - 30$.

Threshold-based regimes are defined as ``treat if both $X_i \leq \psi_i$ for $i=1,2$,'' with $\psi_1 \in [200,600]$ and $\psi_2 \in [40,80]$. The candidate grid $\mathcal{G}_{(\psi_1,\psi_2)}$ is created using increments of 5 units for $\psi_1$ and 1 unit for $\psi_2$. Under this setup, the true optimal threshold is $(\psi_1,\psi_2) = (430,80)$ with value 560.87. Notably, the optimal regime does not tailor treatment based on weight, which is consistent with patterns observed in our application. The sample sizes are identical to those in Simulation 1. For nGCW, we again use $k = 0.5$ and $c = 1$ as a benchmark tuning choice. For nBCW, we use $\lambda_{\text{bias}}=1$ and directional windows $\delta_1 = \{0,2,4,6,8,10\}$ for $X_1$ and $\delta_2 = \{0,1,2\}$ for $X_2$.

Table~\ref{tab:Sim2_t1} reports performance at the estimated optima $(\hat{\psi}_1,\hat{\psi}_2)$ for each method, using the same metrics as in Simulation 1. Across all sample sizes, the mean and median estimates of $\psi_1$ and $\psi_2$ move steadily toward the true optima $(430,80)$, with variability decreasing as $n$ increases. The true value surface in this setting is much steeper in the $\psi_1$ direction and exhibits a flat ridge in the $\psi_2$ direction, with the optimum occurring at the upper boundary of $\psi_2$. As a result, inference for $\psi_2$ is inherently difficult. This feature is reflected in the coverage results: coverage for $\psi_1$ remains close to nominal across all methods, whereas coverage for $\psi_2$ is substantially lower, especially at smaller sample sizes. The weak identifiability along the ridge also makes the estimation more sensitive to small local fluctuations in the estimated value surface. Because the directional window mechanism in nBCW adapts weights locally, it can overreact to these fluctuations, leading to larger threshold RMSE compared with nIPW.


\FloatBarrier
\begin{table}[htbp]
\centering
\caption{Summary of estimated thresholds, RMSE, and coverage for nIPW, nGCW and nBCW based on 500 replications.}
\label{tab:Sim2_t1}
\setlength{\tabcolsep}{3pt}
{\fontsize{10pt}{10pt}\selectfont
\renewcommand{\arraystretch}{1.5}
\begin{tabular}{@{}ccccccccccc@{}}
\toprule
\multirow{2}{*}{$n$} & \multirow{2}{*}{Method} &
\multicolumn{2}{c}{Mean (SD)} &
\multicolumn{2}{c}{Median (IQR)} &
\multicolumn{2}{c}{RMSE} &
\multicolumn{2}{c}{Coverage} \\
\cmidrule(lr){3-4} \cmidrule(lr){5-6} \cmidrule(lr){7-8} \cmidrule(lr){9-10}
& & $\psi_1$ & $\psi_2$ & $\psi_1$ & $\psi_2$ & $\psi_1$ & $\psi_2$ & $\psi_1$ & $\psi_2$ \\
\midrule
200 & nIPW & 384.26 (96.36) & 64.74 (12.66) & 415.00 (146.25) & 68.00 (23.00) & 77.44 & 15.26 & 0.968 & 0.306 \\
    & nGCW & 384.73 (96.26) & 64.88 (12.61) & 415.00 (146.25) & 68.00 (22.00) & 76.73 & 15.12 & 0.972 & 0.304 \\
    & nBCW & 385.56 (110.74) & 64.78 (13.21) & 407.50 (181.25) & 67.50 (24.00) & 92.12 & 15.22 & 0.982 & 0.394 \\
\midrule\midrule
500 & nIPW & 387.25 (87.29) & 64.17 (13.31) & 420.00 (87.50) & 66.00 (25.25) & 63.39 & 15.83 & 0.986 & 0.602 \\
    & nGCW & 386.90 (87.37) & 64.19 (13.32) & 420.00 (91.25) & 66.00 (25.00) & 63.52 & 15.81 & 0.986 & 0.604 \\
    & nBCW & 391.55 (107.14) & 63.74 (13.50) & 425.00 (180.00) & 66.00 (27.00) & 83.55 & 16.26 & 0.986 & 0.622 \\
\midrule\midrule
1000 & nIPW & 401.69 (71.65) & 64.85 (13.55) & 425.00 (35.00) & 67.00 (25.00) & 42.49 & 15.15 & 0.988 & 0.770 \\
     & nGCW & 401.71 (71.64) & 64.85 (13.55) & 425.00 (35.00) & 67.00 (25.00) & 42.43 & 15.15 & 0.988 & 0.766 \\
     & nBCW & 406.72 (87.90) & 64.02 (13.68) & 430.00 (45.00) & 65.00 (27.00) & 56.76 & 15.98 & 0.988 & 0.834 \\
\bottomrule
\end{tabular}
}
\end{table}

Table~\ref{tab:Sim2_t2} reports results averaged over all grid points in $\mathcal{G}_{(\psi_1,\psi_2)}$. Compared to nIPW, both adaptive estimators yield substantial variance reduction across nearly the entire value surface. The effective sample sizes increase correspondingly for nGCW and nBCW, indicating more stable and less extreme weights. For the Spearman rank correlation and regime level metrics, we draw similar conclusion as in Simulation 1. Overall, nGCW provides the most favourable improvement in all measurement.
\FloatBarrier
\begin{table}[htbp]
\centering
\caption{Performance of nIPW, nGCW and nBCW over all regimes. Measurements are reported as mean (standard deviation) across 500 Monte Carlo replications. }
\label{tab:Sim2_t2}
\resizebox{\textwidth}{!}{
{\fontsize{10pt}{10pt}\selectfont
\renewcommand{\arraystretch}{1.5}
\begin{tabular}{cccccccc}
\toprule
$n$ & Method & Var & ESS & Corr & \% Var $\downarrow$ (grid) & POT & Value \\
\midrule
200  & nIPW & 896.91 (124.13) & 66.50 (24.84) & 0.402 (0.304) & ---    & 0.667 (0.196) & 554.14 (4.53) \\
     & nGCW & 779.52 (102.51) & 76.42 (28.03) & 0.404 (0.303) & 100.00 & 0.671 (0.196) & 554.23 (4.55) \\
     & nBCW & 825.84 (93.28)  & 68.12 (23.17) & 0.380 (0.292) & 92.05  & 0.641 (0.201) & 553.51 (4.50) \\
\midrule\midrule
500  & nIPW & 341.68 (48.83) & 157.19 (55.38) & 0.516 (0.245) & ---    & 0.686 (0.216) & 555.12 (4.49) \\
     & nGCW & 312.34 (43.68) & 171.84 (60.22) & 0.517 (0.245) & 100.00 & 0.687 (0.217) & 555.12 (4.51) \\
     & nBCW & 317.56 (38.34) & 161.42 (51.65) & 0.484 (0.236) & 91.24  & 0.642 (0.205) & 554.10 (4.42) \\
\midrule\midrule
1000 & nIPW & 173.90 (23.91) & 301.91 (102.21) & 0.629 (0.215) & ---    & 0.733 (0.228) & 556.52 (4.00) \\
     & nGCW & 162.90 (22.04) & 321.58 (108.63) & 0.630 (0.215) & 100.00 & 0.733 (0.228) & 556.52 (4.00) \\
     & nBCW & 162.45 (19.26) & 309.73 (95.34)  & 0.598 (0.212) & 91.09  & 0.696 (0.228) & 555.81 (4.16) \\
\bottomrule
\end{tabular}}}
\end{table}

The augmented estimators provide further gains in both efficiency and stability, as shown in the supplementary materials~D. Simulation 2 further confirms the efficiency advantages of the adaptive estimators over nIPW across multiple metrics. In particular, nGCW consistently achieves the lowest variance and rMSE while attaining the largest effective sample sizes. nBCW also improves upon nIPW, though the gains are more moderate. Augmentation further strengthens these improvements, yielding additional reductions in variance and improved stability. Overall, the adaptive weighting approaches remain robust and efficient in this HIV-aligned simulation setting.

\section{Analysis of ACTG175}
\label{sec:application}

We apply the proposed methods to the AIDS Clinical Trials Group Protocol 175 (ACTG 175) dataset, an open-access clinical trial dataset originally reported by \citep{hammer1996trial}. ACTG 175 was a randomized, double-blind, placebo-controlled trial conducted in the early 1990s to evaluate the efficacy of various nucleoside antiretroviral regimens in adults living with HIV. The trial enrolled 2,467 participants with baseline CD4 cell counts between 200 and 500 cells/mm\textsuperscript{3}, randomly assigned to one of four daily treatment arms: 1) zidovudine monotherapy, 2) zidovudine with didanosine, 3) zidovudine with zalcitabine, or 4) didanosine monotherapy. Baseline characteristics include demographic and clinical variables such as race, sex, body weight, symptoms of HIV infection, history of antiretroviral use, and Karnofsky performance score, with longitudinal follow-up measurements of CD4 count and other outcomes over a 20-week period. See supplementary materials~E for a summary of the analytic cohort.

For illustration, we restrict attention to a two-arm comparison between the dual-therapy regimens {zidovudine + didanosine} (ZDV+ddI) and {zidovudine + zalcitabine} (ZDV+ddC) as in \citep{duque2021estimation}. After excluding individuals with missing baselines or implausible CD4 values, the analytic cohort includes 1,043 participants (519 on ZDV+ddI; 524 on ZDV+ddC). The known treatment probability is 0.5 by design, but we estimate these probabilities to improve IPW efficiency \citep{henmi2004paradox}. The final propensity score model demonstrates satisfactory covariate overlap between groups, with standardized mean differences (SMDs) below 0.01 after weighting (see supplementary materials~E).

We formally specify the treatment regime family as follows: receive ZDV+ddC if baseline CD4 $\geq \psi_{\text{CD4}}$ cells/mm\textsuperscript{3} and weight $\geq \psi_{\text{W}}$ kg, and otherwise receive ZDV+ddI. We employ a clinically meaningful parameter grid with the CD4 thresholds from 200 to 600 cells/mm\textsuperscript{3} in steps of 5 units, and the weight threshold from 50 to 100 kg in steps of 1 unit. All estimators are evaluated using 500 bootstrap replications and the same performance metrics as in the simulations, except that bias terms are omitted because the true value surface is unknown. For nGCW, we use benchmark configuration $c=1.0$ and $k=0.5$. For nBCW, we use $\lambda_{\text{bias}} = 1$ and candidate windows $\delta$s \{0, 2, 4, 6, 8, 10\} for CD4 and \{0, 1, 2\} for weight.

\begin{figure}
    \centering
    \includegraphics[width=1\linewidth]{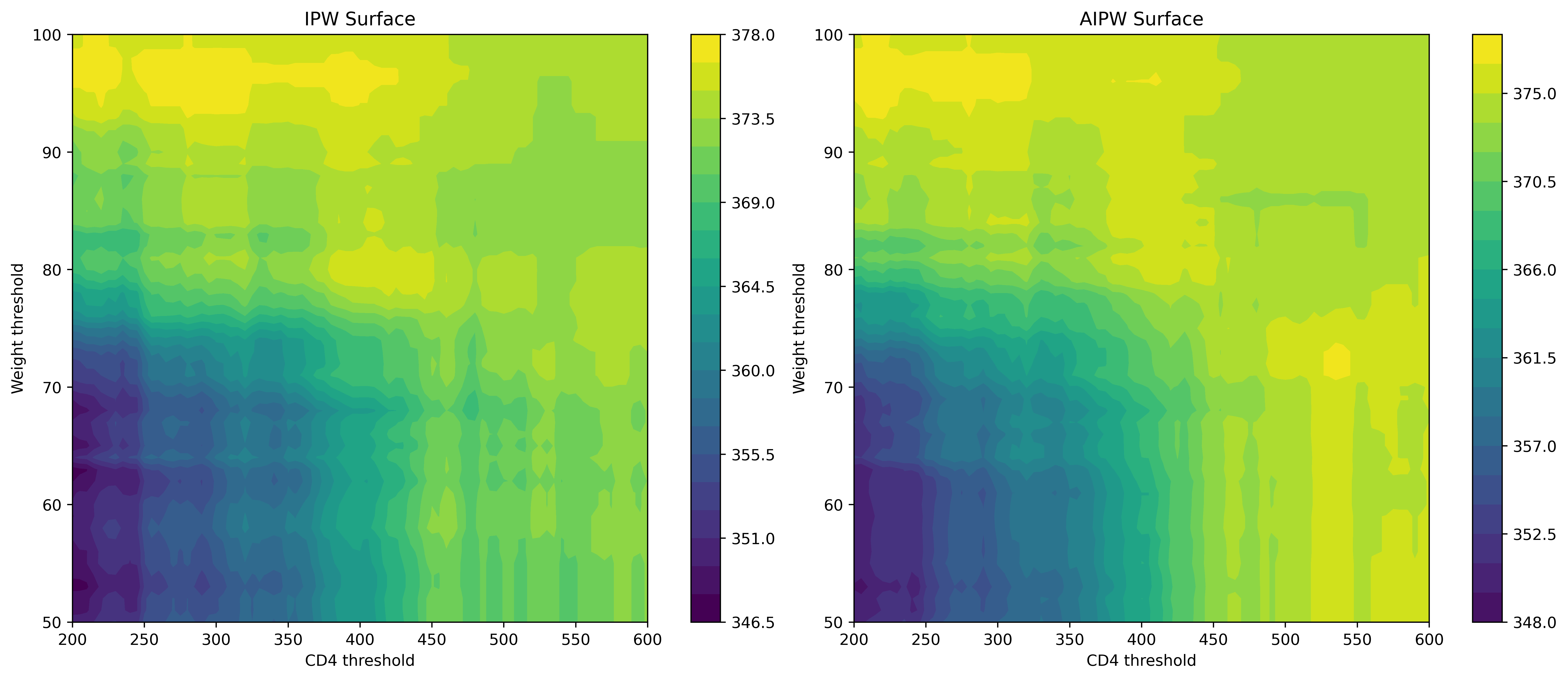}
    \caption{Estimated IPW and AIPW surfaces}
    \label{fig:IPW_AIPW_surf}
\end{figure}

Figure~\ref{fig:IPW_AIPW_surf} displays the estimated IPW and AIPW value surfaces. Under IPW, the optimal regime is identified at $(\psi_{\text{CD4}},\psi_{\text{W}}) = (280, 97)$, suggesting that ZDV+ddI should be recommended for most of the patients unless body weight is massive, aligning with the fact that ddC is associated with higher toxicity and adverse effects \citep{abrams1994comparative}. The AIPW surface shows the same general pattern but with a mild bimodal structure. The estimated optimal regime occurs at $(\psi_{\text{CD4}},\psi_{\text{W}}) = (220, 97)$, suggesting a relatively flat value region for CD4 thresholds between approximately 220 and 280, rather than a fundamentally different treatment recommendation.

Threshold and value summaries are reported in Table~\ref{tab:actg_threshold}. The median CD4 and weight thresholds are highly consistent across methods, with all medians concentrated near the dominant empirical optimum of approximately (280, 97). Mean estimates exhibit greater dispersion, particularly for the augmented estimators due to the bimodality of the estimated surface. Regime values are tightly clustered across all methods, with medians between 377 and 379 and standard deviations around 6 to 8. Together, these results confirm that adaptive weighting methods yield stable and consistent decision rules.

\FloatBarrier
\begin{table}[htbp]
\centering
\caption{Summary of estimated optimal thresholds and value across 500 bootstrap replications. Results are reported as Median (IQR) and Mean (SD).}
\label{tab:actg_threshold}
\setlength{\tabcolsep}{3pt}  
{\fontsize{10pt}{11.5pt}\selectfont
\renewcommand{\arraystretch}{1.5}
\begin{tabular}{lcccccc}
\toprule
\multirow{2}{*}{Method} & 
\multicolumn{2}{c}{$\hat{\psi}_{\text{CD4}}$} & 
\multicolumn{2}{c}{$\hat{\psi}_{\text{W}}$} & 
\multicolumn{2}{c}{$\hat V$} \\ 
\cmidrule(lr){2-3} \cmidrule(lr){4-5} \cmidrule(lr){6-7}
 & Median (IQR) & Mean (SD) & Median (IQR) & Mean (SD) & Median (IQR) & Mean (SD) \\
\midrule
nIPW  & 280.00 (175.00) & 309.12 (88.48) & 96.00 (17.00) & 90.61 (9.76)  & 378.61 (8.09) & 378.93 (6.01) \\
nAIPW & 280.00 (200.00) & 334.52 (119.37) & 91.00 (16.00) & 86.31 (12.86) & 378.06 (8.15) & 378.70 (5.87) \\
\midrule
\midrule
nGCW  & 280.00 (176.25) & 310.03 (88.33) & 96.00 (17.00) & 90.48 (9.85) & 377.28 (7.95) & 377.68 (5.85) \\
nAGCW & 280.00 (200.00) & 332.40 (118.31) & 93.00 (16.00) & 86.70 (12.53) & 376.81 (7.84) & 377.45 (5.72) \\
\midrule
\midrule
nBCW  & 295.00 (120.00) & 329.26 (85.48) & 95.00 (17.00) & 88.55 (9.94) & 378.97 (7.92) & 379.35 (5.89) \\
nABCW & 292.50 (215.00) & 337.18 (123.64) & 89.00 (16.00) & 86.06 (13.00) & 377.98 (8.13) & 378.62 (5.87) \\
\bottomrule
\end{tabular}}
\end{table}

Table~\ref{tab:actg_surface} evaluates performance across the full value surface. nGCW and nAGCW estimators achieve the largest effective sample sizes and the most substantial variance reduction, with improvements at all grid points. nBCW and nABCW estimators also exhibit meaningful improvements over nIPW. Augmentation consistently lowers variance for all methods, and the analytical variance for nAGCW closely matches its Monte Carlo estimate, consistent with the simulation results. {The final column reports the Spearman rank correlation between the estimated mean value surface over all bootstrap replications and the mean n(A)IPW value surface, which is treated as the reference surface. The high correlations indicate that the partial-compatibility estimators stabilize the value surface while largely preserving the relative ordering of candidate treatment regimes. Sensitivity analyses examining alternative hyperparameter choices yield similar conclusions, as shown in supplementary materials~E.}

\FloatBarrier
\begin{table}[htbp]
\centering
\caption{Performance comparison over the value surface across 500 bootstraps, results are reported as mean (SD)}
\label{tab:actg_surface}
{\fontsize{12pt}{10pt}\selectfont
\renewcommand{\arraystretch}{1.5}
\begin{tabular}{lccccc}
\toprule
Method & MC Var & Analytical Var & ESS &  \% grid (Var $\downarrow$) & Corr \\
\midrule
nIPW  & 36.61 (2.69) & -- & 515.21 (8.26)  & -- & -- \\
nAIPW & 34.22 (3.19) & -- & 515.21 (8.26) & -- & --\\
\midrule
\midrule
nGCW  & 34.61 (2.49) & 39.20 (4.20) & 548.30 (8.06) & 100.00 & 1.000\\
nAGCW & 32.49 (2.92) & 32.66 (3.09) & 548.30 (8.06) & 100.00 & 1.000 \\
\midrule
\midrule
nBCW  & 35.62 (3.35) & -- & 527.62 (21.06) & 81.14 & 0.996\\
nABCW & 33.85 (3.59) & -- & 523.96 (17.98) & 64.63 & 0.997\\
\bottomrule
\end{tabular}}
\end{table}

\begin{figure}[H]
    \centering
    \begin{subfigure}{0.48\linewidth}
        \centering
        \includegraphics[width=\linewidth]{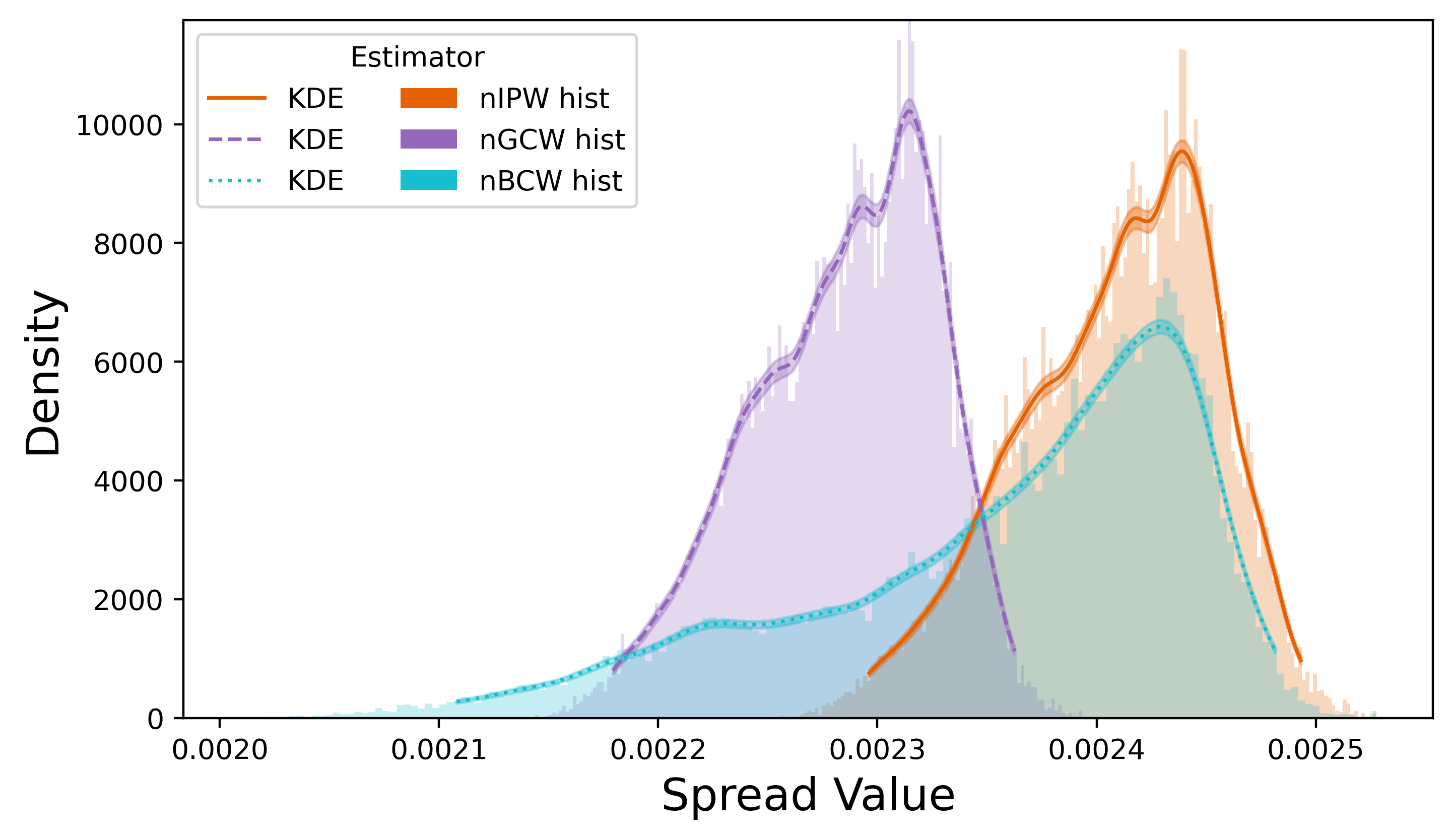}
        \caption{Non-augmented estimators}
        \label{fig:weight_dist_IPW}
    \end{subfigure}
    \hfill
    \begin{subfigure}{0.48\linewidth}
        \centering
        \includegraphics[width=\linewidth]{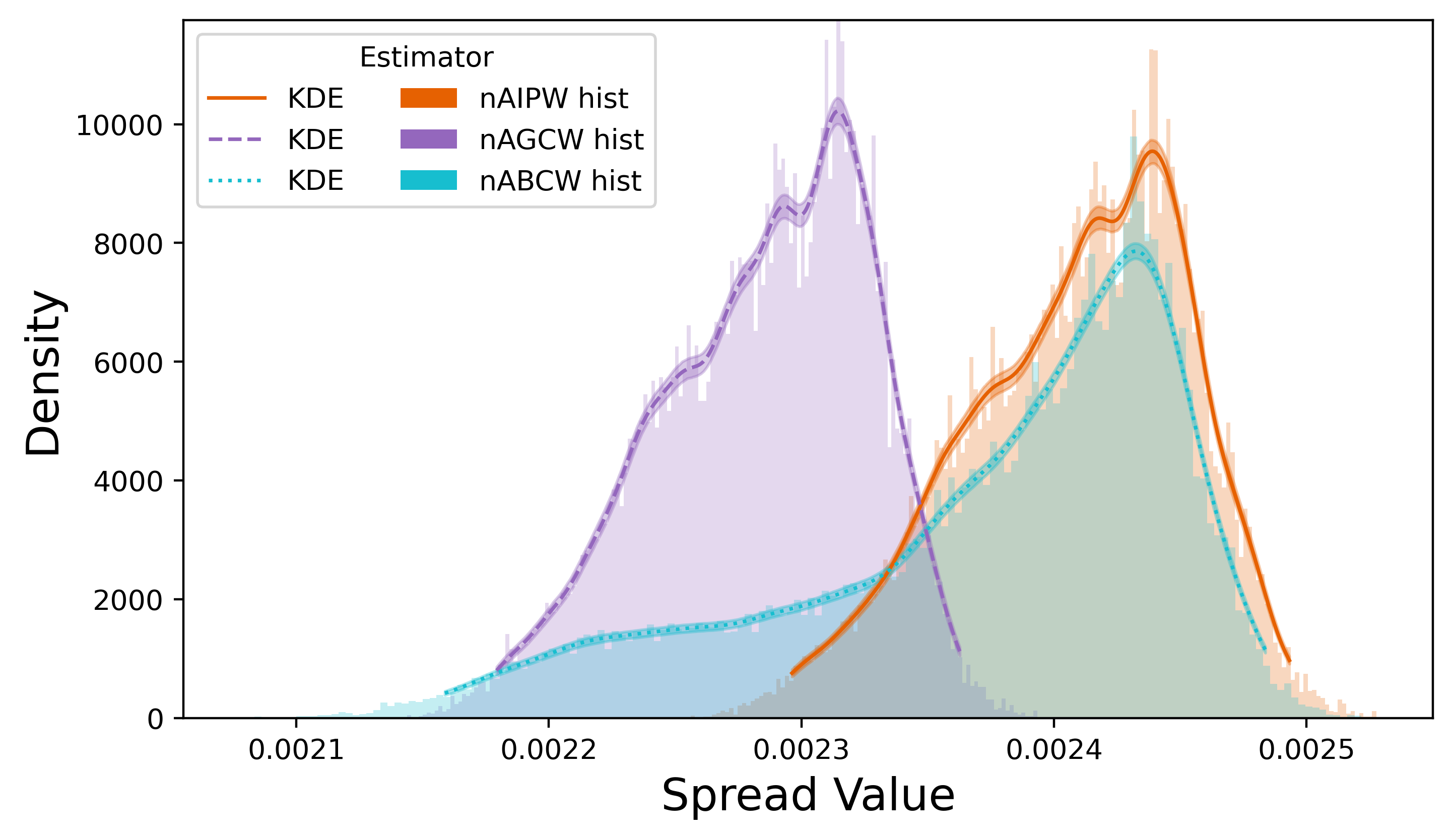}
        \caption{Augmented estimators}
        \label{fig:weight_dist_AIPW}
    \end{subfigure}

    \caption{Comparison of $99^{\text{th}} - 1^{\text{st}}$ percentile weight spread across all estimators.}
    \label{fig:weight_distributions}
\end{figure}

To illustrate weight stabilization directly, Figure~\ref{fig:weight_distributions} summarizes the $99^{\text{th}} - 1^{\text{st}}$ percentile weight range across all grid points using histograms and kernel density estimation (KDE). nGCW and nAGCW estimators show a uniform downward shift of the distribution. 
The BCW estimators also reduce weight magnitudes relative to IPW, but the degree of shrinkage varies across the grid because the window widths change with grid points, leading to a broader overall distribution with a long left tail of smaller spread. This wider spread reflects a locally adaptive heterogeneous weighting rather than instability and is consistent with BCW’s design as an intermediate method between IPW and GCW.

Together, these empirical findings highlight the practical value of our proposed adaptive weighting frameworks in real clinical data settings. All estimators preserved the fundamental objective of DTR estimation by consistently recovering the optimal treatment regime. Beyond this, the generalized compatibility estimators achieved the most substantial gains, indicating efficient borrowing of information across the entire value surface. The BCW estimators produced modest but meaningful improvements and may be preferable when clinical interpretability of inclusion criteria is prioritized. In resource-limited clinical studies where sample sizes or full compatibility are constrained, such efficiency gains are especially valuable for the practical implementation of individualized treatment strategies.

\section{Discussion}
\label{sec:discussion}

We proposed a general framework for improving inverse probability weighting–based estimators in value estimation for dynamic treatment regimes. We introduced two adaptive estimators: the generalized compatibility weighting and the bootstrap compatibility windowing, together with their normalized and augmented extensions. GCW provides a probability-based relaxation of the hard compatibility indicator, allowing individuals with high compatibility propensity to contribute even when their observed treatment deviates from the regime. BCW defines compatibility through covariate proximity and uses a bootstrap procedure to select an optimal window that balances bias and variance. Our proposed methods require minimal modification to existing causal estimators and preserve the identifying assumptions while (i) maintaining consistency and interpretability, (ii) reducing finite-sample variance, and (iii) integrating seamlessly with both IPW and AIPW estimation frameworks.

Simulation studies demonstrated the finite-sample benefits of our approach across a wide range of settings, including multi-stage and multi-threshold regimes and scenarios with low compatibility and no tailored treatment effect. GCW and BCW achieved substantial variance reduction, particularly in regimes with true optima located near decision boundaries where standard IPW and AIPW estimators are most unstable. Augmentation further improved both accuracy and inference, with analytical variances closely matching Monte Carlo estimates. Our application to the ACTG175 dataset illustrated the practical advantages of these stabilized estimators in a real HIV treatment setting. The proposed methods produced substantial gains in effective sample size and significant improvements in weight stability, resulting in more robust and clinically sensible regime identification.

One limitation of our approach is the possibility of small finite-sample bias when the compatibility hyperparameters or window sizes are not chosen properly; however, in the DTR context, this bias has minimal practical impact because the optimal regime is determined by relative rather than absolute value levels, and our simulations show that regime selection remains stable even when slight bias is introduced. In addition, the bootstrap-based selection in BCW can be computationally intensive for fine regime grids or large numbers of bootstrap replicates; nonetheless, computational efficiency may be improved by exploiting the fact that nearby regime points share similar local structure.

Beyond dynamic treatment regimes, the variance inflation induced by binary compatibility indicators also arises in other areas of causal inference and sequential decision-making. Examples include subgroup identification and individualized treatment rule learning \citep{zhao2012estimating}, where sharp subgroup boundaries can produce highly variable estimators, and reinforcement learning/off-policy evaluation \citep{thomas2016data}, where deterministic policies induce indicator-based importance weights. In settings defined by deterministic decision rules, including threshold-based clinical actions, risk-based recommendations, or resource allocation policies \citep{kamath2001model}, 
soft compatibility weighting offers an alternative to improve efficiency while preserving the intended estimand.

Future extensions of this work may explore more flexible forms of partial compliance, such as nonlinear kernels or higher-order borrowing strategies to further reduce variance while preserving asymptotic unbiasedness. Another interesting direction is the identification of subgroups that may follow different optimal treatment regimes. By producing smoother and more stable value surfaces, soft compatibility weighting may facilitate more reliable detection of heterogeneous treatment effects.

In conclusion, we have proposed a broadly applicable and computationally tractable framework for stabilizing IPW- and AIPW-based value estimators through adaptive weighting. Our results demonstrate that modest modifications to the weight structure can yield substantial gains in robustness and efficiency. Given the widespread use of deterministic rule indicators in causal inference, precision medicine, and sequential decision-making, greater attention should be paid to soft compatibility weighting methods to improve estimation and decision quality in a wide range of applications beyond dynamic treatment regimes.


\section{Data Availability Statement}
The data that support the findings of this study are openly available in the \texttt{ACTG175} dataset in the R package \texttt{speff2trial}, available at \url{https://cran.r-project.org/package=speff2trial}. 

\section{Acknowledgements}
This work is supported by a Discovery Grant (DAS) from the Natural Sciences and Engineering Research Council of Canada and funding from the Fonds de recherches du Qu\'ebec - Sant\'e (EEMM). EEMM is a CIHR Canada Research Chair (Tier 1) in Statistical Methods for Precision Medicine. 


\section{Disclosure Statement}
The authors have no conflicts of interests to declare.


\section{Supplementary Materials}
The supplementary materials are divided into five appendices. Appendix A details the causal assumptions. Appendix B provides proofs for the theoretical results in Section 3 of the main paper. Appendices C, D, and E present additional figures and results for Simulations 1, 2, and the real-data application, respectively.

\bibliography{main.bib}

\clearpage
\appendix
\begin{center}
{\LARGE\bfseries Supplementary Materials}
\end{center}
\vspace{1em}

\section{Standard Causal Assumptions}
We impose the following standard assumptions:
\begin{itemize}
    \item[(A1)] Consistency \citep{robins1994correcting}: The observed outcome equals the potential outcome under the treatment sequence that was received. That is, for each individual $i$ and treatment sequence $\bm{a}$, if $\bar{\bm{A}}_i = \bm{a}$, then $Y_i = Y^{\bm{a}}_i$. 
    \item [(A2)] Stable Unit Treatment Value Assumption (SUTVA) \citep{rubin1980randomization}: The potential outcome for any individual is unaffected by the treatment assignment of other individuals. That is, $Y_i(\bm{a}_i, \bm{a}_j) = Y_i(\bm{a}_i)$ for all $\bm{a}_j$, where $\bm{a}_i$ is individual $i$'s treatment and $\bm{a}_{j}$ denotes treatments for others.
    \item[(A3)] Sequential Ignorability \citep{robins1997causal}: At each stage $t$, treatment assignment is conditionally independent of future potential outcomes given the observed history: $Y^{\bm{a}} \perp A_t \mid \bar{\bm{X}}_t, \bar{\bm{A}}_{t-1}$.
    \item[(A4)] Positivity \citep{rubin1978bayesian, rosenbaum1983central}: For each stage $t$ and all possible values of covariates and treatment history, there exists a positive probability of receiving any feasible treatment: $P(A_t = a \mid \bar{\bm{X}}_t = \bar{\bm{x}}_t, \bar{\bm{A}}_{t-1} = \bar{\bm{a}}_{t-1})  > 0$ for all $a \in \mathcal{A}_t$, and all $(\bar{\bm{x}}_t, \bar{\bm{a}}_{t-1})$ are in the support.
\end{itemize}
Additionally, we require the following regularity conditions:
\begin{itemize}
    \item[(A5)] Independent and Identically Distributed Sampling: Observations $\{(Y_i, \bar{\bm{X}}_i, \bar{\bm{A}}_i)\}_{i=1}^n$ are i.i.d. draws from the population distribution.
    \item[(A6)] Bounded Outcomes: The outcome is uniformly bounded: $|Y_i| \leq C < \infty$ almost surely for some constant $C$.
\end{itemize}

\section{Proofs of Lemma and Theorem}
\subsection{Proof of Lemma 1}
\label{appendix:prooflem1}
\begin{proof}
Recall the definition $w_t^{\text{G}} = \dfrac{m_t}{D_t}$
where
$$m_{t} := m_{t}(\bar{\bm{X}}_{t}, \bar{\bm A}_{t-1}, A_t) = \mathbb{I}(A_{t} = d_t) + \gamma_{t} \cdot \mathbb{I}(A_{t} \neq d_t),$$
$$D_{t} := D_{t}(\bar{\bm{X}}_{t},\bar{\bm{A}}_{t-1}) = P(A_{t} = d_{t} \mid \bar{\bm{X}}_{t}, \bar{\bm{A}}_{t-1}) + \gamma_{t} \cdot P(A_{t} \neq d_{t} \mid \bar{\bm{X}}_{t}, \bar{\bm{A}}_{t-1}).$$
For any $t = 1,\, \ldots \, T$, we have for weight $w_t^{\text{G}}$, 
\begin{align*}
\mathbb{E}[w_{t}^{\text{G}}] 
&= \mathbb{E}\left[ \mathbb{E}[\frac{m_t}{D_t} \mid \bar{\bm X}_{t}, \bar{\bm A}_{t-1}]\right]\\
&= \mathbb{E}\left[ \frac{1}{D_t}\mathbb{E}[m_t \mid \bar{\bm X}_{t}, \bar{\bm A}_{t-1}]\right] \\
&= \mathbb{E}\left[ \frac{1}{D_t}\left(1 \cdot P(A_t = d_t \mid \bar{\bm X}_{t}, \bar{\bm A}_{t-1}) + \gamma_t \cdot P(A_t \neq d_t \mid \bar{\bm X}_{t}, \bar{\bm A}_{t-1}) \right)\right] \\
&= \mathbb{E}\left[ \frac{D_t}{D_t}\right] = \mathbb{E}[1] = 1.
\end{align*}
The above proves that the expectation of $w_t^{\text{G}}$ is 1 for each stage $t$, or a stronger statement, $\mathbb{E}\left[ w_t^{\text{G}} \mid \bar{\bm X}_{t}, \bar{\bm A}_{t-1} \right]=1$ \\
If we consider $\bar{w}_t^{\text{G}}$, the product of weights up to stage $t$, then we have
\begin{align*}
\mathbb{E}[\bar{w}_{t}^{\text{G}}] &= \mathbb{E}\left[w_t^{\text{G}} \left\{ \prod_{s=1}^{t-1} w_s^{\text{G}} \right\} \right]\\
&= \mathbb{E}\left[\mathbb{E} \left[w_t^{\text{G}} \left\{ \prod_{s=1}^{t-1} w_s^{\text{G}} \right\} \, \middle| \, \bar{\bm X}_t, \bar{\bm A}_{t-1}  \right] \right] \\
&= \mathbb{E}\left[ \left\{ \prod_{s=1}^{t-1} w_s^{\text{G}}\right\}\mathbb{E} \left[w_t^{\text{G}} \mid \bar{\bm X}_t, \bar{\bm A}_{t-1}  \right] \right] \\
&= \mathbb{E}\left[ \left\{ \prod_{s=1}^{t-1} w_s^{\text{G}}\right\} 1 \right]= \mathbb{E}[\bar{w}_{t-1}^{\text{G}}].
\end{align*}
Recursively, we have
$\mathbb{E}[\bar{w}_{t}^{\text{G}}] = \mathbb{E}[\bar{w}_{t-1}^{\text{G}}] = \mathbb{E}[\bar{w}_{t-2}^{\text{G}}] = \ldots = \mathbb{E}[w_1^{\text{G}}] = 1.$
\end{proof}

\newpage
\subsection{Proof of Lemma 2}
\label{appendix:prooflem2}
\begin{proof}
Recall the generalized adherence weighting estimator 
$$\widehat{V}_{\text{GCW}} = \dfrac{1}{n} \sum_{i=1}^n \left\{\prod_{t=1}^T w_{t,i}^{\text{G}} \right\}  Y_i = \dfrac{1}{n} \sum_{i=1}^n \bar{w}_{T,i}^{\text{G}}  Y_i.$$
Recall the definition that we omit the subscript $T$ when it represents full covariates or full treatment trajectory.\\
Under (A1)–(A5), we have
\begin{align*}
\mathbb{E}\left[\widehat{V}_\text{GCW}(d)\right]
&= \mathbb{E}\left[\bar{w}_T^{\text{G}} Y\right]\\
&= \mathbb{E}\left[\left\{\prod_{t=1}^T w_{t}^{\text{G}} \right\} Y\right]\\
&= \mathbb{E} \left[ \mathbb{E}\left[\left\{\prod_{t=1}^T w_{t}^{\text{G}} \right\} Y  \, \middle| \, \bar{\bm X}\right] \right] \\
&= \mathbb{E} \left[ \sum_{\bar{\bm a}} P(\bar{\bm A} = \bar{\bm a} \mid \bar{\bm X}) \left\{\prod_{t=1}^T \frac{m_t(\bar{\bm X}_t, \bar{\bm a}_{t-1}, a_t)}{D_t(\bar{\bm X}_t, \bar{\bm a}_{t-1})}\right\} \mathbb{E}[Y\mid \bar{\bm X},\bar{\bm a}]\right] \\
&= \mathbb{E} \left[ \sum_{\bar{\bm a}} \left\{\prod_{t=1}^T P(A_t = a_t \mid \bar{\bm X}_t, \bar{\bm a}_{t-1}) \frac{m_t(\bar{\bm X}_t, \bar{\bm a}_{t-1}, a_t)}{D_t(\bar{\bm X}_t, \bar{\bm a}_{t-1})}\right\} \mathbb{E}[Y\mid \bar{\bm X},\bar{\bm a}]\right]. \tag{*}
\end{align*}
Define $p_t := P(A_t = d_t \mid \bar{\bm X}_t, \bar{\bm a}_{t-1})$, then define
\[
F_t(\bar{\bm X}_t, \bar{\bm a}_{t-1}, a_t) =
\begin{cases}
\dfrac{p_t}{D_t}, & a_t = d_t, \\[1.2em]
\dfrac{\gamma_t\, P\!\left(A_t = a_t \,\middle|\, \bar{\bm X}_t, \bar{\bm a}_{t-1}\right)}{D_t}, & a_t \neq d_t.
\end{cases}
\]
Then we can rewrite (*) as
\begin{align*}
    (*) &= \mathbb{E} \left[ \sum_{\bar{\bm a}} \left\{\prod_{t=1}^T F_t(\bar{\bm X}_t,\bar{\bm a}_{t-1}, a_t) \right\} \mathbb{E}[Y\mid \bar{\bm X},\bar{\bm a}]\right]  \\
    &= \mathbb{E} \left[ \left\{\prod_{t=1}^T F_t(\bar{\bm X}_t, \bar{\bm d}_{t-1}, d_t) \right\} \mathbb{E}[Y\mid \bar{\bm X}, \bar{\bm d}] + \sum_{\bar{\bm a} \neq \bar{\bm d}}  \left\{\prod_{t=1}^T F_t(\bar{\bm X}_t,\bar{\bm a}_{t-1}, a_t) \right\} \mathbb{E}[Y\mid \bar{\bm X},\bar{\bm a}] \right] \\
    &= \mathbb{E} \left[ \left\{\prod_{t=1}^T \frac{p_t}{D_t} \right\} \mathbb{E}[Y\mid \bar{\bm X}, \bar{\bm d}] + \sum_{\bar{\bm a} \neq \bar{\bm d}}  \left\{\prod_{t=1}^T F_t(\bar{\bm X}_t,\bar{\bm a}_{t-1}, a_t) \right\} \mathbb{E}[Y\mid \bar{\bm X},\bar{\bm a}] \right].  \tag{**}
\end{align*}
Define $$\theta(\bar{\bm X}) = \left\{\prod_{t=1}^T \frac{p_t}{D_t} \right\},$$ and notice that 
$$
\sum_{a_t} F_t(\bar{\bm X}_t, \bar{\bm a}_{t-1}, a_t) = \frac{p_t}{D_t} + \frac{\gamma_t}{D_t} \sum_{a_t \neq d_t} P(A_t =a_t \mid \bar{\bm X}_t, \bar{\bm a}_{t-1}) = \frac{p_t + \gamma_t(1-p_t)}{D_t} = 1.$$
Thus, we have
\begin{align*}
    &\sum_{\bar{\bm a} \neq \bar{\bm d}}  \left\{\prod_{t=1}^T F_t(\bar{\bm X}_t,\bar{\bm a}_{t-1}, a_t) \right\}\\
    &= \sum_{\bar{\bm a}}  \left\{\prod_{t=1}^T F_t(\bar{\bm X}_t,\bar{\bm a}_{t-1}, a_t) \right\} - \left\{\prod_{t=1}^T \frac{p_t}{D_t} \right\} \\
    &= \left\{\sum_{a_1} \sum_{a_2} \ldots \sum_{a_T} F_1(X_1, a_1) F_2(\bar{\bm X}_2, a_1, a_2) \ldots F_T(\bar{\bm X}_T, \bar{\bm a}_{T-1}, a_T) \right\} - \theta(\bar{\bm X})\\
    &= \left\{\left( \sum_{a_1} \sum_{a_2} \ldots F_1(X_1, a_1) F_2(\bar{\bm X}_2, a_1, a_2) \ldots \right) \sum_{a_T} F_T(\bar{\bm X}_T, \bar{\bm a}_{T-1}, a_T) \right\}- \theta(\bar{\bm X})\\
    &=\left\{ \left( \sum_{a_1} \sum_{a_2} \ldots F_1(X_1, a_1) F_2(\bar{\bm X}_2, a_1, a_2) \ldots \right)  \cdot 1 \right\} - \theta(\bar{\bm X})\\
    &= 1-\theta(\bar{\bm X}) \quad \text{by the recursive marginalization.}
\end{align*}
Then we can write (**) as
\begin{align*}
    &\mathbb{E} \left[ \theta(\bar{\bm X}) \mathbb{E}[Y^d \mid \bar{\bm X}] + \left\{\sum_{\bar{\bm a} \neq \bar{\bm d}} \prod_{t=1}^T F_t(\bar{\bm X}_t,\bar{\bm a}_{t-1}, a_t) \right\} \sum_{\bar{\bm a} \neq \bar{\bm d}} \frac{\prod_{t=1}^T F_t(\bar{\bm X}_t,\bar{\bm a}_{t-1}, a_t) \mathbb{E}[Y\mid \bar{\bm X},\bar{\bm a}]}{\sum_{\bar{\bm a} \neq \bar{\bm d}} \prod_{t=1}^T F_t(\bar{\bm X}_t,\bar{\bm a}_{t-1}, a_t)} \right]\\[1.2em]
    =\,& \mathbb{E} \left[\theta(\bar{\bm X}) \mathbb{E}[Y^d \mid \bar{\bm X}] + (1- \theta(\bar{\bm X})) \mathbb{E}[Y^{\neg d} \mid \bar{\bm X}] \right],
\end{align*}
where $$\mathbb{E}[Y^{\neg d} \mid \bar{\bm X}] =\sum_{\bar{\bm a} \neq \bar{\bm d}} \frac{\prod_{t=1}^T F_t(\bar{\bm X}_t,\bar{\bm a}_{t-1}, a_t) \mathbb{E}[Y\mid \bar{\bm X},\bar{\bm a}]}{\sum_{\bar{\bm a} \neq \bar{\bm d}} \prod_{t=1}^T F_t(\bar{\bm X}_t,\bar{\bm a}_{t-1}, a_t)}.$$
Therefore, the bias of the GCW estimator is defined as 
\begin{align*}
\mathbb{E}\!\left[\widehat{V}_\text{GCW}(d)\right] - \mathbb{E}[Y^d]
&= \mathbb{E}\left[
\theta(\bar{\bm X})\,\mathbb{E}[Y^{d}\mid \bar{\bm X}]
+ \big(1-\theta(\bar{\bm X})\big)\,\mathbb{E}[Y^{\neg d}\mid \bar{\bm X}]
- \mathbb{E}[Y^{d}\mid \bar{\bm X}]
\right] \\
&= \mathbb{E}\left[
\big(1-\theta(\bar{\bm X})\big)\,
\big(\mathbb{E}[Y^{\neg d}\mid \bar{\bm X}] - \mathbb{E}[Y^{d}\mid \bar{\bm X}]\big)
\right].
\end{align*}
By the definition of $\theta(\bar{\bm X})$, we have $\theta(\bar{\bm X}) = 1$ iff $\gamma_t=0, \, \forall t$. That is, we recover the IPW estimator as $\gamma_t \to 0$, which is known to be unbiased. 
\end{proof}

\newpage
\subsection{Proof of Theorem 1}
\begin{proof}
\medskip
\noindent\textbf{Consistency of the unnormalized estimator.}
Let $\gamma_t$ be chosen that for all $\bar{\bm X}$,
\[
1- \frac{p_t}{D_t} \leq \epsilon_t = \frac{\epsilon_n}{T}, 
\qquad \epsilon_n=c\,n^{-k}\to 0,
\]
where $p_t = P(A_t = d_t \mid \bar{\bm X}_t, \bar{\bm a}_{t-1})$, $D_t = p_t + \gamma_t (1-p_t)$ and $c>0,\,k>0$. Then, 
$$1-\theta(\bar{\bm X}) \leq \sum_{t=1}^T (1-\frac{p_t}{D_t}) \leq T \cdot \frac{\epsilon_n}{T} = \epsilon_n.$$
By Lemma~\ref{appendix:prooflem2},
\[
\mathbb{E}\!\left[\widehat{V}_\text{GCW}(d)\right]-\mathbb{E}[Y^d]
=\mathbb{E}\left[
\big(1-\theta(\bar{\bm X})\big)\,
\Big(\mathbb{E}[Y^{\neg d}\mid \bar{\bm X}]-\mathbb{E}[Y^{d}\mid \bar{\bm X}]\Big)
\right].
\]
Given that outcomes are uniformly bounded (A6), we have $ | \mathbb{E}[Y^{\neg d}\mid \bar{\bm X}]-\mathbb{E}[Y^{d}\mid \bar{\bm X}] | \leq C_Y$ for some finite constant $C_Y >0$. Since $1-\theta(\bar{\bm X})\le \epsilon_n$ by construction,
\begin{equation*}
    \left| \mathbb{E}\!\left[\widehat{V}_\text{GCW}(d)\right]-\mathbb{E}[Y^d] \right| = \left| \mathbb{E}\!\left[\bar{w}_T^{\text{G}} Y\right]-\mathbb{E}[Y^d] \right|
\;\le\; C_Y \cdot \epsilon_n \;\xrightarrow[n\to\infty]{}\; 0.
\tag{*}
\end{equation*}
By the Weak Law of Large Numbers (WLLN), we have:
\begin{equation*}
    \widehat{V}_{\text{GCW}}(d) = \frac{1}{n} \sum_{i=1}^n \bar{w}_{T,i}^{\text{G}} Y_i \xrightarrow{p} \mathbb{E}[\bar{w}_{T}^{\text{G}} Y].
\tag{**}
\end{equation*}
Using the triangle inequality, we write:
\begin{align*}
\left| \widehat{V}_{\text{GCW}}(d) - \mathbb{E}[Y^d] \right|
&\leq \underbrace{\left| \widehat{V}_{\text{GCW}}(d) - \mathbb{E}[\bar{w}_{T}^{\text{G}} Y] \right|}_{\xrightarrow{} \text{ 0 by (**)}}
+ \underbrace{\left| \mathbb{E}[\bar{w}_{T}^{\text{G}} Y] - \mathbb{E}[Y^d] \right|}_{\to \text{ 0 by (*)}}.
\end{align*}
Therefore, the full expression converges in probability to zero:
\[
\left| \widehat{V}_{\text{GCW}}(d) - \mathbb{E}[Y^d] \right| \xrightarrow{p} 0,
\]
which implies:
\[
\widehat{V}_{\text{GCW}}(d) \xrightarrow{p} \mathbb{E}[Y^d].
\]

\medskip
\noindent\textbf{Consistency of the normalized estimator.}
\[
\widehat{V}_{\text{nGCW}}(d) = \frac{\sum_{i=1}^n \bar{w}_{T,i}^{\text{G}} Y_i}{\sum_{i=1}^n \bar{w}_{T,i}^{\text{G}}}.
\]
By Lemma~\ref{appendix:prooflem1}, we have $\mathbb{E}[\bar{w}_{T}] = 1$, so by WLLN:
\[
\frac{1}{n} \sum_{i=1}^n \bar{w}_{T,i}^{\text{G}} \xrightarrow{p} 1
\quad \text{and} \quad
\frac{1}{n} \sum_{i=1}^n \bar{w}_{T,i}^{\text{G}} Y_i \xrightarrow{p} \mathbb{E}[\bar{w}_{T}^{\text{G}} Y].
\]
Then by Slutsky’s Theorem,
\[
\widehat{V}_{\text{nGCW}}(d) = \frac{\frac{1}{n} \sum_{i=1}^n \bar{w}_{T,i}^{\text{G}} Y_i}{\frac{1}{n} \sum_{i=1}^n \bar{w}_{T,i}^{\text{G}}} = \frac{\sum_{i=1}^n \bar{w}_{T,i}^{\text{G}} Y_i}{\sum_{i=1}^n \bar{w}_{T,i}^{\text{G}}} \xrightarrow{p} \mathbb{E}[\bar{w}_T^{\text{G}} Y] .
\]
Again by (*), $\mathbb{E}[\bar{w}_T^{\text{G}} Y] \xrightarrow{p} \mathbb{E}[Y^d]$. Thus, we have
\[
\widehat V_{\text{nGCW}}(d)\;\xrightarrow{p}\; \mathbb{E}[Y^d].
\]
\end{proof}

\newpage
\subsection{Proof of Theorem 2}
\begin{proof}
We prove for the normalized IPW and GCW estimators to match what we used in the simulations.\\
Denote the normalized IPW estimator as
\[
\widehat{\phi}_{\text{nIPW}} = \widehat{V}_{\text{nIPW}} 
= \frac{\sum_{i=1}^n \bar{w}_{T,i}^{\text{I}} Y_i}{\sum_{i=1}^n \bar{w}_{T,i}^{\text{I}}}, \quad \text{where} \quad 
\bar{w}_{T,i}^{\text{I}} = \prod_{t=1}^T w_{t,i}^{\text{I}}, \quad w_{t,i}^{\text{I}}= \frac{\mathbb{I}(A_{t,i}=d_{t,i})}{P(A_{t,i} \mid \bar{\bm X}_{t,i}, \bar{\bm A}_{t-1,i})}.
\]
Denote the normalized GCW estimator as
\[
\widehat{\phi}_{\text{nGCW}} = \widehat{V}_{\text{nGCW}} 
= \frac{\sum_{i=1}^n \bar{w}_{T,i}^{\text{G}} Y_i}{\sum_{i=1}^n \bar{w}_{T,i}^{\text{G}}}, 
\quad \text{where} \quad 
\bar{w}_{T,i}^{\text{G}} = \prod_{t=1}^T w_{t,i}^{\text{G}}, \quad w_{t,i}^{\text{G}} = \frac{m_{t,i}}{D_{t,i}}.
\]
We analyze the asymptotic variance of both estimators using estimating equation theory.  
Let \(\phi\) denote the true regime value. Define the estimating function:
\[
\sum_{i=1}^n U_i = \sum_{i=1}^n \bar{w}_{T,i} (Y_i - \phi) = 0.
\]
The asymptotic variance is
\[
\operatorname{Var}(\widehat{\phi}) = \frac{1}{n} \cdot \frac{I}{J^2},
\]
where
\[
I = \mathbb{E}[U_i^2] = \mathbb{E}[\bar{w}_{T,i}^2 (Y_i - \phi)^2], 
\qquad 
J = \mathbb{E}\!\left[-\frac{\partial}{\partial \phi} U_i\right] = \mathbb{E}[\bar{w}_{T,i}].
\]
For both estimators, \(\mathbb{E}[\bar{w}_{T,i}] = 1\) by construction, so the asymptotic variance reduces to
\[
\operatorname{Var}(\widehat{\phi}) = \frac{1}{n} I =\frac{1}{n} \cdot \mathbb{E}[\bar{w}_{T,i}^2 (Y_i - \phi)^2].
\]

\newpage
\noindent\textbf{Variance of the nIPW estimator.}
Recall our definition $p_t = P(A_{t}=d_t \mid \bar{\bm X}_{t}, \bar{\bm a}_{t-1})$, $\phi = \mathbb{E}[Y^d]$, then
\begin{align*}
    I_{\text{IPW}} 
    &= \mathbb{E}\left[ \bar{w}_{T}^{\text{I,2}} (Y - \phi)^2 \right] \\
    &= \mathbb{E}\left[ \left\{\prod_{t=1}^T w_{t}^{\text{I}}\right\}^2 (Y-\phi)^2  \right] \\
    &= \mathbb{E} \left[ \mathbb{E} \left[ \left\{\prod_{t=1}^T w_{t}^{\text{I,2}}\right\} (Y-\phi)^2  \;\middle|\; \bar{\bm X} \right]\right]\\
    &= \mathbb{E} \left[ \sum_{\bar{\bm a}} P(\bar{\bm A} = \bar{\bm a} \mid \bar{\bm X}) \left\{\prod_{t=1}^T w_{t}^{\text{I,2}}\right\} \mathbb{E}[(Y-\phi)^2 \mid \bar{\bm X}, \bar{\bm a}] \right]\\
    &= \mathbb{E} \left[ \sum_{\bar{\bm a}} \left\{\prod_{t=1}^T P(A_t = a_t \mid \bar{\bm X}_t, \bar{\bm a}_{t-1}) \left(\frac{\mathbb{I}(a_{t}=d_{t})}{P(A_{t}=a_t \mid \bar{\bm X}_{t}, \bar{\bm a}_{t-1})}\right)^2 \right\} \mathbb{E}[(Y-\phi)^2 \mid \bar{\bm X}, \bar{\bm a}] \right]\\
    &= \mathbb{E}\left[\left\{\prod_{t=1}^T \frac{1}{P(A_{t}=d_t \mid \bar{\bm X}_{t}, \bar{\bm a}_{t-1})} \right\} \mathbb{E}[(Y-\phi)^2 \mid \bar{\bm X}, \bar{\bm d}] \right]\\
    &= \mathbb{E}\left[\left\{\prod_{t=1}^T \frac{1}{p_t} \right\} \sigma_d^2(\bar{\bm X}) \right]
\end{align*}
where $\sigma_d^2(\bar{\bm X}) = \mathbb{E}[(Y-\phi)^2 \mid \bar{\bm X}, \bar{\bm d}].$

\newpage
\noindent\textbf{Variance of the nGCW estimator.}  
Recall $p_t = P(A_{t}=d_t \mid \bar{\bm X}_{t}, \bar{\bm a}_{t-1})$ and $\tilde{\phi} = \mathbb{E} \left[\theta(\bar{\bm X}) \mathbb{E}[Y^d \mid \bar{\bm X}] + (1- \theta(\bar{\bm X})) \mathbb{E}[Y^{\neg d} \mid \bar{\bm X}] \right]$.
\begin{align*}
    I_{\text{nGCW}}  &= \mathbb{E}\left[ \bar{w}_{T}^{\text{G,2}} (Y - \tilde{\phi})^2 \right] \\
    &= \mathbb{E} \left[ \sum_{\bar{\bm a}} P(\bar{\bm A} = \bar{\bm a} \mid \bar{\bm X}) \left\{\prod_{t=1}^T w_{t}^{\text{G,2}}\right\} \mathbb{E}[(Y-\tilde{\phi})^2 \mid \bar{\bm X}, \bar{\bm a}] \right]\\
    &= \mathbb{E} \left[ \sum_{\bar{\bm a}} \left\{\prod_{t=1}^T P(A_t = a_t \mid \bar{\bm X}_t, \bar{\bm a}_{t-1}) \left(\frac{m_t(\bar{\bm X}_t, \bar{\bm a}_{t-1}, a_t)}{D_t(\bar{\bm X}_t, \bar{\bm a}_{t-1})}\right)^2 \right\} \mathbb{E}[(Y-\tilde{\phi})^2 \mid \bar{\bm X}, \bar{\bm a}] \right].\tag{*}
\end{align*}
Now define
\[
R_t(\bar{\bm X}_t, \bar{\bm a}_{t-1}, a_t) =
\begin{cases}
\dfrac{p_t}{D_t^2}, & a_t = d_t, \\[1.2em]
\dfrac{\gamma_t^2\, P\!\left(A_t = a_t \,\middle|\, \bar{\bm X}_t, \bar{\bm a}_{t-1}\right)}{D_t^2}, & a_t \neq d_t.
\end{cases}
\]
Then we can write (*) as
\begin{align*}
    (*) &= \mathbb{E} \left[ \sum_{\bar{\bm a}} \left\{\prod_{t=1}^T R_t(\bar{\bm X}_t,\bar{\bm a}_{t-1}, a_t) \right\} \mathbb{E}[(Y-\tilde{\phi})^2 \mid \bar{\bm X}, \bar{\bm a}] \right]  \\
    &= \mathbb{E} \Bigg[ \left\{\prod_{t=1}^T R_t(\bar{\bm X}_t, \bar{\bm d}_{t-1}, d_t) \right\} \mathbb{E}[(Y-\tilde{\phi})^2 \mid \bar{\bm X}, \bar{\bm d}] \\[-0.25em]
    &\qquad + \sum_{\bar{\bm a} \neq \bar{\bm d}}  \left\{\prod_{t=1}^T R_t(\bar{\bm X}_t,\bar{\bm a}_{t-1}, a_t) \right\} \mathbb{E}[(Y-\tilde{\phi})^2 \mid \bar{\bm X}, \bar{\bm a}] \Bigg] \\
    &= \mathbb{E} \left[ \left\{\prod_{t=1}^T \frac{p_t}{D_t^2} \right\} \mathbb{E}[(Y-\tilde{\phi})^2 \mid \bar{\bm X}, \bar{\bm d}] + \underbrace{\sum_{\bar{\bm a} \neq \bar{\bm d}}  \left\{\prod_{t=1}^T R_t(\bar{\bm X}_t,\bar{\bm a}_{t-1}, a_t) \right\} \mathbb{E}[(Y-\tilde{\phi})^2 \mid \bar{\bm X}, \bar{\bm a}]}_{(**)} \right]. 
\end{align*}
Notice that 
$$
\sum_{a_t} R_t(\bar{\bm X}_t, \bar{\bm a}_{t-1}, a_t) = \frac{p_t}{D_t^2} + \frac{\gamma_t^2}{D_t^2} \sum_{a_t \neq d_t} P(A_t =a_t \mid \bar{\bm X}_t, \bar{\bm a}_{t-1}) = \frac{p_t + \gamma_t^2(1-p_t)}{D_t^2}. $$
Thus, we can write
\begin{align*}
    \sum_{\bar{\bm a} \neq \bar{\bm d}}  \left\{\prod_{t=1}^T R_t(\bar{\bm X}_t,\bar{\bm a}_{t-1}, a_t) \right\} &= \sum_{\bar{\bm a}}  \left\{\prod_{t=1}^T R_t(\bar{\bm X}_t,\bar{\bm a}_{t-1}, a_t) \right\} - \prod_{t=1}^T \frac{p_t}{D_t^2}\\
    &= \prod_{t=1}^T \sum_{{a_t}} R_t(\bar{\bm X}_t,\bar{\bm a}_{t-1}, a_t) - \prod_{t=1}^T \frac{p_t}{D_t^2}\\
    &= \prod_{t=1}^T \frac{p_t + \gamma_t^2(1-p_t)}{D_t^2} - \prod_{t=1}^T \frac{p_t}{D_t^2}.
\end{align*}
Now (**) can be written as
\begin{align*}
    (**) &= \left\{\sum_{\bar{\bm a} \neq \bar{\bm d}} \prod_{t=1}^T R_t(\bar{\bm X}_t,\bar{\bm a}_{t-1}, a_t) \right\} \sum_{\bar{\bm a} \neq \bar{\bm d}} \frac{\prod_{t=1}^T R_t(\bar{\bm X}_t,\bar{\bm a}_{t-1}, a_t) \mathbb{E}[(Y-\tilde{\phi})^2 \mid \bar{\bm X}, \bar{\bm a}]}{\sum_{\bar{\bm a} \neq \bar{\bm d}} \prod_{t=1}^T R_t(\bar{\bm X}_t,\bar{\bm a}_{t-1}, a_t)}\\
    &= \left\{ \prod_{t=1}^T \frac{p_t + \gamma_t^2(1-p_t)}{D_t^2} - \prod_{t=1}^T \frac{p_t}{D_t^2} \right\} \tilde{\sigma}^2_{\neg d}(\bar{\bm X}).
\end{align*}
Thus, we can write $I_{\text{nGCW}}$ as
\begin{align*}
    I_{\text{nGCW}} &= \mathbb{E}\left[ \left\{\prod_{t=1}^T \frac{p_t}{D_t^2} \right\} \mathbb{E}[(Y-\tilde{\phi})^2 \mid \bar{\bm X}, \bar{\bm d}] + \left\{ \prod_{t=1}^T \frac{p_t + \gamma_t^2(1-p_t)}{D_t^2} - \prod_{t=1}^T \frac{p_t}{D_t^2} \right\} \tilde{\sigma}^2_{\neg d}(\bar{\bm X}) \right] \\
    &= \mathbb{E}\left[ \left\{\prod_{t=1}^T \frac{p_t}{D_t^2} \right\} \tilde{\sigma}^2_{d}(\bar{\bm X})+ \left\{ \prod_{t=1}^T \frac{p_t + \gamma_t^2(1-p_t)}{D_t^2} - \prod_{t=1}^T \frac{p_t}{D_t^2} \right\} \tilde{\sigma}^2_{\neg d}(\bar{\bm X}) \right],
\end{align*}
where
$$\tilde{\sigma}^2_{d}(\bar{\bm X}) = \mathbb{E}[(Y-\tilde{\phi})^2 \mid \bar{\bm X}, \bar{\bm d}],$$
$$\tilde{\sigma}^2_{\neg d}(\bar{\bm X}) = \sum_{\bar{\bm a} \neq \bar{\bm d}} \frac{\prod_{t=1}^T R_t(\bar{\bm X}_t,\bar{\bm a}_{t-1}, a_t) \mathbb{E}[(Y-\tilde{\phi})^2 \mid \bar{\bm X}, \bar{\bm a}]}{\sum_{\bar{\bm a} \neq \bar{\bm d}} \prod_{t=1}^T R_t(\bar{\bm X}_t,\bar{\bm a}_{t-1}, a_t)}. $$

\newpage
\noindent\textbf{Difference in asymptotic variances.}  
The variance reduction is
\begin{align*}
    &\Delta \operatorname{Var} = \frac{1}{n} \Big( I_{\text{nIPW}} - I_{\text{nGCW}} \Big) \\
    &= \dfrac{1}{n} \cdot \mathbb{E} \Bigg[ \left\{\prod_{t=1}^T \frac{1}{p_t} \right\} \sigma_d^2(\bar{\bm X}) - \left\{\prod_{t=1}^T \frac{p_t}{D_t^2} \right\} \tilde{\sigma}^2_{d}(\bar{\bm X}) - \left\{ \prod_{t=1}^T \frac{p_t + \gamma_t^2(1-p_t)}{D_t^2} - \prod_{t=1}^T \frac{p_t}{D_t^2} \right\} \tilde{\sigma}^2_{\neg d}(\bar{\bm X}) \Bigg].
\end{align*}
Asymptotically, we know from Theorem 1 that the nGCW estimator is consistent. Then, $\tilde{\phi} \to \phi$ as $n \to \infty$. Thus, we have $\tilde{\sigma}_d^2 \to \sigma_d^2$ and $\tilde{\sigma}_{\neg d}^2 \to \sigma_{\neg d}^2$ as well as $n \to \infty$.\\
We can rewrite the difference in asymptotic variances as
\[
\Delta \operatorname{Var}
= \frac{1}{n} \cdot \mathbb{E}\left[
\left( \prod_{t=1}^T \frac{1}{p_t}   - \prod_{t=1}^T \frac{p_t}{D_t^2} \right) \sigma_d^2(\bar{\bm X}) 
- \left\{ \prod_{t=1}^T \frac{p_t + \gamma_t^2(1-p_t)}{D_t^2} - \prod_{t=1}^T \frac{p_t}{D_t^2} \right\} \sigma_{\neg d}^2(\bar{\bm X}) 
\right].
\]
Suppose \(\epsilon_n = c \cdot n^{-k}\) with \(c > 0\), \(k > 0\), $\epsilon_t = \epsilon_n / T $ and choose
\[
\gamma_t = \frac{\epsilon_t p_t}{(1 - \epsilon_t)(1 - p_t)}.
\]
Then,
\begin{align*}
    &(1) \quad D_t = p_t + \gamma_t (1 - p_t) = p_t + \dfrac{\epsilon_t p_t}{(1-\epsilon_t)(1-p_t)} (1-p_t) =  \frac{p_t}{1 - \epsilon_t}. \\
    &(2) \quad \prod_{t=1}^T \frac{1}{p_t}   - \prod_{t=1}^T \frac{p_t}{D_t^2}
    = \prod_{t=1}^T \frac{1}{p_t} - \prod_{t=1}^T \frac{(1-\epsilon_t)^2}{p_t}
    = \prod_{t=1}^T \frac{1}{p_t} (1-\prod_{t=1}^T (1-\epsilon_t)^2).\\
    &(3) \quad 
        \prod_{t=1}^T \frac{p_t + \gamma_t^2(1-p_t)}{D_t^2} 
        = \prod_{t=1}^T \left\{p_t + 
        \dfrac{\epsilon_t^2 p_t^2}{(1-\epsilon_t)^2(1-p_t)^2} (1-p_t)\right\} 
        \frac{(1-\epsilon_t)^2}{p_t^2} \\
    &\hphantom{(3) \quad {}} 
        = \prod_{t=1}^T \left(\frac{(1-\epsilon_t)^2}{p_t} + 
        \frac{\epsilon_t^2}{1-p_t}\right).
\end{align*}
We now approximate the term with $\epsilon_t$, and keep only the dominating term. Firstly,
\begin{align*}
    1-\prod_{t=1}^T (1-\epsilon_t)^2 &= 1-\prod_{t=1}^T (1-2\epsilon_t+\epsilon_t^2)\\
    &= 1-\prod_{t=1}^T (1-\delta_t) \quad \text{where } \delta_t = 2\epsilon_t-\epsilon_t^2\\
    &\approx 1-(1-\sum_{t=1}^T \delta_t+ O(\delta_T^2))\\
    &= \sum_{t=1}^T \delta_t
    = 2\sum_{t=1}^T \epsilon_t - \sum_{t=1}^T \epsilon_t^2\\
    &= 2\sum_{t=1}^T \frac{\epsilon_n}{T} -\sum_{t=1}^T \frac{\epsilon_n^2}{T^2}\\
    &= 2\epsilon_n - \frac{\epsilon_n^2}{T} \approx 2\epsilon_n.
\end{align*}
Secondly,
\begin{align*}
    &\prod_{t=1}^T \frac{p_t + \gamma_t^2(1-p_t)}{D_t^2} - \prod_{t=1}^T \frac{p_t}{D_t^2} \\
    &= \prod_{t=1}^T \left(\frac{(1-\epsilon_t)^2}{p_t} + \frac{\epsilon_t^2}{1-p_t}\right) - \prod_{t=1}^T \frac{(1-\epsilon_t)^2}{p_t}\\
    &= \prod_{t=1}^T (A_t + B_t) - \prod_{t=1}^T A_t \\
    &\approx \prod_{t=1}^T A_t + \sum_{t=1}^T B_t \prod_{j \neq t}^T A_j + O(B_t^2) - \prod_{t=1}^T A_t\\
    & \approx \sum_{t=1}^T \left\{\frac{\epsilon_t^2}{1-p_t} \prod_{j \neq t}^T \frac{(1-\epsilon_j)^2}{p_j}\right\} \\
    &= \sum_{t=1}^T \left\{\frac{\epsilon_t^2}{1-p_t} \prod_{j=1}^T \frac{(1-\epsilon_j)^2}{p_j} \frac{p_t}{(1-\epsilon_t)^2}\right\} \\
    &= \sum_{t=1}^T \left\{\epsilon_t^2\frac{p_t}{1-p_t} \left(\prod_{j=1}^T\frac{1}{p_j} \right)\frac{1}{(1-\epsilon_t)^2}\prod_{j=1}^T (1-\epsilon_j)^2 \right\}.
\end{align*}
We know 
$$\prod_{j=1}^T (1-\epsilon_j)^2 \approx 1-2\epsilon_n \quad \text{and} \quad (1-\epsilon_t)^{-2} \approx 1+2\epsilon_t.
$$
Thus,
\begin{align*}
    \sum_{t=1}^T \left\{\epsilon_t^2\frac{p_t}{1-p_t} \left(\prod_{j=1}^T\frac{1}{p_j} \right)\frac{1}{(1-\epsilon_t)^2}\prod_{j=1}^T (1-\epsilon_j)^2 \right\} &= \sum_{t=1}^T \frac{\epsilon_n^2}{T^2} \frac{p_t}{1-p_t} \left(\prod_{j=1}^T\frac{1}{p_j} \right) (1+2\epsilon_t)(1-2\epsilon_n) \\
    &\approx \left(\prod_{j=1}^T\frac{1}{p_j} \right) \frac{\epsilon_n^2}{T^2} \sum_{t=1}^T \frac{p_t}{1-p_t}.
\end{align*}
Now we can rewrite $\Delta \operatorname{Var}$ with simplified term as
\begin{align*}
    \Delta \operatorname{Var} &= \frac{1}{n}\mathbb{E}\left[ \left\{\prod_{t=1}^T \frac{1}{p_t}\right\} 2\epsilon_n \sigma_d^2(\bar{\bm X}) -  \left\{\prod_{j=1}^T \frac{1}{p_j}\right\} \left( \sum_{t=1}^T \frac{p_t}{1-p_t} \right)\frac{\epsilon_n^2}{T^2} \sigma_{\neg d}^2(\bar{\bm X}) \right] \\
    &= \frac{1}{n}\mathbb{E}\left[  \left\{\prod_{t=1}^T \frac{1}{p_t}\right\} \left( 2\epsilon_n \sigma_d^2(\bar{\bm X}) - \frac{\epsilon_n^2}{T^2} \left( \sum_{t=1}^T \frac{p_t}{1-p_t} \right) \sigma_{\neg d}^2(\bar{\bm X}) \right) \right].
\end{align*}
Under assumptions (A1)-(A6), we should have $p_t \in (0,1)$ and $\sigma_d^2, \sigma_{\neg d}^2$ finite. \\
Using \(\epsilon_n = c  n^{-k} = O(n^{-k})\), we have 
$$\Delta \operatorname{Var} = \frac{1}{n} (O(n^{-k}) - O(n^{-2k})) = O(n^{-k-1}),$$
which means for sufficiently small $\epsilon_n$, the leading term will dominate the negative component, making the difference in variance asymptotically positive, implying that the nGCW estimator achieves variance reduction in the limit. 
\end{proof}

\newpage
\subsection{Proof of Theorem 3}
\begin{proof}
Recall the definition of AGCW estimator,
$$\widehat{V}_{\text{AGCW}}(d)  = \frac{1}{n} \sum_{i=1}^n  \left\{Q_1^d + \sum_{t=1}^T \bar{w}_{t,i}^{\text{G}} (Q_{t+1}^d - Q_t^d)\right\} \xrightarrow{p} \mathbb{E}[Y^d].
$$
Also, recall the definition of Q-functions,
\[
Q_t^d:= \mathbb{E}\left[ Q_{t+1}^d \mid \bar{\bm X}_t, \bar{\bm A}_{t-1}, d_t \right].
\]
It is possible to decompose the bias term like in Lemma 2 and prove the bias term converges to 0 asymptotically when either the propensity or outcome model is correct. 

An easier statement would be asymptotically $w_t^{\text{GCW}} \to w_t^{\text{IPW}}$, so $\mathbb{E}\left[\widehat{V}_{\text{AGCW}}\right] \to \mathbb{E}\left[\widehat{V}_{\text{AIPW}}\right]$, which is unbiased if either propensity or outcome model is correctly specified. Thus, AGCW is also doubly robust.

For the normalized AGCW, consistency follows immediately after using the weak law of large numbers and Slutsky's Theorem with $\mathbb{E}[\bar{w}_t] = 1$ by Lemma 1.

\end{proof}

\newpage
\subsection{Proof of Theorem 4}
\begin{proof}
We can use estimating equations to quantify the variances of AIPW and AGCW, like in Theorem 2.\\
Denote the normalized AIPW estimator as
\[
\widehat{\phi}_{\text{nAIPW}} = \widehat{V}_{\text{nAIPW}} 
= \frac{1}{n}\sum_{i=1}^n Q_1^d + \sum_{t=1}^T \frac{\sum_{i=1}^n \bar{w}_{t,i}^{\text{I}}(Q_{t+1}^d - Q_t^d)}{\sum_{i=1}^n \bar{w}_{t,i}^{\text{I}}}.
\]
Denote the normalized AGCW estimator as
\[
\widehat{\phi}_{\text{nAGCW}} = \widehat{V}_{\text{nAGCW}} 
= \frac{1}{n}\sum_{i=1}^n Q_1^d + \sum_{t=1}^T \frac{\sum_{i=1}^n \bar{w}_{t,i}^{\text{G}}(Q_{t+1}^d - Q_t^d)}{\sum_{i=1}^n \bar{w}_{t,i}^{\text{G}}}.
\]
Let \(\phi\) denote the true regime value. Define the estimating function:
\[
U_i = \sum_{t=1}^T \frac{ \bar{w}_{t,i} (Q^d_{t+1} - Q^d_t)}{\sum_{i=1}^n \bar{w}_{t,i}} + \frac{1}{n} (Q^d_1 - \phi),
\]
with $\sum_{i=1}^n U_i = 0$.\\
The asymptotic variance is
\[
\operatorname{Var}(\widehat{\phi}) = \frac{1}{n} \cdot \frac{I}{J^2},
\]
where
\[
I = \mathbb{E}[U_i^2], 
\qquad 
J = \mathbb{E}\!\left[-\frac{\partial}{\partial \phi} U_i\right] = \frac{1}{n}.
\]
Since $J = 1/n$, the asymptotic variance reduces to
\[
\operatorname{Var}(\widehat{\phi}) = n I.
\]
We first expand $I$ in a general form. Define 
$$S_t := \sum_{i=1}^n \bar{w}_{t,i}, \quad \Delta_t := Q_{t+1}^d - Q_t^d.$$
\begin{align*}
    \mathbb{E}\left[U_i^2\right] &= \mathbb{E}\left[ \left( \sum_{t=1}^T \frac{ \bar{w}_{t,i} (Q^d_{t+1} - Q^d_t)}{\sum_{i=1}^n \bar{w}_{t,i}} + \frac{1}{n} (Q^d_1 - \phi)\right) ^ 2\right] \\
    &= \mathbb{E}\left[ \left( \sum_{t=1}^T \frac{ \bar{w}_{t,i} \Delta_t}{S_t} + \frac{1}{n} (Q^d_1 - \phi)\right)^ 2\right] \\
    &= \mathbb{E}\left[ \frac{1}{n^2} (Q^d_1 - \phi)^2 + \frac{2(Q_1^d - \phi)}{n} \sum_{t=1}^T \frac{\bar{w}_t \Delta_t}{S_t} + \left( \sum_{t=1}^T \frac{\bar{w}_t \Delta_t}{S_t}\right)^2\right] \\
    &= \mathbb{E}\left[ \underbrace{\frac{1}{n^2} (Q^d_1 - \phi)^2}_{(A)} + \underbrace{\frac{2(Q_1^d - \phi)}{n} \sum_{t=1}^T \frac{\bar{w}_t \Delta_t}{S_t}}_{(B)} + \underbrace{\sum_{t=1}^T \frac{\bar{w}_t^2 \Delta_t^2}{S_t^2}}_{(C)} +\underbrace{ 2 \sum_{1\leq \ell < t \leq T} \frac{\bar{w}_\ell \Delta_\ell \bar{w}_t \Delta_t}{S_\ell S_t}}_{(D)}
   \right].
\end{align*}

\noindent\textbf{Term C under nAIPW and nAGCW}  \mbox{}\\
The decomposition of term C follows immediately from Theorem 2. So we have for nAIPW:
\[ \sum_{t=1}^T \mathbb{E}\left[\frac{\bar{w}_t^{\text{I},2} \Delta_t^2}{S_t^{\text{I},2}} \right] = \sum_{t=1}^T \mathbb{E} \left[\frac{1}{S_t^{\text{I},2}} \left\{ \prod_{\ell=1}^t \frac{1}{p_\ell}\right\} \tau_{t,d}^2 \right], \]
and for nAGCW:
\[
\sum_{t=1}^T \mathbb{E}\left[\frac{\bar{w}_t^{\text{G},2} \Delta_t^2}{S_t^{\text{G},2}} \right] = \sum_{t=1}^T \mathbb{E}\left[ \frac{1}{S_t^{\text{G},2}} \left(\left\{ \prod_{\ell=1}^t \frac{1}{p_\ell}\right\} \tau_{t,d}^2 + \left\{ \prod_{\ell=1}^t \frac{p_\ell + \gamma_\ell^2(1-p_\ell)}{D_\ell^2} - \prod_{\ell=1}^t \frac{p_\ell}{D_\ell^2} \right\} \tau_{t,\neg d}^2\right)\right],
\]
where $\tau_{t,d}^2 = \mathbb{E}[\Delta_t^2 \mid \bar{\bm X_t}, \bar{\bm d}_t]$.

\noindent\textbf{Difference in Variance}  \mbox{}\\
As $n \to \infty$, $\tilde{\phi} \to \phi$ and recall $\epsilon_n = c n^{-k}$. Then we look at the differences between terms (A), (B), (C) and (D).\\
Term (A):
$$A_{\text{nAIPW}} - A_{\text{nAGCW}} = \frac{1}{n^2} (Q^d_1 - \phi)^2 - \frac{1}{n^2} (Q^d_1 - \tilde{\phi})^2 \to 0. $$
Term (B):
$$B_{\text{nAIPW}} - B_{\text{nAGCW}} = \mathbb{E}\left[ \frac{2(Q_1^d - \phi)}{n} \sum_{t=1}^T \left(\frac{\bar{w}_t^{\text{I}}}{S^{\text{I}}_t}- \frac{\bar{w}_t^{\text{G}}}{S^{\text{G}}_t} \right) \Delta_t\right] = O(\frac{\epsilon_n}{n^2}) = O(n^{-k-2}).$$
Term (C):
$$C_{\text{nAIPW}} - C_{\text{nAGCW}} = O(\frac{1}{n^2}) (O(n^{-k})-O(n^{-2k})) \, \text{ by Theorem 2.}$$
Term (D): 
$$D_{\text{nAIPW}} - D_{\text{nAGCW}} = \mathbb{E}\left[2 \sum_{1\leq \ell < t \leq T} \left(\frac{\bar{w}_\ell^{\text{I}} \bar{w}_t^{\text{I}}}{S_\ell^{\text{I}} S_t^{\text{I}}} - \frac{\bar{w}_\ell^{\text{G}} \bar{w}_t^{\text{G}}}{S_\ell^{\text{G}} S_t^{\text{G}}} \right)\Delta_\ell \Delta_t \right] = O(\frac{\epsilon_n^2}{n^2}) = O(n^{-2k-2}).$$
By Theorem 2, the difference in term (C) is positive, but the signs of B and D are not deterministic. Although the dominant terms ((B) and (C)) share the same big-O rate, their constant coefficients differ.

For term (B), the contributing constant is $c_1 = (Q^d_1-\phi)\sum_{t=1}^T \Delta_t$, which grows at most linearly with T. In contrast, the constant associated with term (C), $c_2 = \left\{\prod_{t=1}^T p_t^{-1} \right\} \sigma_d^2$ grows exponentially in T due to the product over inverse propensities. Consequently, $c_2$ dominates $c_1$, implying that the overall difference in I is positive for sufficiently small $\epsilon_n$. Hence, we have $$I_{\text{nAIPW}} - I_{\text{nAGCW}} = O(n^{-k-2}) > 0,$$
and therefore
$$\Delta\operatorname{Var}
= \operatorname{Var}(\widehat\phi_{\text{nAIPW}}) -
\operatorname{Var}(\widehat\phi_{\text{nAGCW}})
= n(I_{\text{nAIPW}}-I_{\text{nAGCW}})
= O(n^{-k-1}),
$$
which is asymptotically positive under regularity conditions and vanishes as $n \to \infty$.

Analogous to the classical efficiency gain achieved by the AIPW estimator over the IPW estimator, the AGCW estimator achieves improved efficiency whenever the outcome model is correctly specified and the standard causal identification assumptions hold.

\end{proof}

\newpage
\subsection{Proof of Theorem 5}
\begin{proof}
\noindent\textbf{Consistency of the unnormalized estimator.}
Let $d$ be a fixed threshold regime and $d^{(\delta)}$ be its windowed version with $\delta=(\delta_1,\ldots,\delta_T)$.\\
The unnormalized BCW estimator is
\[
\widehat{V}_{\text{BCW}}(d) = \dfrac{1}{n} \sum_{i=1}^n \left\{\prod_{t=1}^T w_{t,i}^{\text{B}}\right\} Y_i, \text{ where } w_{t,i}^{\text{B}} = \frac{\mathbb{I}\left( A_{t,i} = d^{(\delta_{\text{opt}})}_{t,i} \right)}{P\left(A_{t,i} \mid \bar{\bm X}_{t,i}, \bar{\bm A}_{t-1, i}\right)},
\]
and $\delta_{\mathrm{opt}}$ is the bootstrap-selected window.

By (A7), $\|\delta_{\mathrm{opt}}\|_\infty \to 0$ in probability.
Hence, for any $i$,
\[
\mathbb{I}\left( A_{t,i} = d^{(\delta_{\text{opt}})}_{t,i} \right)
\;\xrightarrow{p}\;
\mathbb{I}\left( A_{t,i} = d_{t,i} \right),
\]
because the two indicators differ only when the covariates used at stage $t$ fall within a band of width $\|\delta_{\mathrm{opt}}\|_\infty$, whose probability goes to $0$. \\
Then, define
\[
S_i^{(boot)}:= \left\{\prod_{t=1}^T \frac{\mathbb{I}\left( A_{t,i} = d^{(\delta_{\text{opt}})}_{t,i} \right)}{P\left(A_{t,i} \mid \bar{\bm X}_{t,i}, \bar{\bm A}_{t-1, i}\right)} \right\}\,Y_i,
\qquad
S_i:= \left\{\prod_{t=1}^T
\frac{\mathbb{I}\left( A_{t,i} = d_{t,i} \right)}{P\left(A_{t,i} \mid \bar{\bm X}_{t,i}, \bar{\bm A}_{t-1, i}\right)}\right\}\,Y_i.
\]
By the stage-wise indicator convergence and boundedness of the factors, Slutsky’s Theorem gives 
$$S_i^{(boot)} \xrightarrow{p} S_i \quad \text{for each $i$.}$$
By i.i.d.\ sampling (A5) and the Weak Law of Large Numbers,
\[
\frac{1}{n}\sum_{i=1}^n S_i^{(boot)} \;\xrightarrow{p}\; \mathbb E[S_i].
\]
By the standard IPW identification assumptions (A1)–(A4), we have $\mathbb E[S_i] = \mathbb E[Y^{d}]$. Thus,
$$\widehat V_{\mathrm{BCW}} \xrightarrow{p} \mathbb E[Y^{d}].$$

\noindent\textbf{Consistency of the normalized estimator.}
Define
\[
\widehat V_{\mathrm{nBCW}}(d)
=\frac{\sum_{i=1}^n \bar{w}_{T,i}^{\text{B}} Y_i}{\sum_{i=1}^n \bar{w}_{T,i}^{\text{B}}} \xrightarrow{p} \mathbb{E}[Y^d],
\qquad
\bar{w}_{T,i}^{\text{B}}:=\prod_{t=1}^T w_{t,i}^{\text{B}}.
\]
With
\[
\frac{1}{n}\sum_{i=1}^n S_i^{(boot)} \xrightarrow{p} \mathbb E[Y^{d}],
\qquad
\frac{1}{n}\sum_{i=1}^n \bar{w}_{T,i}^{\text{B}}
\xrightarrow{p} 1,
\]
we have
$$\widehat V_{\mathrm{nBCW}} \xrightarrow{p} \mathbb E[Y^{d}],$$
by Slutsky's Theorem.

\subsubsection*{Variance of the nBCW estimator with fixed window size}
For a fixed window $\delta$, we can use the estimating function to derive the variance of the nBCW estimator
\[
\sum_{i=1}^n U_i = \sum_{i=1}^n \bar{w}_{T,i}^{B} (Y_i - \phi) = 0,
\]
where \(\phi\) denotes the true regime value.

The asymptotic variance is
\[
\operatorname{Var}(\widehat{\phi}) = \frac{1}{n} \cdot \frac{I}{J^2},
\]
where
\[
I = \mathbb{E}[U_i^2] = \mathbb{E}[\bar{w}_{T,i}^{B,2} (Y_i - \phi)^2], 
\qquad 
J = \mathbb{E}\!\left[-\frac{\partial}{\partial \phi} U_i\right] = \mathbb{E}[\bar{w}_{T,i}^B].
\]
Thus, the asymptotic variance is
\[
\operatorname{Var}\{\widehat V_{\mathrm{nBCW}}(d;\delta)\}
=
\frac{1}{n}
\frac{E\!\left[
\{\bar w^{\text{B}}_{T}(\delta)\}^2
\{Y-\phi(\delta)\}^2
\right]}{\mathbb{E}[\bar{w}_{T,i}^B]^2}.
\]

\subsubsection*{Limiting distribution with bootstrap-selected window}
We now return to the adaptive estimator with $\delta_{\mathrm{opt}}$.  
We have$\|\delta_{\mathrm{opt}}\|_\infty \xrightarrow{p} 0$. Assume additionally that
$\sqrt n\,\|\delta_{\mathrm{opt}}\|_\infty \xrightarrow{p} 0$. The probability of BCW and IPW weights differ vanishes as $\|\delta_{\mathrm{opt}}\|_\infty \to 0$, so
\[
\widehat V_{\mathrm{nBCW}}(d)-\widehat V_{\mathrm{nIPW}}(d)=o_p(n^{-1/2}).
\]
Since the nIPW estimator satisfies
\[
\sqrt n\{\widehat V_{\mathrm{nIPW}}(d)-E[Y^d]\}
\xrightarrow{d}
N(0,\sigma^2_{\mathrm{IPW}}),
\]
Slutsky’s theorem yields
\[
\sqrt n\{\widehat V_{\mathrm{nBCW}}(d)-E[Y^d]\}
\xrightarrow{d}
N(0,\sigma^2_{\mathrm{IPW}}).
\]
Thus the nBCW estimator is asymptotically equivalent to the nIPW estimator.

\subsubsection*{Discussion of the variance expression}
Unlike the nGCW estimator, where variance reduction can be shown analytically, the difference between nBCW and nIPW variances does not have a fixed sign. Nevertheless, the enlargement of the compatible set typically stabilizes the weight distribution, which explains the empirical variance reduction observed in finite samples.
\end{proof}

\newpage
\section{Additional Results for Simulation 1}

\subsection{True Value Surface}
The true value surface obtains the optimal thresholds $(\psi_1, \psi_2) = (350, 450)$, with a value of 1212.22. This surface serves as the benchmark for evaluating the accuracy of the estimated value surfaces obtained from different weighting methods.

\FloatBarrier
\begin{figure}[htbp]
    \centering
    \includegraphics[width=0.8\linewidth]{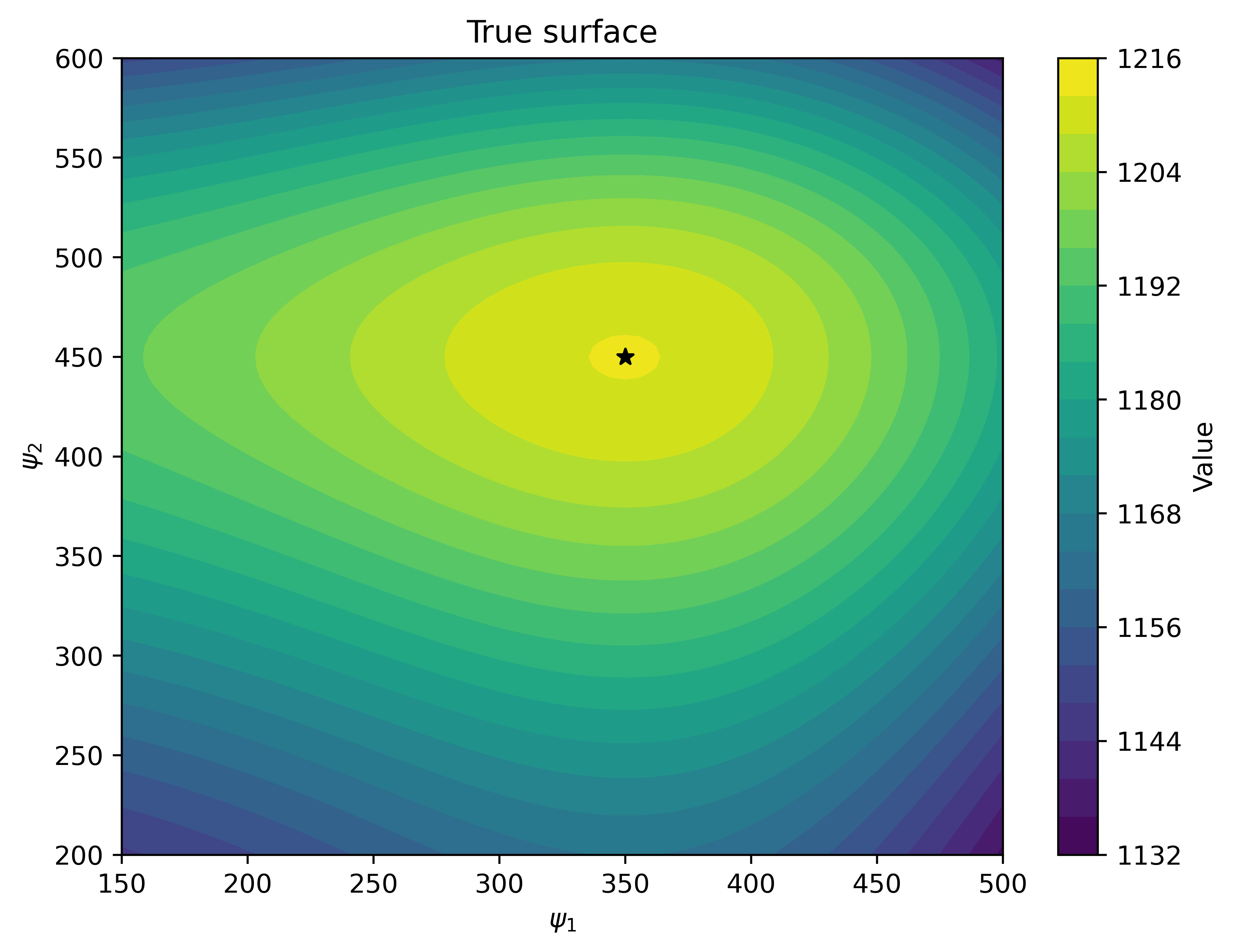}
    \caption{Simulation 1: True value surface}
\end{figure}

\subsection{Estimated Value Surfaces}
We present estimated value surfaces from normalized IPW, GCW, BCW estimators and normalized AIPW, AGCW, ABCW estimators for sample sizes $n = 200$, $500$, and $1000$. As the sample size increases, all estimators yield smoother and more accurate approximations of the true surface, with estimated optima converging toward the true threshold. Augmentation further stabilizes the surfaces by reducing local irregularity and noise.

\FloatBarrier
\begin{figure}[htbp]
    \centering
    
    \includegraphics[width=0.98\linewidth]{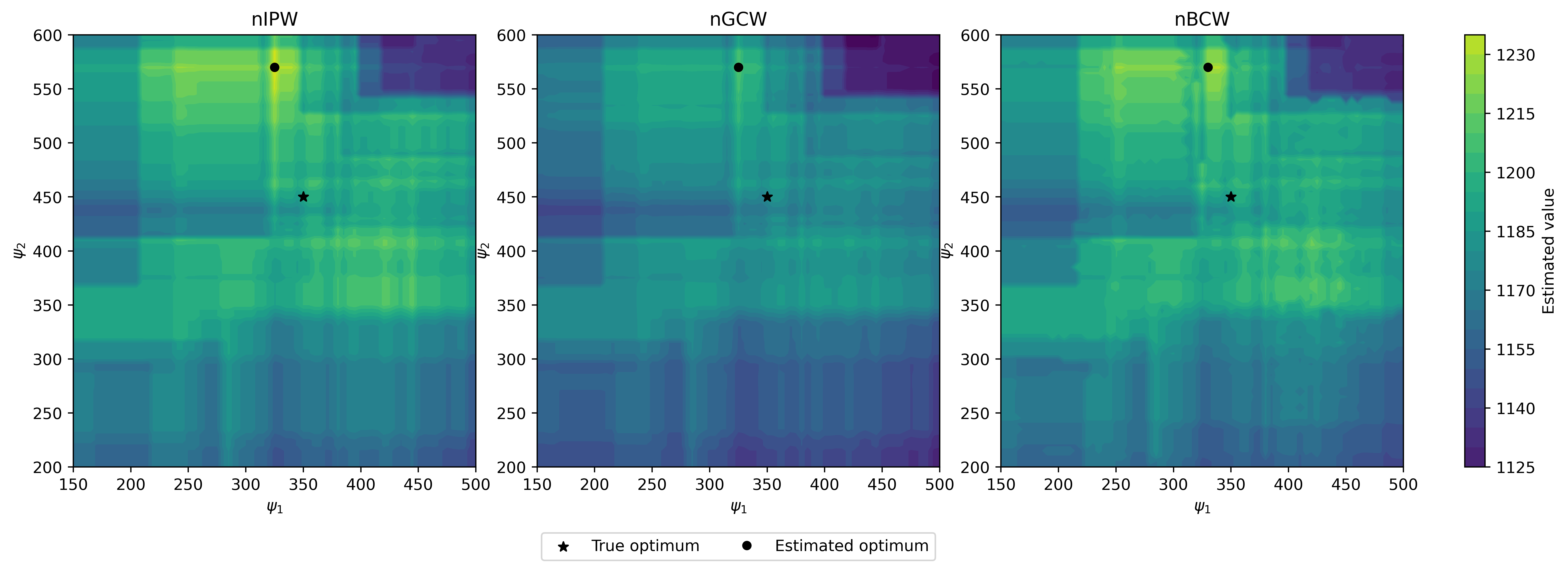}
    \scriptsize{\caption*{(a) Estimated value surfaces for $n=200$}}
    \vspace{0.6em}
    
    \includegraphics[width=0.98\linewidth]{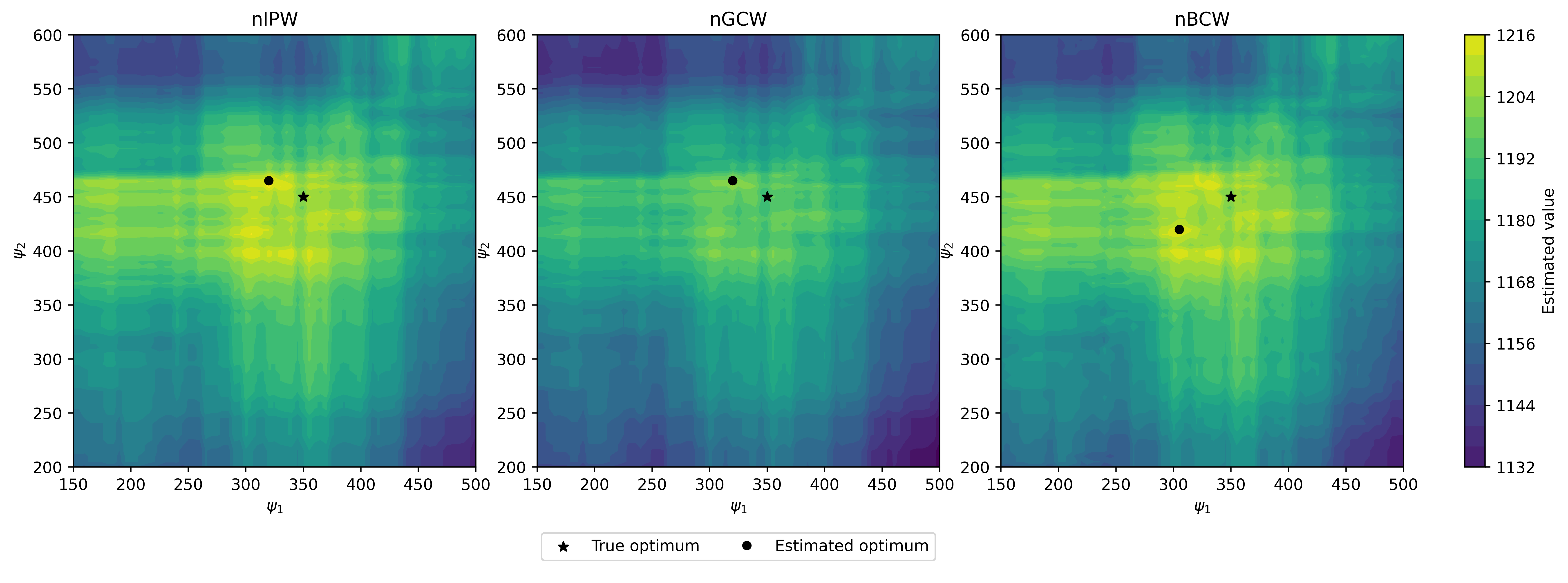}
    \scriptsize{\caption*{(b) Estimated value surfaces for $n=500$}}
    \vspace{0.6em}

    \includegraphics[width=0.98\linewidth]{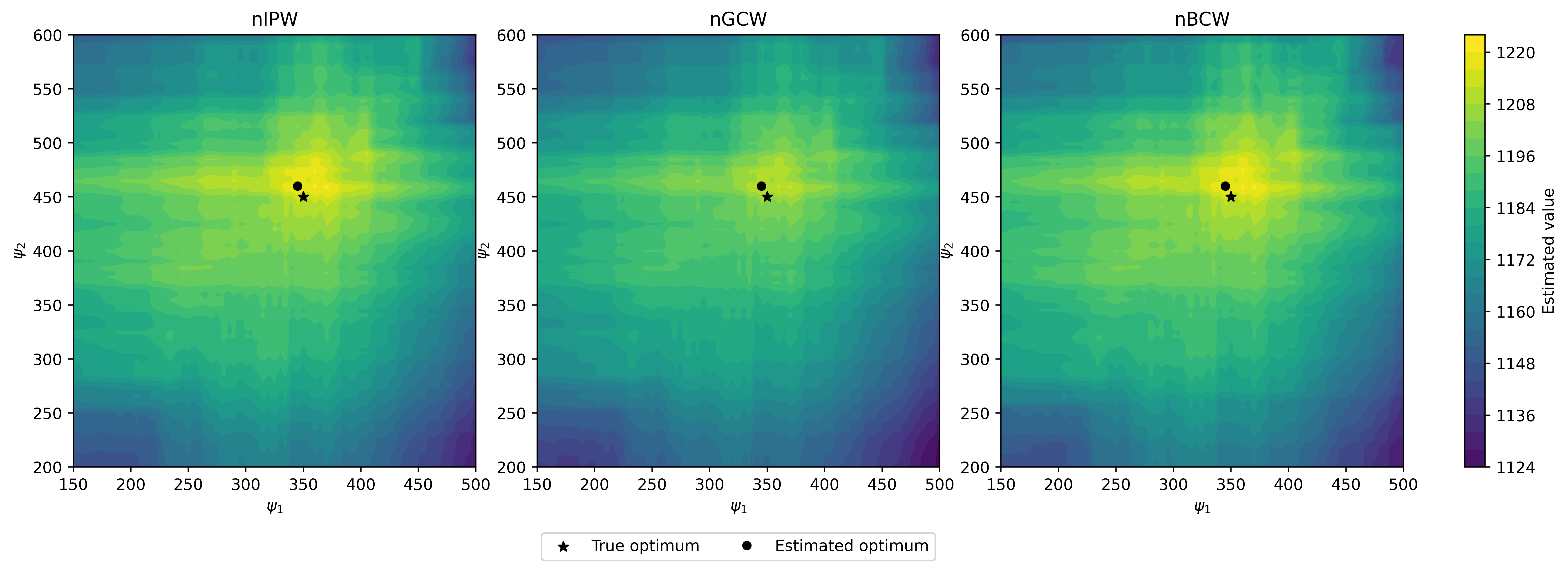}
    \scriptsize{\caption*{(c) Estimated value surfaces for $n=1000$}}

    \caption{Simulation 1: estimated value surfaces of non-augmented estimators for sample sizes $n=200$, $n=500$, and $n=1000$.}

\end{figure}

\FloatBarrier
\begin{figure}[htbp]
    \centering
    
    \includegraphics[width=0.98\linewidth]{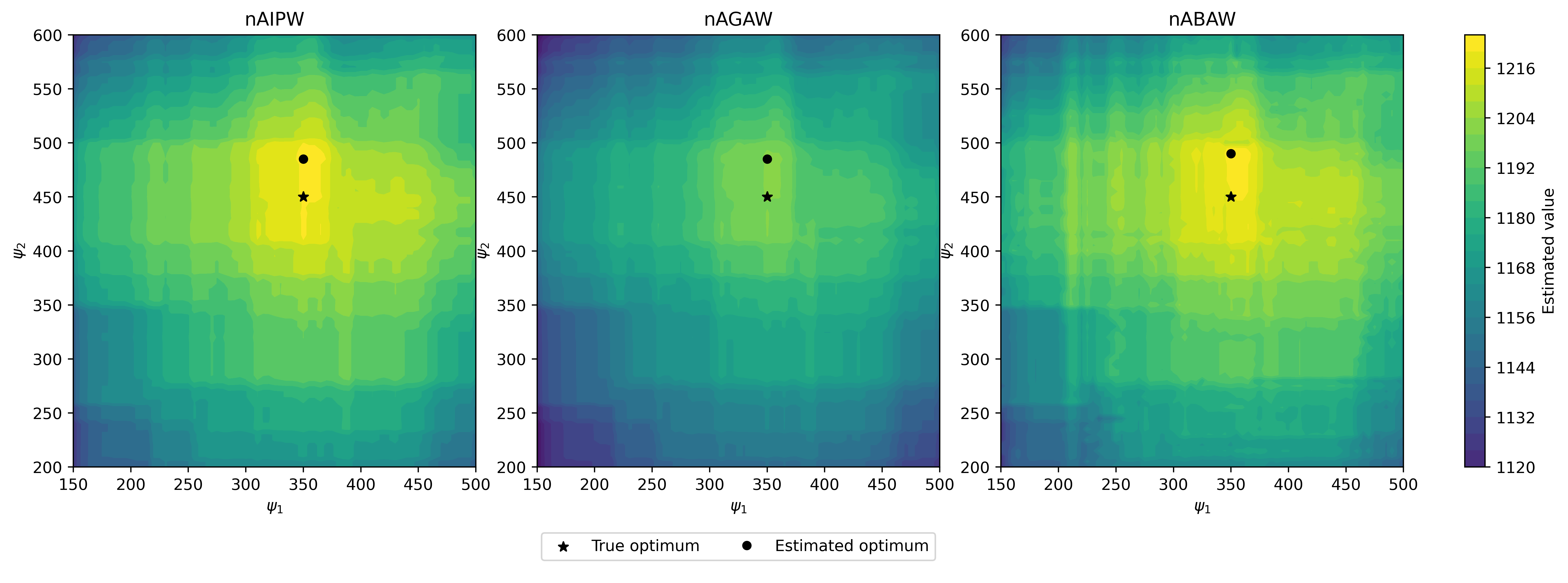}
    \scriptsize{\caption*{(a) Estimated value surfaces for $n=200$}}
    \vspace{0.6em}
    
    \includegraphics[width=0.98\linewidth]{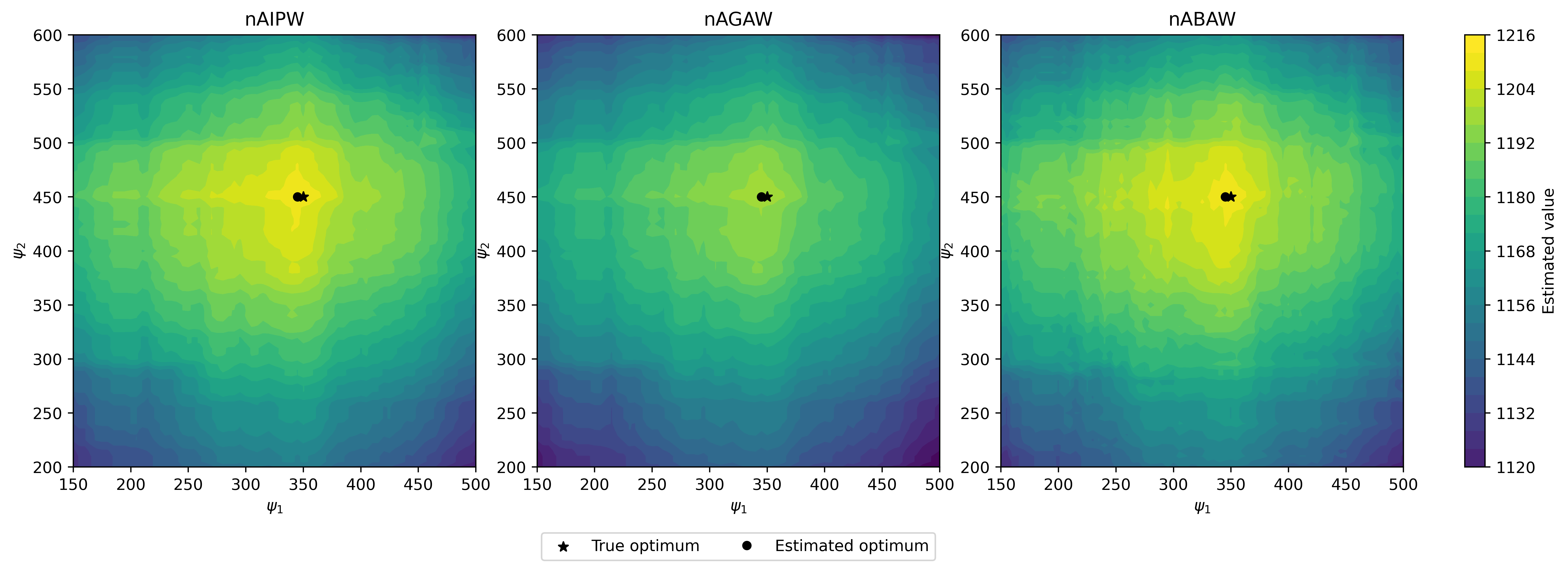}
    \scriptsize{\caption*{(b) Estimated value surfaces for $n=500$}}
    \vspace{0.6em}

    \includegraphics[width=0.98\linewidth]{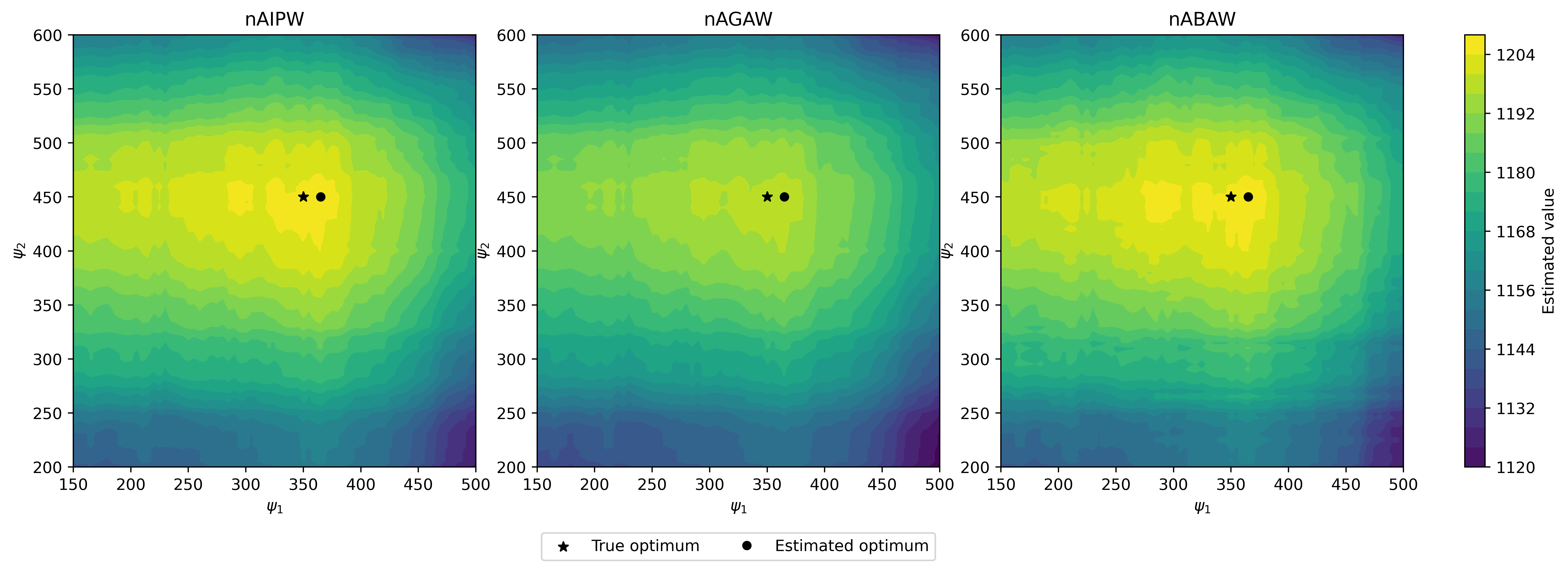}
    \scriptsize{\caption*{(c) Estimated value surfaces for $n=1000$}}

    \caption{Simulation 1: estimated value surfaces of augmented estimators for sample sizes $n=200$, $n=500$, and $n=1000$.}

\end{figure}

\FloatBarrier
\subsection{Results for Normalized Augmented Estimators}
We report additional results for the normalized augmented estimators using the same metrics as in the main paper. Tables~\ref{tab:Sim1_t3} and~\ref{tab:Sim1_t4} present the threshold-level and surface-level results, respectively.

\begin{table}[htbp]
\centering
\caption{Summary of estimated thresholds, RMSE, and coverage for nAIPW, nAGCW and nABCW based on 500 replications.}
\label{tab:Sim1_t3}
\setlength{\tabcolsep}{3pt}
{\fontsize{10pt}{10pt}\selectfont
\renewcommand{\arraystretch}{1.5}
\begin{tabular}{@{}ccccccccccc@{}}
\toprule
\multirow{2}{*}{$n$} & \multirow{2}{*}{Method} &
\multicolumn{2}{c}{Mean (SD)} &
\multicolumn{2}{c}{Median (IQR)} &
\multicolumn{2}{c}{RMSE} &
\multicolumn{2}{c}{Coverage} \\
\cmidrule(lr){3-4} \cmidrule(lr){5-6} \cmidrule(lr){7-8} \cmidrule(lr){9-10}
& & $\psi_1$ & $\psi_2$ & $\psi_1$ & $\psi_2$ & $\psi_1$ & $\psi_2$ & $\psi_1$ & $\psi_2$ \\
\midrule
200 & nAIPW  & 333.34 (57.92) & 450.77 (23.86) & 345.00 (60.00) & 450.00 (30.00) & 42.34 & 18.49 & 0.966 & 0.990 \\
    & nAGCW  & 332.72 (57.00) & 450.81 (23.58) & 345.00 (60.00) & 450.00 (30.00) & 41.16 & 18.35 & 0.962 & 0.972 \\
    & nABCW  & 328.05 (59.53) & 450.78 (27.29) & 340.00 (70.00) & 455.00 (40.00) & 45.13 & 22.04 & 0.968 & 0.996 \\
\midrule\midrule
500 & nAIPW  & 341.62 (38.05) & 451.45 (17.69) & 345.00 (40.00) & 450.00 (20.00) & 28.08 & 13.95 & 0.972 & 0.982 \\
    & nAGCW  & 343.09 (36.12) & 451.41 (17.63) & 345.00 (40.00) & 450.00 (25.00) & 26.89 & 13.95 & 0.978 & 0.978 \\
    & nABCW  & 338.69 (38.99) & 450.91 (18.83) & 345.00 (45.00) & 450.00 (25.00) & 30.11 & 15.15 & 0.974 & 0.992 \\
\midrule\midrule
1000 & nAIPW & 345.04 (27.81) & 449.38 (13.69) & 345.00 (40.00) & 450.00 (20.00) & 22.18 & 10.98 & 0.976 & 0.980 \\
     & nAGCW & 344.65 (27.39) & 449.50 (13.57) & 345.00 (40.00) & 450.00 (20.00) & 21.91 & 10.92 & 0.968 & 0.982 \\
     & nABCW & 344.77 (29.46) & 449.07 (14.70) & 350.00 (40.00) & 450.00 (20.00) & 23.35 & 11.51 & 0.968 & 0.992 \\
\bottomrule
\end{tabular}
}
\end{table}

\FloatBarrier
\begin{table}[htbp]
\centering
\caption{Performance of nAIPW, nAGCW and nABCW over all regimes. Measurements are reported as mean (standard deviation) across 500 Monte Carlo replications.}
\label{tab:Sim1_t4}
\resizebox{\textwidth}{!}{
{\fontsize{10pt}{10pt}\selectfont
\renewcommand{\arraystretch}{1.5}
\begin{tabular}{cccccccc}
\toprule
$n$ & Method & Var & ESS & Corr & \% Var $\downarrow$ (grid) & POT & Value \\
\midrule
200  & nAIPW & 282.74 (43.81) & 76.80 (17.49) & 0.909 (0.047) & ---    & 0.901 (0.065) & 1139.39 (4.33) \\
     & nAGCW & 246.96 (38.23) & 87.97 (19.85) & 0.913 (0.046) & 100.00 & 0.904 (0.066) & 1139.46 (4.37) \\
     & nABCW & 279.05 (43.26) & 77.35 (17.59) & 0.907 (0.047) & 80.42  & 0.894 (0.070) & 1138.88 (4.69) \\
\midrule\midrule
500  & nAIPW & 105.04 (17.16) & 192.69 (44.28) & 0.960 (0.022) & ---    & 0.928 (0.047) & 1141.01 (2.28) \\
     & nAGCW & 97.99 (15.67) & 210.24 (48.18) & 0.961 (0.021) & 100.00 & 0.930 (0.045) & 1141.09 (2.13) \\
     & nABCW & 103.73 (17.06) & 193.92 (44.51) & 0.959 (0.021) & 75.80  & 0.924 (0.048) & 1140.84 (2.20) \\
\midrule\midrule
1000 & nAIPW & 57.87 (8.99) & 384.85 (89.01) & 0.980 (0.012) & ---    & 0.942 (0.035) & 1141.71 (1.20) \\
     & nAGCW & 54.89 (8.47) & 409.62 (94.60) & 0.980 (0.012) & 100.00 & 0.943 (0.034) & 1141.73 (1.19) \\
     & nABCW & 57.23 (8.98) & 387.40 (89.19) & 0.980 (0.012) & 67.48  & 0.940 (0.036) & 1141.62 (1.29) \\
\bottomrule
\end{tabular}}}
\end{table}

\subsection{Sensitivity Analysis}
{
We evaluated the sensitivity of the proposed methods to the tuning parameters governing the bias--variance trade-off in the partial-compatibility estimators. In addition to the main analysis, we considered two alternative configurations: \textit{Config 1}, with $\lambda_{\text{bias}} = 2$, $c = 0.5$, and $k = 0.5$, and \textit{Config 2}, with $\lambda_{\text{bias}} = 0$, $c = 2$, and $k = 0.5$. We fixed $k$ and varied only $c$, because together these parameters control the borrowing strength in the GCW estimators. \textit{Config 1} represents a more conservative setting with less borrowing, as the larger $\lambda_{\text{bias}}$ places greater emphasis on the bias term while the smaller $c$ imposes a tighter bias tolerance. By contrast, \textit{Config 2} represents a more aggressive setting with greater borrowing.}

{Table~\ref{tab:sens_threshold} summarizes the estimated optimal thresholds, threshold RMSE, and coverage, while Table~\ref{tab:sens_surface} reports regime-surface-level performance, including variance, effective sample size, rank correlation, and external-population performance measures for nIPW, nGCW and nBCW. Across the two configurations, the estimated optimal regimes were broadly similar, indicating that the choice of tuning parameters had little effect on the identified optimal thresholds, provided the tuning parameters were not chosen too aggressively. However, configuration 2, which allows more borrowing, consistently achieved greater variance reduction and larger gains in ESS. This again confirms that increased borrowing can improve the stability of the estimated value surface without sacrificing identification of the optimal treatment rule. Table~\ref{tab:sens_threshold_aug} and Table~\ref{tab:sens_surface_aug} report the corresponding results for the augmented estimators, and the conclusions are similar.}
\FloatBarrier
\begin{table}[htbp]
\centering
\caption{Sensitivity analysis of nIPW, nGCW and nBCW for threshold-level performance under two tuning configurations. Measurements are reported as mean (standard deviation) across 500 Monte Carlo replications.}
\label{tab:sens_threshold}
\resizebox{\textwidth}{!}{
{\fontsize{10pt}{10pt}\selectfont
\renewcommand{\arraystretch}{1.5}
\begin{tabular}{llcccccccc}
\toprule
Config & Method & \multicolumn{2}{c}{Mean (SD)} & \multicolumn{2}{c}{Median (IQR)} & \multicolumn{2}{c}{RMSE} & \multicolumn{2}{c}{Coverage} \\
\cmidrule(lr){3-4} \cmidrule(lr){5-6} \cmidrule(lr){7-8} \cmidrule(lr){9-10}
& & $\psi_1$ & $\psi_2$ & $\psi_1$ & $\psi_2$ & $\psi_1$ & $\psi_2$ & $\psi_1$ & $\psi_2$ \\
\midrule
\multicolumn{10}{l}{\textit{$n=200$}} \\
Config 1
& nIPW & 351.59 (73.00) & 485.70 (57.05) & 355.00 (105.00) & 485.00 (75.00) & 59.51 & 54.44 & 0.964 & 0.894 \\
& nGCW & 352.00 (70.39) & 485.43 (55.36) & 355.00 (101.25) & 485.00 (75.00) & 56.90 & 53.41 & 0.962 & 0.878 \\
& nBCW & 352.13 (69.34) & 484.36 (54.64) & 357.50 (95.00) & 485.00 (70.00) & 55.83 & 52.48 & 0.968 & 0.896 \\
\cmidrule(lr){2-10}
Config 2
& nIPW & 351.59 (73.00) & 485.70 (57.05) & 355.00 (105.00) & 485.00 (75.00) & 59.51 & 54.44 & 0.964 & 0.894 \\
& nGCW & 350.30 (63.88) & 479.88 (48.93) & 355.00 (86.25) & 480.00 (61.25) & 50.18 & 46.26 & 0.974 & 0.902 \\
& nBCW & 340.98 (72.28) & 483.84 (53.18) & 350.00 (105.00) & 480.00 (60.00) & 56.38 & 50.66 & 0.980 & 0.904 \\
\midrule
\midrule
\multicolumn{10}{l}{\textit{$n=500$}} \\
Config 1
& nIPW & 347.46 (56.63) & 472.71 (42.03) & 350.00 (80.00) & 475.00 (55.00) & 45.08 & 38.23 & 0.980 & 0.906 \\
& nGCW & 347.38 (55.04) & 472.14 (41.05) & 350.00 (75.00) & 470.00 (55.00) & 43.44 & 37.18 & 0.982 & 0.906 \\
& nBCW & 346.88 (56.05) & 471.53 (42.08) & 350.00 (75.00) & 470.00 (55.00) & 44.08 & 37.79 & 0.980 & 0.910 \\
\cmidrule(lr){2-10}
Config 2
& nIPW & 347.46 (56.63) & 472.71 (42.03) & 350.00 (80.00) & 475.00 (55.00) & 45.08 & 38.23 & 0.980 & 0.906 \\
& nGCW & 345.72 (49.89) & 469.33 (38.24) & 350.00 (65.00) & 470.00 (50.00) & 39.32 & 34.45 & 0.982 & 0.914 \\
& nBCW & 328.16 (49.47) & 466.74 (35.66) & 330.00 (45.00) & 470.00 (40.00) & 39.24 & 31.40 & 0.986 & 0.922 \\
\midrule
\midrule
\multicolumn{10}{l}{\textit{$n=1000$}} \\
Config 1
& nIPW & 346.25 (43.42) & 465.39 (33.19) & 350.00 (55.00) & 465.00 (45.00) & 34.15 & 29.31 & 0.988 & 0.954 \\
& nGCW & 346.68 (42.08) & 465.18 (33.31) & 350.00 (55.00) & 465.00 (45.00) & 33.32 & 29.34 & 0.988 & 0.950 \\
& nBCW & 346.66 (42.85) & 465.17 (33.19) & 350.00 (55.00) & 465.00 (45.00) & 33.74 & 29.25 & 0.990 & 0.954 \\
\cmidrule(lr){2-10}
Config 2
& nIPW & 346.25 (43.42) & 465.39 (33.19) & 350.00 (55.00) & 465.00 (45.00) & 34.15 & 29.31 & 0.988 & 0.954 \\
& nGCW & 346.24 (40.81) & 464.28 (30.60) & 350.00 (55.00) & 465.00 (40.00) & 32.32 & 27.18 & 0.988 & 0.956 \\
& nBCW & 326.85 (37.36) & 456.24 (32.38) & 320.00 (50.00) & 450.00 (41.25) & 35.61 & 25.40 & 0.946 & 0.978 \\
\bottomrule
\end{tabular}}}
\end{table}

\FloatBarrier
\begin{table}[htbp]
\centering
\caption{Sensitivity analysis of nIPW, nGCW and nBCW for value-surface-level performance under two tuning configurations. Measurements are reported as mean (standard deviation) across 500 Monte Carlo replications.}
\label{tab:sens_surface}
\resizebox{\textwidth}{!}{
{\fontsize{10pt}{10pt}\selectfont
\renewcommand{\arraystretch}{1.5}
\begin{tabular}{llcccccc}
\toprule
Config & Method & Var & ESS & Corr & \% Var $\downarrow$ (grid) & POT & Value \\
\midrule
\multicolumn{8}{l}{\textit{$n=200$}} \\
Config 1
& nIPW & 504.14 (237.94) & 76.80 (17.49) & 0.728 (0.102) & --- & 0.743 (0.189) & 1118.76 (33.65) \\
& nGCW & 447.57 (206.34) & 82.33 (18.69) & 0.742 (0.098) & 100.00 & 0.751 (0.185) & 1119.96 (32.31) \\
& nBCW & 489.04 (224.36) & 77.28 (17.30) & 0.738 (0.099) & 81.31 & 0.757 (0.179) & 1120.99 (30.79) \\
\cmidrule(lr){2-8}
Config 2
& nIPW & 504.14 (237.94) & 76.80 (17.49) & 0.728 (0.102) & --- & 0.743 (0.189) & 1118.76 (33.65) \\
& nGCW & 321.08 (132.59) & 99.35 (21.92) & 0.781 (0.087) & 100.00 & 0.790 (0.160) & 1126.02 (24.95) \\
& nBCW & 473.49 (216.54) & 78.36 (17.48) & 0.711 (0.106) & 97.76 & 0.767 (0.177) & 1121.99 (30.63) \\
\midrule
\midrule
\multicolumn{8}{l}{\textit{$n=500$}} \\
Config 1
& nIPW & 190.06 (87.84) & 192.69 (44.28) & 0.865 (0.056) & --- & 0.829 (0.125) & 1131.90 (17.35) \\
& nGCW & 176.88 (80.54) & 201.39 (46.24) & 0.871 (0.054) & 100.00 & 0.835 (0.123) & 1132.51 (16.81) \\
& nBCW & 186.85 (85.18) & 193.36 (43.95) & 0.869 (0.054) & 63.71 & 0.833 (0.122) & 1132.31 (16.98) \\
\cmidrule(lr){2-8}
Config 2
& nIPW & 190.06 (87.84) & 192.69 (44.28) & 0.865 (0.056) & --- & 0.829 (0.125) & 1131.90 (17.35) \\
& nGCW & 143.83 (61.67) & 228.25 (51.88) & 0.887 (0.048) & 100.00 & 0.854 (0.106) & 1134.73 (12.78) \\
& nBCW & 178.17 (80.93) & 196.75 (44.39) & 0.841 (0.061) & 98.75 & 0.869 (0.097) & 1135.98 (11.61) \\
\midrule
\midrule
\multicolumn{8}{l}{\textit{$n=1000$}} \\
Config 1
& nIPW & 104.18 (42.60) & 384.85 (89.01) & 0.922 (0.035) & --- & 0.878 (0.084) & 1137.24 (8.45) \\
& nGCW & 98.98 (39.82) & 397.15 (91.82) & 0.925 (0.034) & 100.00 & 0.879 (0.084) & 1137.32 (8.42) \\
& nBCW & 102.94 (41.46) & 385.87 (88.50) & 0.924 (0.035) & 50.22 & 0.879 (0.083) & 1137.31 (8.39) \\
\cmidrule(lr){2-8}
Config 2
& nIPW & 104.18 (42.60) & 384.85 (89.01) & 0.922 (0.035) & --- & 0.878 (0.084) & 1137.24 (8.45) \\
& nGCW & 85.36 (32.34) & 434.95 (100.05) & 0.932 (0.031) & 100.00 & 0.887 (0.074) & 1138.20 (6.16) \\
& nBCW & 97.43 (38.18) & 393.14 (89.06) & 0.896 (0.041) & 99.03 & 0.892 (0.065) & 1138.63 (6.21) \\
\bottomrule
\end{tabular}}}
\end{table}

\FloatBarrier
\begin{table}[htbp]
\centering
\caption{Sensitivity analysis of nAIPW, nAGCW and nABCW for threshold-level performance under two tuning configurations. Measurements are reported as mean (standard deviation) across 500 Monte Carlo replications.}
\label{tab:sens_threshold_aug}
\resizebox{\textwidth}{!}{
{\fontsize{10pt}{10pt}\selectfont
\renewcommand{\arraystretch}{1.5}
\begin{tabular}{llcccccccc}
\toprule
Config & Method & \multicolumn{2}{c}{Mean (SD)} & \multicolumn{2}{c}{Median (IQR)} & \multicolumn{2}{c}{RMSE} & \multicolumn{2}{c}{Coverage} \\
\cmidrule(lr){3-4} \cmidrule(lr){5-6} \cmidrule(lr){7-8} \cmidrule(lr){9-10}
& & $\psi_1$ & $\psi_2$ & $\psi_1$ & $\psi_2$ & $\psi_1$ & $\psi_2$ & $\psi_1$ & $\psi_2$ \\
\midrule
\multicolumn{10}{l}{\textit{$n=200$}} \\
Config 1
& nAIPW & 330.98 (60.28) & 448.30 (23.28) & 345.00 (62.50) & 450.00 (30.00) & 43.77 & 18.26 & 0.967 & 0.986 \\
& nAGCW & 331.40 (58.81) & 448.88 (23.38) & 345.00 (60.00) & 450.00 (30.00) & 42.60 & 18.47 & 0.967 & 0.981 \\
& nABCW & 328.54 (57.30) & 454.00 (23.67) & 340.00 (80.00) & 455.00 (25.00) & 44.69 & 18.77 & 0.938 & 1.000 \\
\cmidrule(lr){2-10}
Config 2
& nAIPW & 338.00 (52.06) & 452.15 (21.70) & 345.00 (55.00) & 455.00 (25.00) & 37.85 & 17.23 & 0.908 & 1.000 \\
& nAGCW & 337.46 (55.78) & 451.77 (23.24) & 350.00 (50.00) & 455.00 (30.00) & 39.77 & 19.15 & 0.954 & 0.985 \\
& nABCW & 322.09 (63.89) & 447.53 (26.70) & 340.00 (82.50) & 450.00 (35.00) & 50.33 & 21.44 & 0.949 & 0.995 \\
\midrule
\midrule
\multicolumn{10}{l}{\textit{$n=500$}} \\
Config 1
& nAIPW & 335.93 (38.44) & 449.47 (18.06) & 340.00 (42.50) & 450.00 (25.00) & 29.00 & 14.13 & 0.987 & 0.987 \\
& nAGCW & 335.60 (37.69) & 449.07 (17.93) & 340.00 (40.00) & 450.00 (25.00) & 28.27 & 14.13 & 0.987 & 0.987 \\
& nABCW & 332.50 (51.67) & 453.75 (14.19) & 345.00 (56.25) & 455.00 (20.00) & 37.33 & 11.58 & 1.000 & 0.983 \\
\cmidrule(lr){2-10}
Config 2
& nAIPW & 332.25 (55.25) & 454.92 (14.10) & 345.00 (47.50) & 455.00 (20.00) & 37.92 & 11.42 & 1.000 & 0.967 \\
& nAGCW & 328.17 (52.65) & 452.75 (17.45) & 345.00 (51.25) & 455.00 (20.00) & 36.17 & 13.58 & 1.000 & 0.950 \\
& nABCW & 320.80 (41.69) & 447.73 (21.19) & 325.00 (60.00) & 450.00 (30.00) & 36.67 & 17.47 & 0.947 & 0.987 \\
\midrule
\midrule
\multicolumn{10}{l}{\textit{$n=1000$}} \\
Config 1
& nAIPW & 346.75 (26.60) & 450.88 (13.82) & 350.00 (40.00) & 450.00 (20.00) & 21.17 & 11.04 & 0.992 & 0.967 \\
& nAGCW & 345.62 (26.03) & 451.21 (13.89) & 350.00 (35.00) & 450.00 (20.00) & 21.13 & 11.21 & 0.992 & 0.967 \\
& nABCW & 345.08 (29.05) & 449.10 (14.84) & 350.00 (40.00) & 450.00 (20.00) & 23.16 & 11.64 & 0.970 & 0.990 \\
\cmidrule(lr){2-10}
Config 2
& nAIPW & 345.04 (27.81) & 449.38 (13.69) & 345.00 (40.00) & 450.00 (20.00) & 22.18 & 10.98 & 0.976 & 0.980 \\
& nAGCW & 344.69 (27.30) & 450.00 (13.54) & 345.00 (40.00) & 450.00 (20.00) & 21.89 & 10.80 & 0.966 & 0.978 \\
& nABCW & 339.04 (33.37) & 451.08 (14.68) & 340.00 (45.00) & 450.00 (20.00) & 26.79 & 12.00 & 0.958 & 0.975 \\
\bottomrule
\end{tabular}}}
\end{table}

\FloatBarrier
\begin{table}[htbp]
\centering
\caption{Sensitivity analysis of nAIPW, nAGCW and nABCW for value-surface-level performance under two tuning configurations. Measurements are reported as mean (standard deviation) across 500 Monte Carlo replications.}
\label{tab:sens_surface_aug}
\resizebox{\textwidth}{!}{
{\fontsize{10pt}{10pt}\selectfont
\renewcommand{\arraystretch}{1.5}
\begin{tabular}{llcccccc}
\toprule
Config & Method & Var & ESS & Corr & \% Var $\downarrow$ (grid) & POT & Value \\
\midrule
\multicolumn{8}{l}{\textit{$n=200$}} \\
Config 1
& nAIPW & 272.92 (48.53) & 76.92 (17.54) & 0.907 (0.049) & --- & 0.903 (0.067) & 1139.36 (4.62) \\
& nAGCW & 254.19 (45.45) & 82.47 (18.74) & 0.910 (0.048) & 100.00 & 0.904 (0.067) & 1139.48 (4.57) \\
& nABCW & 323.39 (53.14) & 77.38 (17.42) & 0.909 (0.052) & 50.25 & 0.896 (0.063) & 1139.18 (4.24) \\
\cmidrule(lr){2-8}
Config 2
& nAIPW & 327.17 (53.59) & 76.92 (17.33) & 0.911 (0.053) & --- & 0.910 (0.056) & 1140.14 (3.45) \\
& nAGCW & 262.85 (39.20) & 99.44 (21.57) & 0.917 (0.049) & 100.00 & 0.903 (0.059) & 1139.62 (3.67) \\
& nABCW & 268.39 (47.47) & 77.75 (17.69) & 0.900 (0.051) & 69.88 & 0.890 (0.069) & 1138.64 (4.71) \\
\midrule
\midrule
\multicolumn{8}{l}{\textit{$n=500$}} \\
Config 1
& nAIPW & 96.82 (19.06) & 192.01 (44.19) & 0.960 (0.020) & --- & 0.928 (0.046) & 1141.03 (2.42) \\
& nAGCW & 92.71 (17.83) & 200.68 (46.16) & 0.961 (0.020) & 100.00 & 0.930 (0.046) & 1141.09 (2.41) \\
& nABCW & 105.39 (18.99) & 192.85 (44.68) & 0.959 (0.023) & 45.09 & 0.917 (0.060) & 1140.23 (3.44) \\
\cmidrule(lr){2-8}
Config 2
& nAIPW & 106.59 (19.28) & 191.89 (44.55) & 0.959 (0.023) & --- & 0.917 (0.063) & 1140.10 (3.83) \\
& nAGCW & 90.73 (15.68) & 227.40 (52.21) & 0.962 (0.022) & 100.00 & 0.921 (0.061) & 1140.25 (3.83) \\
& nABCW & 95.04 (18.94) & 194.44 (44.97) & 0.952 (0.022) & 64.23 & 0.914 (0.055) & 1140.35 (3.10) \\
\midrule
\midrule
\multicolumn{8}{l}{\textit{$n=1000$}} \\
Config 1
& nAIPW & 59.34 (9.33) & 385.62 (89.09) & 0.980 (0.011) & --- & 0.944 (0.034) & 1141.77 (1.03) \\
& nAGCW & 57.66 (9.05) & 397.89 (91.87) & 0.980 (0.011) & 100.00 & 0.943 (0.033) & 1141.78 (0.97) \\
& nABCW & 57.34 (9.00) & 386.95 (89.07) & 0.980 (0.012) & 59.59 & 0.940 (0.036) & 1141.63 (1.26) \\
\cmidrule(lr){2-8}
Config 2
& nAIPW & 57.87 (8.99) & 384.85 (89.01) & 0.980 (0.012) & --- & 0.942 (0.035) & 1141.71 (1.20) \\
& nAGCW & 52.86 (8.02) & 434.95 (100.05) & 0.981 (0.011) & 100.00 & 0.943 (0.034) & 1141.72 (1.19) \\
& nABCW & 58.21 (9.30) & 390.99 (90.18) & 0.975 (0.013) & 64.61 & 0.933 (0.037) & 1141.27 (1.74) \\
\bottomrule
\end{tabular}}}
\end{table}

\subsection{Comparaison between Analytical and Monte Carlo Variances}
Table~\ref{tab:Sim1_t5} compares the analytical variance for nGCW and nAGCW with Monte Carlo variance estimates across all sample sizes and all configurations. The reference configuration is $(\lambda_{\text{bias}}, c, k) = (1, 1, 0.5)$ used in the main text.
\FloatBarrier
\begin{table}[htbp]
\centering
\caption{Comparison of Monte Carlo and analytical variance of nGCW and nAGCW across tuning configurations. Measurements are reported as mean (standard deviation). The reference configuration is $(\lambda_{\text{bias}}, c, k) = (1, 1, 0.5)$.}
\label{tab:Sim1_t5}
\resizebox{\textwidth}{!}{
\begin{tabular}{ccccccccccccc}
\toprule
\multirow{2}{*}{Config} &
\multirow{2}{*}{Variance} &
\multicolumn{2}{c}{n=200} &
\multicolumn{2}{c}{n=500} &
\multicolumn{2}{c}{n=1000} \\
\cmidrule(lr){3-4} \cmidrule(lr){5-6} \cmidrule(lr){7-8}
& & nGCW & nAGCW & nGCW & nAGCW & nGCW & nAGCW \\
\midrule

\multirow{2}{*}{Reference}
& Monte Carlo
& 399.00 (178.52) & 246.96 (38.23)
& 164.84 (73.77)  & 97.99 (15.67)
& 94.12 (37.19)   & 54.89 (8.47) \\
& Analytical
& 484.59 (192.38)    & 257.36 (36.63)
& 206.82 (85.25)   & 105.21 (15.57)
& 106.71 (45.24)   & 53.27 (8.05) \\
\midrule

\multirow{2}{*}{Config 1}
& Monte Carlo
& 447.57 (206.34) & 254.19 (45.45)
& 176.88 (80.54)  & 92.71 (17.83)
& 98.98 (39.82)   & 57.66 (9.05) \\
& Analytical
& 526.16 (216.77) & 264.92 (39.53)
& 218.78 (91.89)  & 108.51 (16.50)
& 111.23 (47.70)  & 54.59 (8.39) \\
\midrule

\multirow{2}{*}{Config 2}
& Monte Carlo
& 321.08 (132.59) & 262.85 (39.20)
& 143.83 (61.67)  & 90.73 (15.68)
& 85.36 (32.34)   & 52.86 (8.02) \\
& Analytical
& 428.39 (151.34) & 270.32 (32.21)
& 188.05 (73.26)  & 105.35 (13.82)
& 99.05 (40.67)   & 52.13 (7.45) \\
\bottomrule
\end{tabular}}
\end{table}

\FloatBarrier
\subsection{Distribution of Effective Sample Size}
We use effective sample size (ESS) to assess the stability and efficiency of the weighting schemes. To illustrate ESS gains from adaptive weighting, we plot the differences in mean ESS between nGCW/nBCW and nIPW, using histograms with overlaid kernel density estimations (KDEs) and 95\% confidence intervals (CIs). Across all sample sizes, nGCW estimator consistently achieves higher ESS, indicating more stable weights. The nBCW estimator also exhibits ESS gains, although with greater heterogeneity across the threshold grid due to their localized weighting mechanism. For augmented versions, ESS are similar and thus omitted.

\FloatBarrier
\begin{figure}[htbp]
    \centering

    \begin{minipage}{0.48\linewidth}
        \centering
        \includegraphics[width=\linewidth]{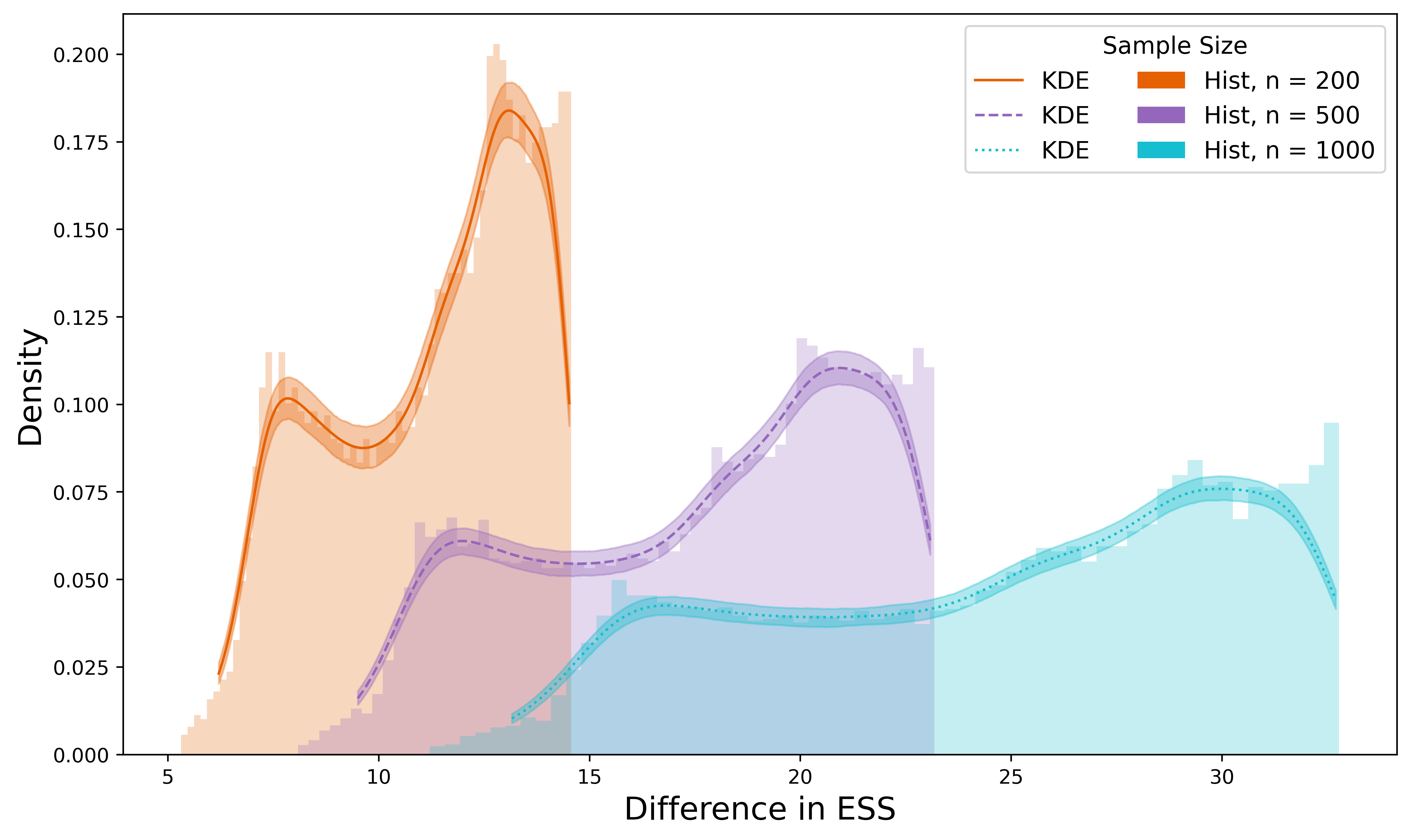}
        \scriptsize (a) Mean: nGCW $-$ nIPW
    \end{minipage}
    \hfill
    \begin{minipage}{0.48\linewidth}
        \centering
        \includegraphics[width=\linewidth]{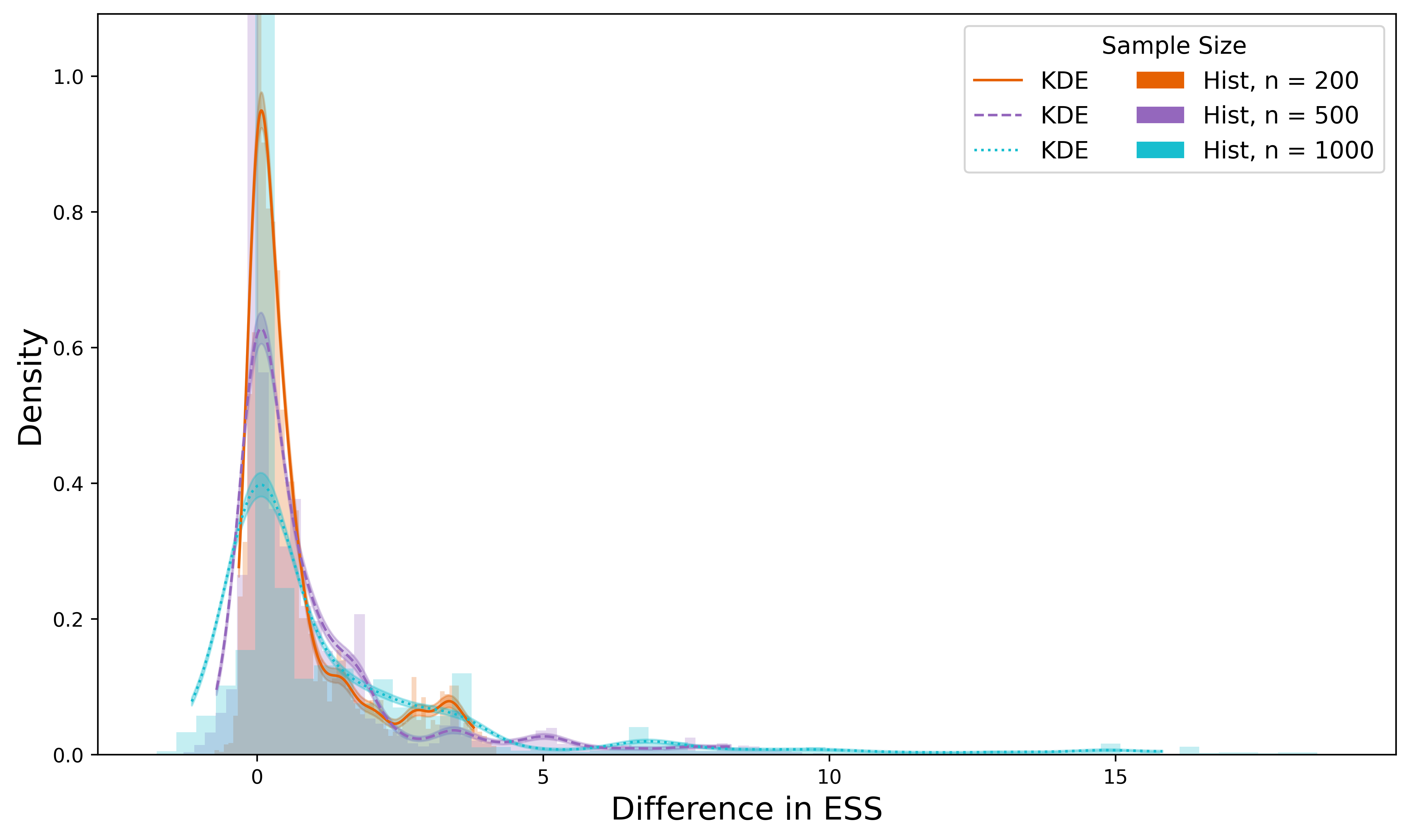}
        \scriptsize (b) Mean: nBCW $-$ nIPW
    \end{minipage}

    \vspace{0.8em}

    \caption{Simulation 1: Differences in mean ESS for nGCW and nBCW relative to nIPW across sample sizes. Each panel displays the distribution of differences in ESS, visualized using histograms, KDE curves, and 95\% confidence intervals. Positive values indicate improved ESS.}
\end{figure}

\newpage
\section{Additional Results for Simulation 2}
The results follow the same structure as in Simulation 1: we present the true value surface, the estimated value surfaces for non-augmented and augmented estimators, and then the differences in effective sample size relative to nIPW and nAIPW, respectively.

\subsection{True Value Surface}
The true value surface obtains the optimal thresholds $(\psi_1, \psi_2) = (430, 80)$, with a value of 560.87. This surface serves as the benchmark for evaluating the accuracy of the estimated value surfaces obtained from different weighting methods.

\FloatBarrier
\begin{figure}[htbp]
    \centering
    \includegraphics[width=0.8\linewidth]{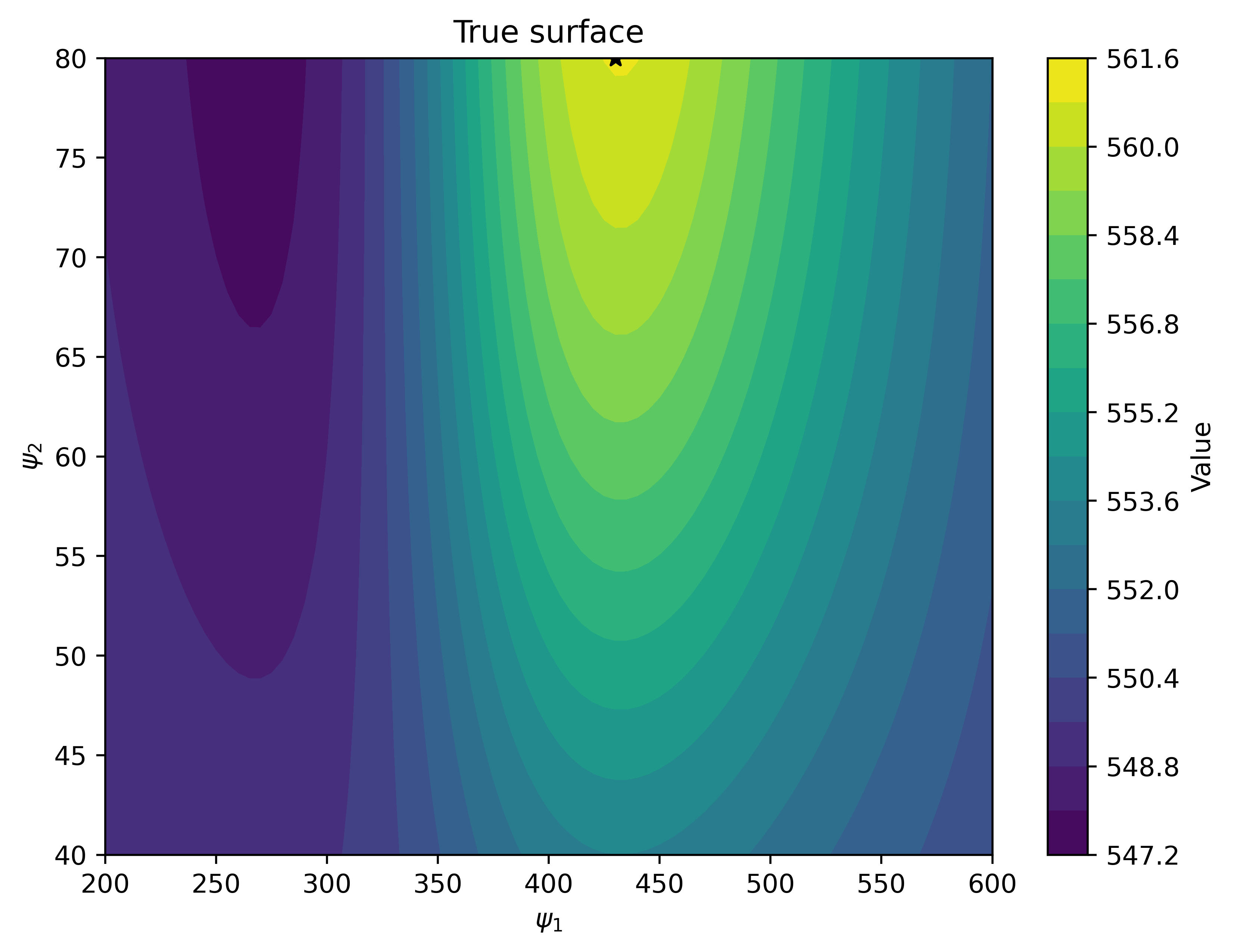}
    \caption{Simulation 2: True value surface}
    \label{fig:placeholder}
\end{figure}

\subsection{Estimated Value Surfaces}
We follow the same plotting structure as in Simulation 1.
\FloatBarrier
\begin{figure}[htbp]
    \centering
    
    \includegraphics[width=0.98\linewidth]{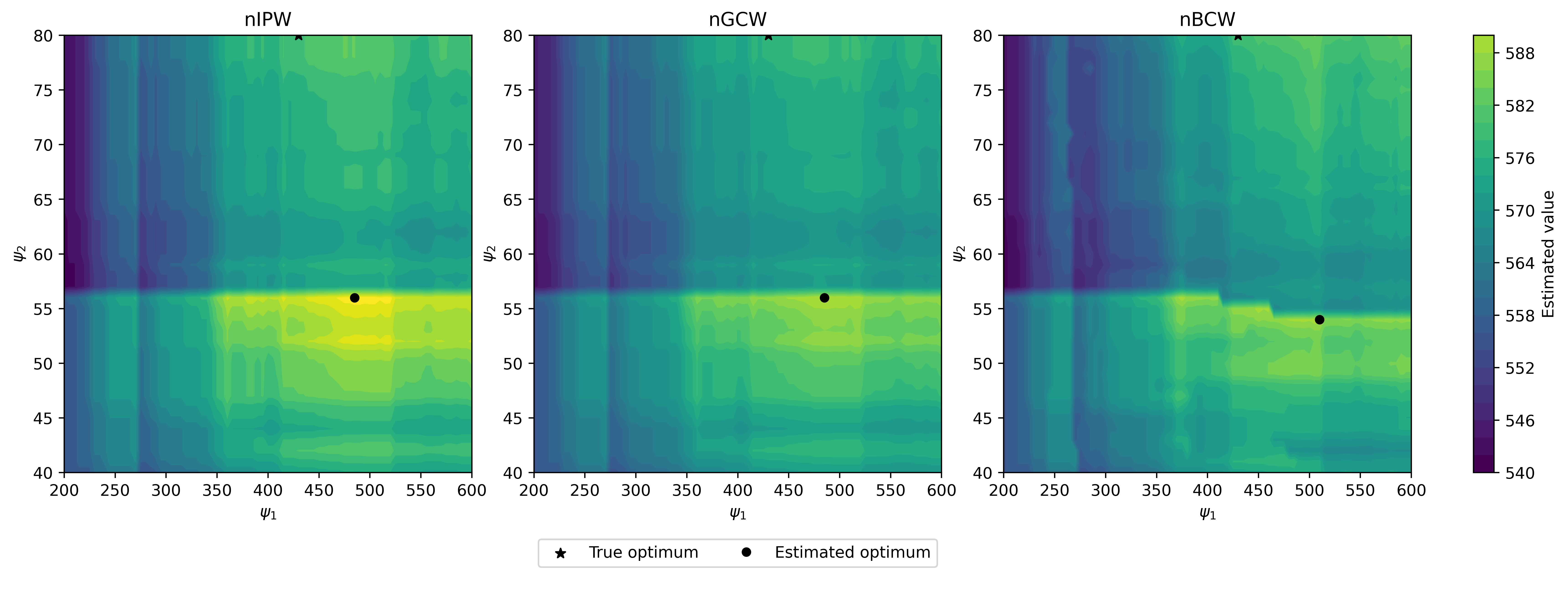}
    \scriptsize{\caption*{(a) Estimated value surfaces for $n=200$}}
    \vspace{0.5em}
    
    \includegraphics[width=0.98\linewidth]{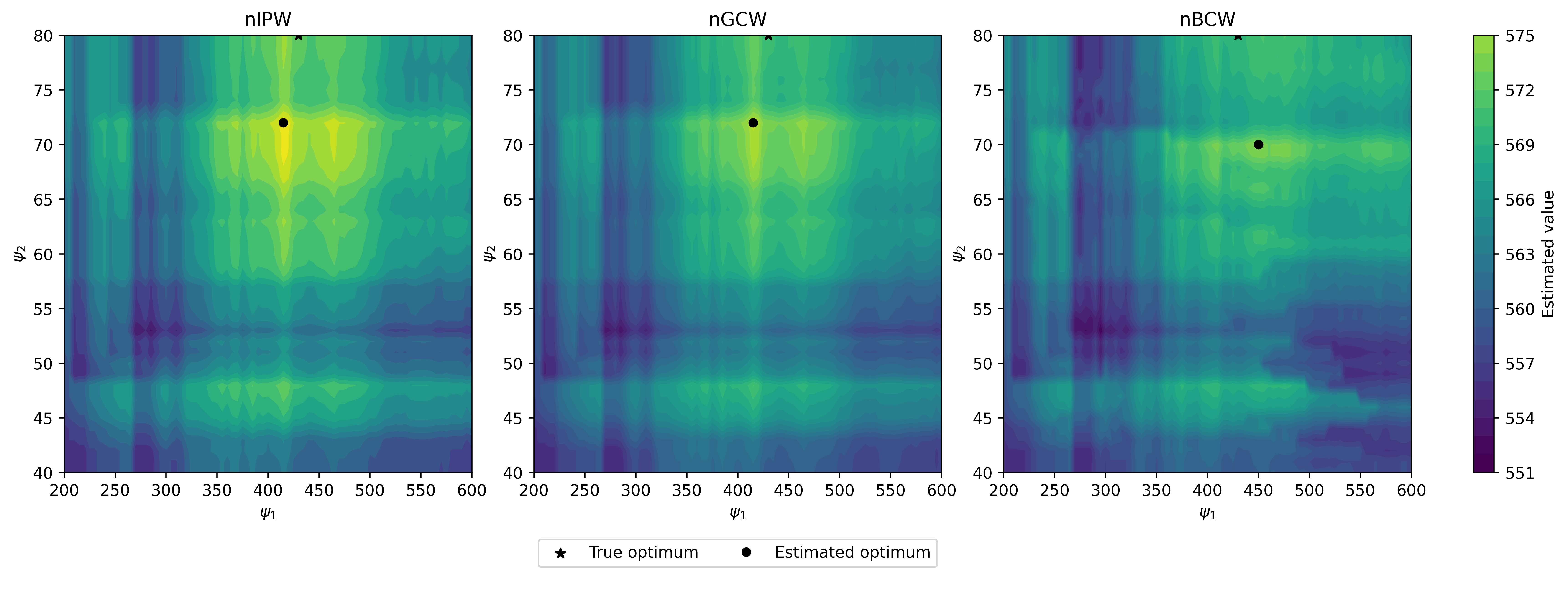}
    \scriptsize{\caption*{(b) Estimated value surfaces for $n=500$}}
    \vspace{0.5em}

    \includegraphics[width=0.98\linewidth]{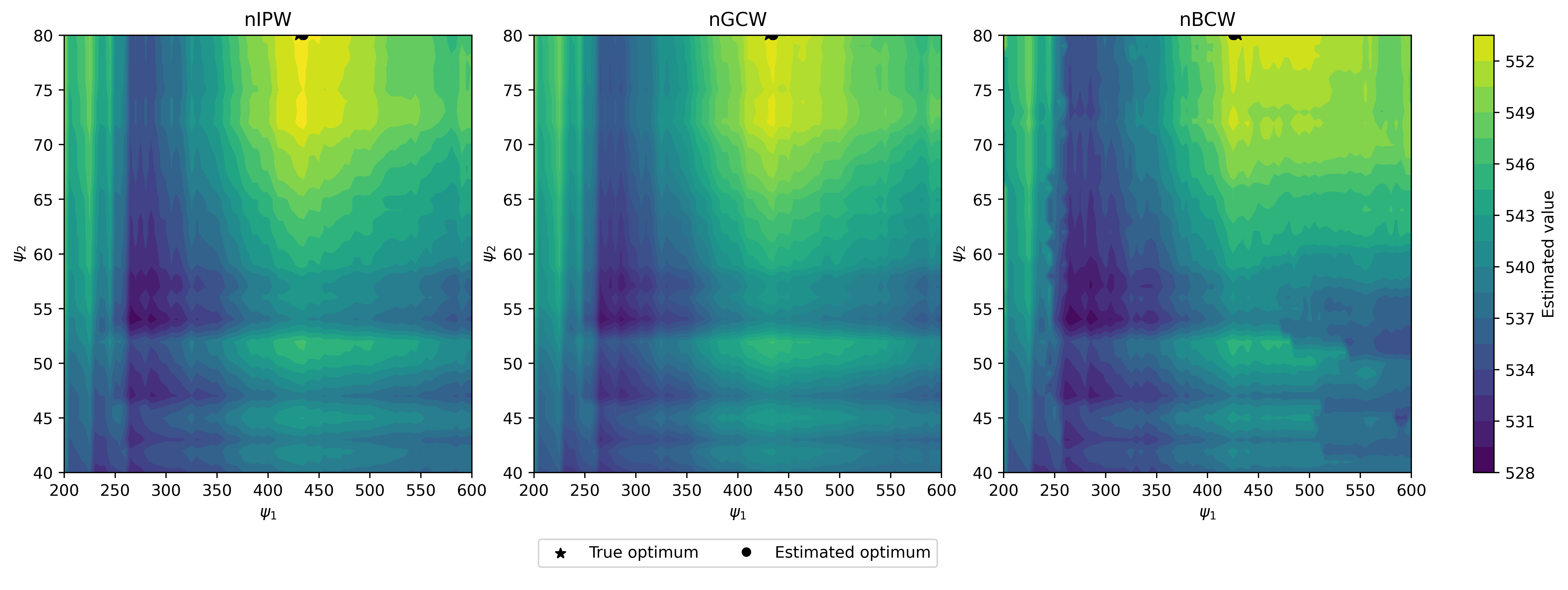}
    \scriptsize{\caption*{(c) Estimated value surfaces for $n=1000$}}

    \caption{Simulation 2: estimated value surfaces of non-augmented estimators for sample sizes $n=200$, $n=500$, and $n=1000$.}
\end{figure}

\begin{figure}[htbp]
    \centering
    \includegraphics[width=0.97\linewidth]{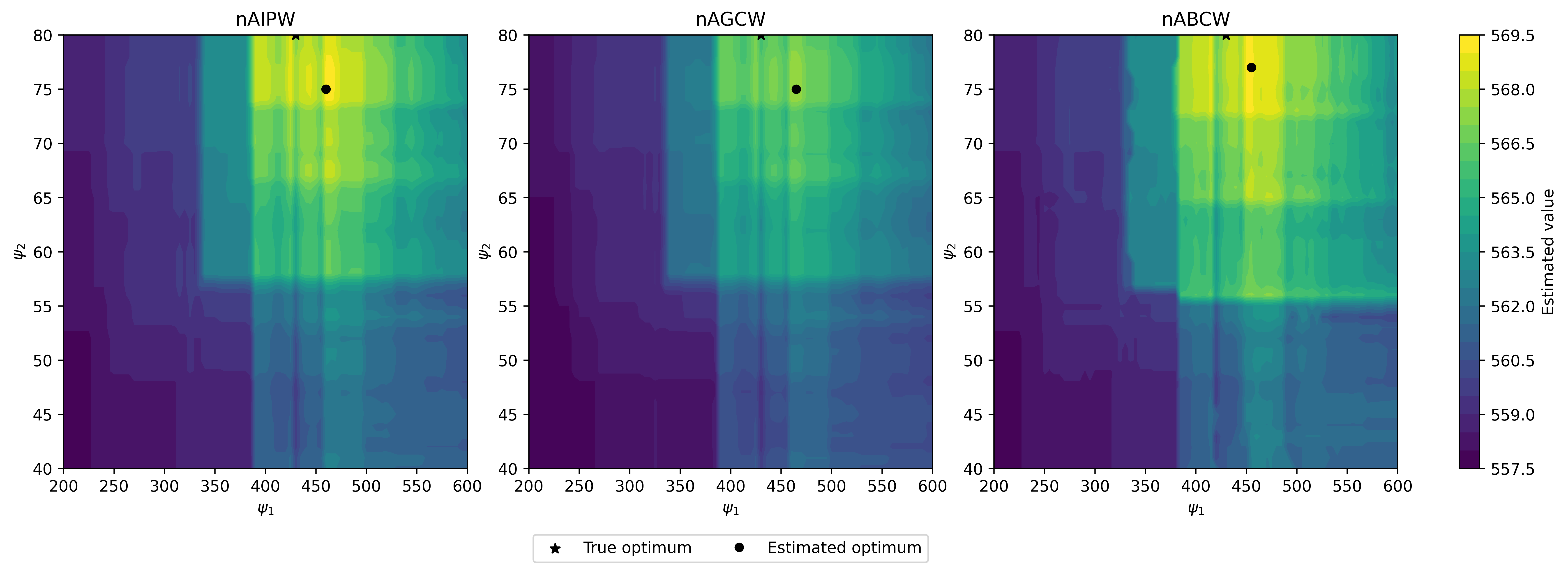}
    \scriptsize{\caption*{(a) Estimated value surfaces for $n=200$}}
    \vspace{0.5em}
    
    \includegraphics[width=0.97\linewidth]{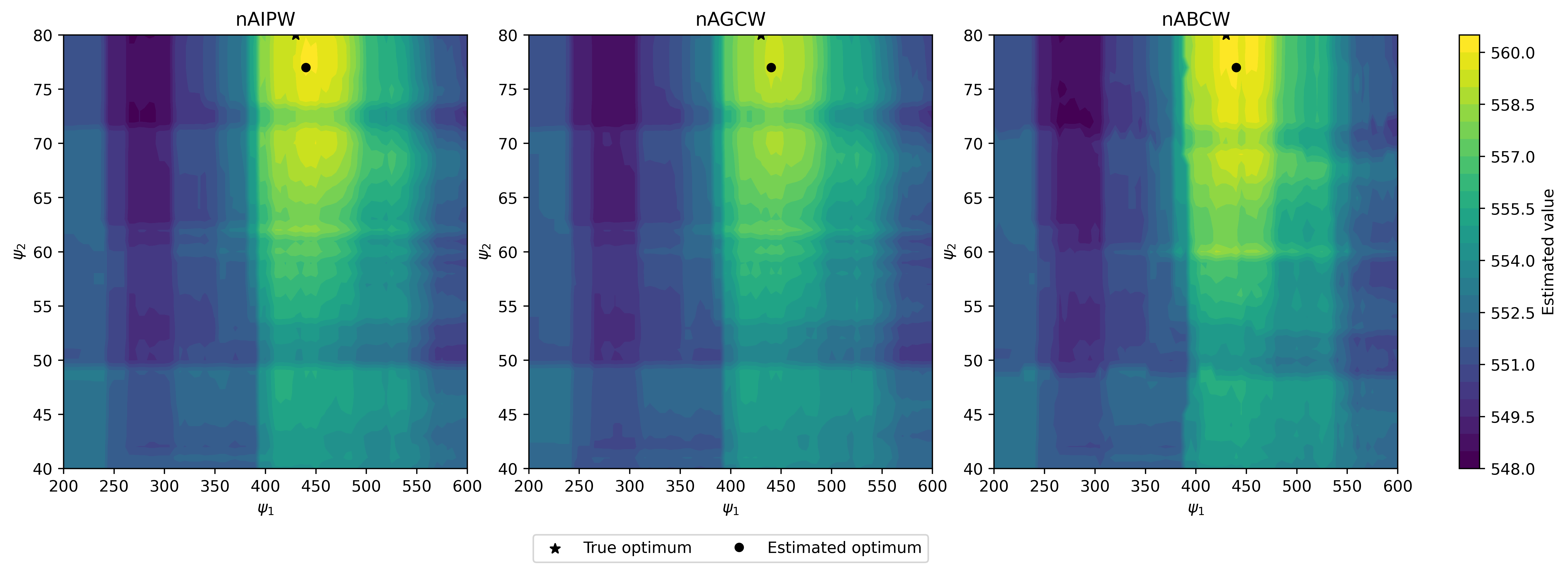}
    \scriptsize{\caption*{(b) Estimated value surfaces for $n=500$}}
    \vspace{0.5em}

    \includegraphics[width=0.97\linewidth]{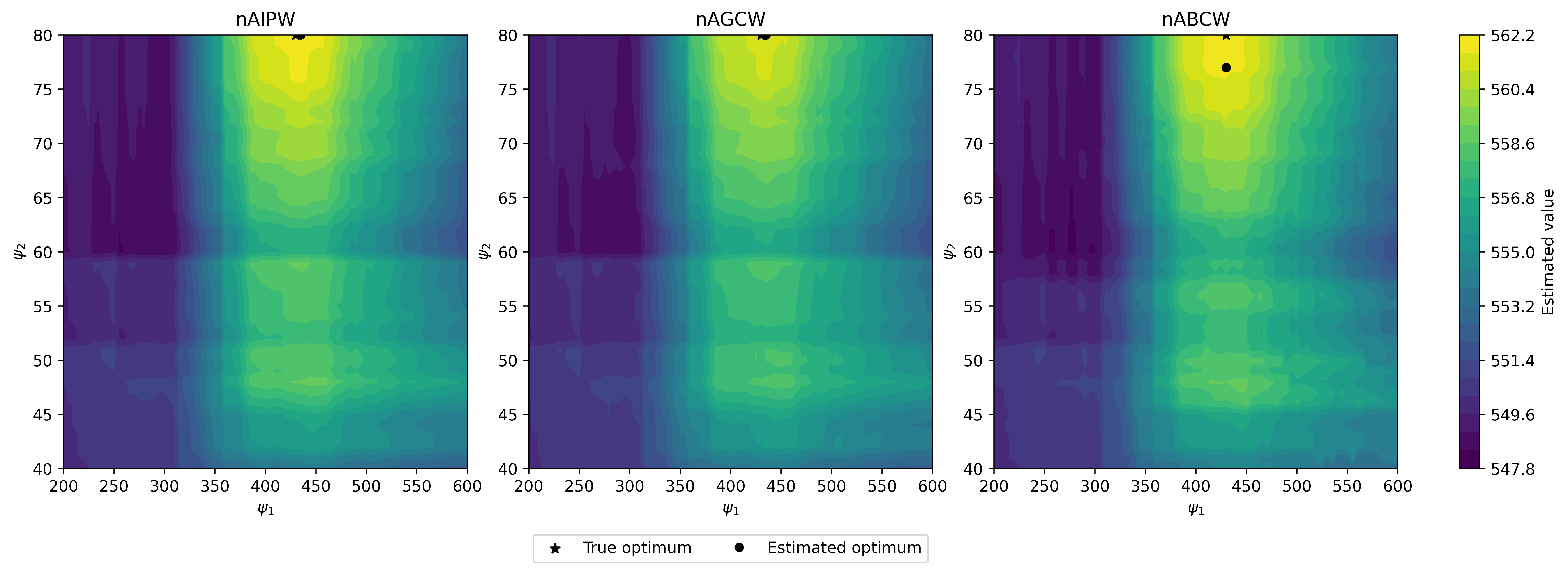}
    \scriptsize{\caption*{(c) Estimated value surfaces for $n=1000$}}

    \caption{Simulation 2: estimated value surfaces of augmented estimators for sample sizes $n=200$, $n=500$, and $n=1000$.}
\end{figure}

\FloatBarrier
\subsection{Results for Normalized Augmented Estimators}
We report additional results for the normalized augmented estimators using the same metrics as in the main paper. Tables~\ref{tab:Sim2_t3} and~\ref{tab:Sim2_t4} present the threshold-level and surface-level results, respectively.
\begin{table}[htbp]
\centering
\caption{Summary of estimated thresholds, RMSE, and coverage for nAIPW, nAGCW and nABCW based on 500 replications.}
\label{tab:Sim2_t3}
\setlength{\tabcolsep}{3pt}
{\fontsize{10pt}{10pt}\selectfont
\renewcommand{\arraystretch}{1.5}
\begin{tabular}{@{}ccccccccccc@{}}
\toprule
\multirow{2}{*}{$n$} & \multirow{2}{*}{Method} &
\multicolumn{2}{c}{Mean (SD)} &
\multicolumn{2}{c}{Median (IQR)} &
\multicolumn{2}{c}{RMSE} &
\multicolumn{2}{c}{Coverage} \\
\cmidrule(lr){3-4} \cmidrule(lr){5-6} \cmidrule(lr){7-8} \cmidrule(lr){9-10}
& & $\psi_1$ & $\psi_2$ & $\psi_1$ & $\psi_2$ & $\psi_1$ & $\psi_2$ & $\psi_1$ & $\psi_2$ \\
\midrule
200 & nAIPW  & 432.80 (30.09) & 71.70 (10.38) & 435.00 (25.00) & 77.00 (11.00) & 18.90 & 8.30 & 0.966 & 0.380 \\
    & nAGCW  & 432.79 (30.98) & 71.65 (10.52) & 435.00 (26.25) & 77.00 (11.00) & 19.47 & 8.35 & 0.966 & 0.380 \\
    & nABCW  & 424.78 (30.70) & 70.58 (10.07) & 425.00 (26.25) & 74.50 (12.00) & 20.16 & 9.42 & 0.986 & 0.274 \\
\midrule\midrule
500 & nAIPW  & 433.81 (13.13) & 75.41 (7.67) & 435.00 (15.00) & 79.00 (6.00) & 10.55 & 4.59 & 0.974 & 0.728 \\
    & nAGCW  & 433.79 (13.12) & 75.33 (7.76) & 435.00 (15.00) & 79.00 (6.00) & 10.53 & 4.67 & 0.974 & 0.730 \\
    & nABCW  & 428.00 (15.21) & 74.81 (7.53) & 425.00 (25.00) & 78.00 (6.00) & 12.28 & 5.19 & 0.978 & 0.610 \\
\midrule\midrule
1000 & nAIPW & 432.83 (9.29) & 77.49 (4.50) & 432.50 (15.00) & 80.00 (3.00) & 7.55 & 2.51 & 0.964 & 0.908 \\
     & nAGCW & 432.87 (9.31) & 77.37 (4.84) & 432.50 (15.00) & 80.00 (3.00) & 7.57 & 2.63 & 0.964 & 0.908 \\
     & nABCW & 425.80 (10.78) & 76.96 (4.47) & 425.00 (20.00) & 79.00 (4.00) & 9.58 & 3.04 & 0.982 & 0.902 \\
\bottomrule
\end{tabular}
}
\end{table}

\FloatBarrier
\begin{table}[htbp]
\centering
\caption{Performance of nAIPW, nAGCW and nABCW over all regimes. Measurements are reported as mean (standard deviation) across 500 Monte Carlo replications.}
\label{tab:Sim2_t4}
\resizebox{\textwidth}{!}{
{\fontsize{10pt}{10pt}\selectfont
\renewcommand{\arraystretch}{1.5}
\begin{tabular}{cccccccc}
\toprule
$n$ & Method & Var & ESS & Corr & \% Var $\downarrow$ (grid) & POT & Value \\
\midrule
200  & nAIPW & 153.57 (9.44) & 66.50 (24.84) & 0.756 (0.208) & ---    & 0.862 (0.159) & 558.75 (2.15) \\
     & nAGCW & 147.42 (8.07) & 76.42 (28.03) & 0.753 (0.209) & 100.00 & 0.859 (0.162) & 558.72 (2.21) \\
     & nABCW & 152.49 (9.10) & 67.39 (23.68) & 0.756 (0.205) & 84.52  & 0.848 (0.156) & 558.50 (2.13) \\
\midrule\midrule
500  & nAIPW & 61.75 (3.25) & 157.19 (55.38) & 0.889 (0.096) & ---    & 0.924 (0.105) & 559.62 (1.21) \\
     & nAGCW & 60.11 (2.94) & 171.84 (60.22) & 0.888 (0.096) & 100.00 & 0.923 (0.106) & 559.61 (1.23) \\
     & nABCW & 61.36 (3.06) & 159.76 (52.64) & 0.888 (0.096) & 79.77  & 0.916 (0.105) & 559.48 (1.23) \\
\midrule\midrule
1000 & nAIPW & 30.99 (1.52) & 301.91 (102.21) & 0.944 (0.039) & ---    & 0.958 (0.053) & 560.00 (0.58) \\
     & nAGCW & 30.45 (1.42) & 321.58 (108.63) & 0.943 (0.039) & 100.00 & 0.956 (0.059) & 559.98 (0.66) \\
     & nABCW & 30.81 (1.45) & 306.75 (97.08) & 0.944 (0.039) & 79.70  & 0.949 (0.051) & 559.87 (0.59) \\
\bottomrule
\end{tabular}}}
\end{table}

\subsection{Comparaison Between Analytical and Monte Carlo Variances}
Table~\ref{tab:Sim2_t5} compares the analytical variance for nGCW and nAGCW with Monte Carlo variance estimates across sample sizes.

\FloatBarrier
\begin{table}[htbp]
\centering
\caption{Comparison of analytical variance and Monte Carlo variance of nGCW and nAGCW. Measurements reported as mean (standard deviation).}
\label{tab:Sim2_t5}
\resizebox{\textwidth}{!}{
\begin{tabular}{ccccccc}
\toprule
\multirow{2}{*}{Method} & 
\multicolumn{2}{c}{n=200} & 
\multicolumn{2}{c}{n=500} & 
\multicolumn{2}{c}{n=1000} \\ 
\cmidrule(lr){2-3} \cmidrule(lr){4-5} \cmidrule(lr){6-7} 
 & nGCW & nAGCW & nGCW & nAGCW & nGCW & nAGCW \\
\midrule
Monte-Carlo & 320.27 (18.61) & 147.42 (8.07) & 128.77 (8.39) & 60.11 (2.94) & 61.79 (4.05) & 30.45 (1.42)\\
Analytical  & 312.69 (17.59) & 143.63 (6.66) & 125.03 (8.36) & 59.57 (3.13) & 60.68 (4.15) & 30.50 (1.59)\\
\bottomrule
\end{tabular}}
\end{table}

\subsection{Distribution of Effective Sample Size}
We follow the same plotting structure as in Simulation 1.
\FloatBarrier
\begin{figure}[htbp]
    \centering

    \begin{minipage}{0.48\linewidth}
        \centering
        \includegraphics[width=\linewidth]{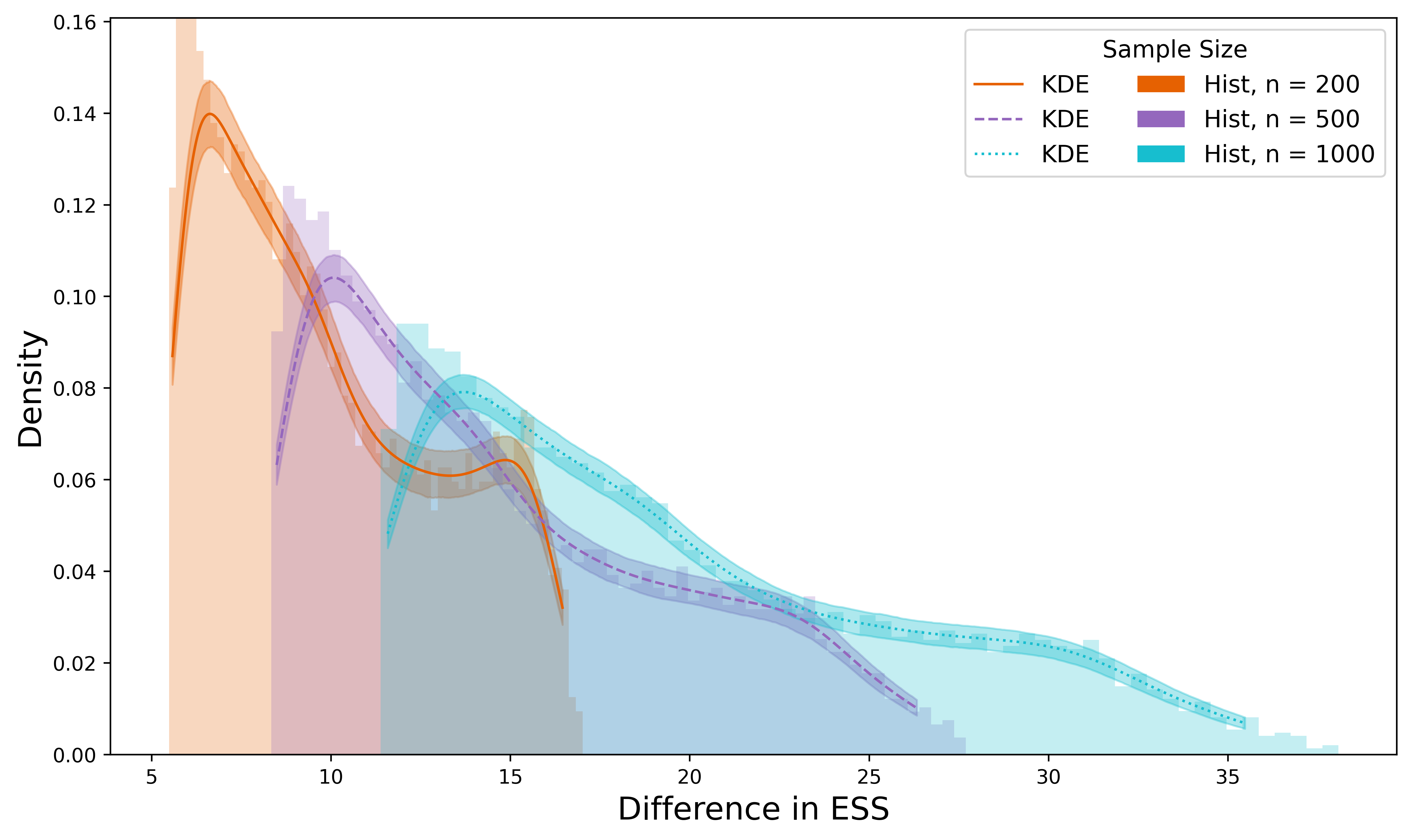}
        \scriptsize (a) Mean: nGCW $-$ nIPW
    \end{minipage}
    \hfill
    \begin{minipage}{0.48\linewidth}
        \centering
        \includegraphics[width=\linewidth]{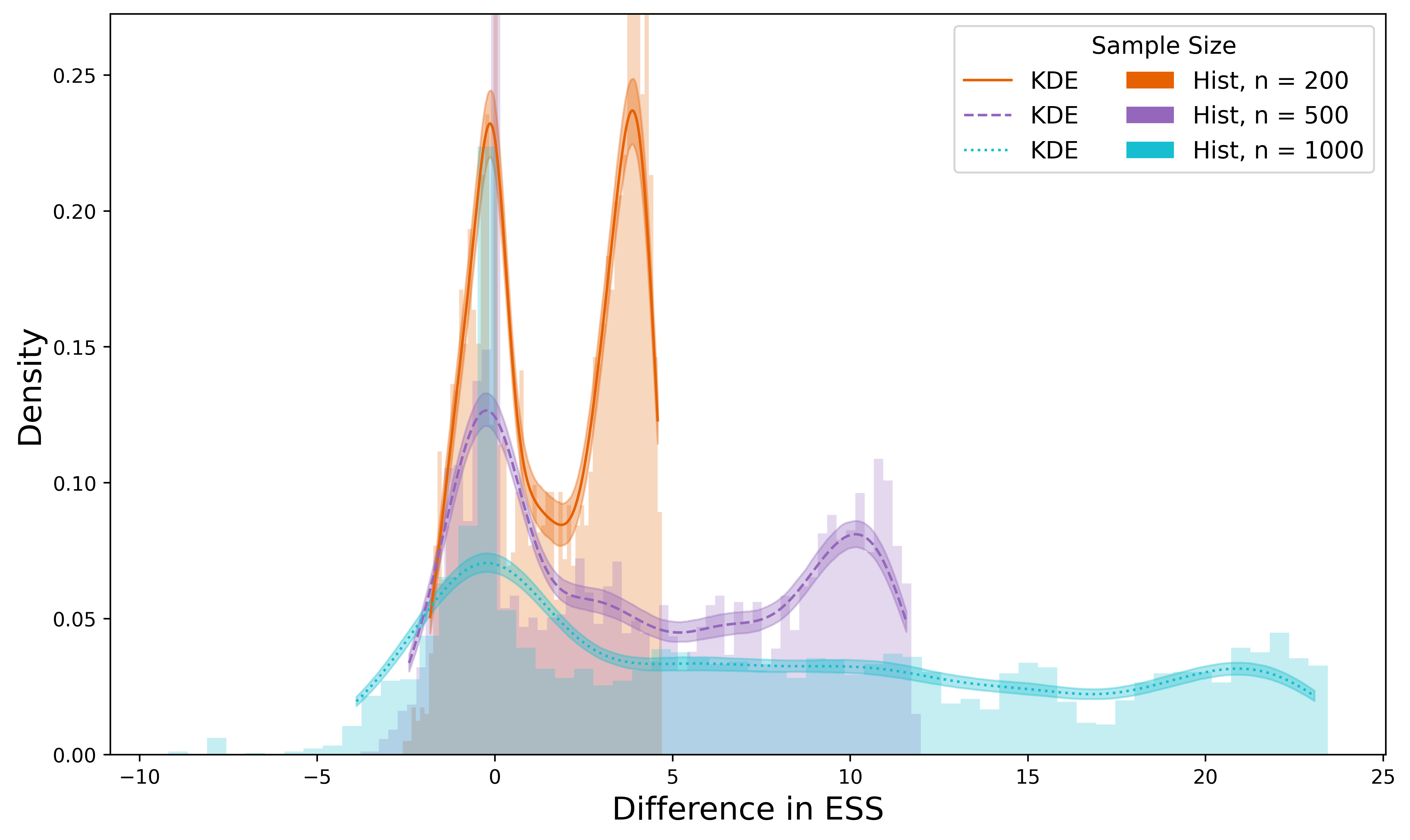}
        \scriptsize (b) Mean: nBCW $-$ nIPW
    \end{minipage}

    \vspace{0.8em}

    \caption{Simulation 2: Differences in mean ESS for nGCW and nBCW relative to nIPW across sample sizes. Each panel displays the distribution of differences in ESS, visualized using histograms, KDE curves, and 95\% confidence intervals. Positive values indicate improved ESS.}
\end{figure}

\newpage 
\section{Additional Results for ACTG175}
\subsection{Details of the ACTG175 Dataset}
In Table~\ref{tab:actg175-baseline}, we report the baseline characteristics of the patients stratified by treatments. ``ZDV+ddI'' refers to zidovudine plus didanosine, and ``ZDV+ddC'' refers to zidovudine plus zalcitabine.

\FloatBarrier
\begin{table}[htbp]
\centering
\caption{Baseline characteristics by treatment group. Continuous variables are summarized using mean (standard deviation), and categorical variables using number (percentage). SMD denotes the standardized mean difference. ARV history indicates prior antiretroviral treatment.}
\label{tab:actg175-baseline}
\begin{tabular}{llccc}
\toprule
\textbf{Variable} & \textbf{Level} & \textbf{ZDV+ddI} & \textbf{ZDV+ddC} & \textbf{SMD} \\
\midrule
\textbf{n} & & 519 & 524 & \\
\midrule
\multirow{2}{*}{\textbf{Race (\%)}} 
& 0 (while) & 383 (73.8) & 374 (71.4) & \multirow{2}{*}{0.054} \\
& 1 (non-white) & 136 (26.2) & 150 (28.6) & \\
\midrule
\multirow{2}{*}{\textbf{Sex (\%)}} 
& 0 (female) & 87 (16.8) & 89 (17.0) & \multirow{2}{*}{0.006} \\
& 1 (male) & 432 (83.2) & 435 (83.0) & \\
\midrule
\multirow{2}{*}{\textbf{Symptom (\%)}} 
& 0 (Asymptomatic) & 423 (81.5) & 435 (83.0) & \multirow{2}{*}{0.040} \\
& 1 (Symptomatic) & 96 (18.5) & 89 (17.0) & \\
\midrule
\multirow{2}{*}{\textbf{ARV history (\%)}} 
& 0 (naive) & 212 (40.8) & 212 (40.5) & \multirow{2}{*}{0.008} \\
& 1 (experienced) & 307 (59.2) & 312 (59.5) & \\
\midrule
\textbf{CD4 at baseline} & & 350.74 (127.83) & 352.77 (115.51) & 0.017 \\
\midrule
\textbf{Age} & & 35.17 (8.64) &	35.43 (8.80) & 0.030\\
\midrule
\textbf{Karnofsky score} & & 95.51 (5.77) & 95.71 (5.91) & 0.033 \\
\midrule
\textbf{Weight (kg)} & & 74.92 (13.62) & 74.71 (13.20) & 0.015 \\
\bottomrule
\end{tabular}
\end{table}

\subsection{Propensity Score Overlapping and SMD}
\label{appendix:PS}
\begin{figure}[htbp]
    \centering
    \begin{minipage}{0.48\linewidth}
        \centering
        \includegraphics[width=\linewidth]{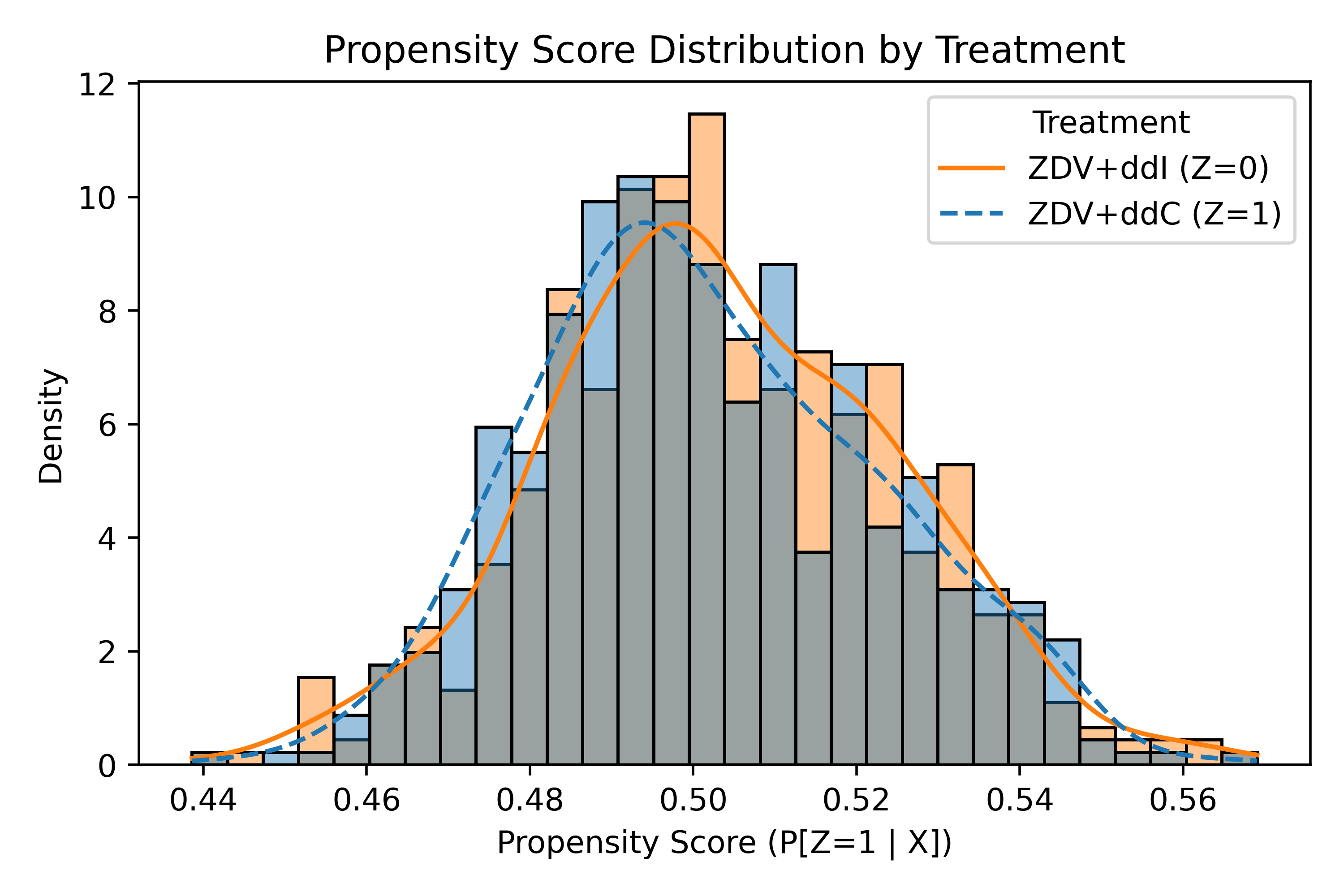}
        \scriptsize (a) Propensity score distribution
    \end{minipage}
    \hfill
    \begin{minipage}{0.48\linewidth}
        \centering
        \includegraphics[width=\linewidth]{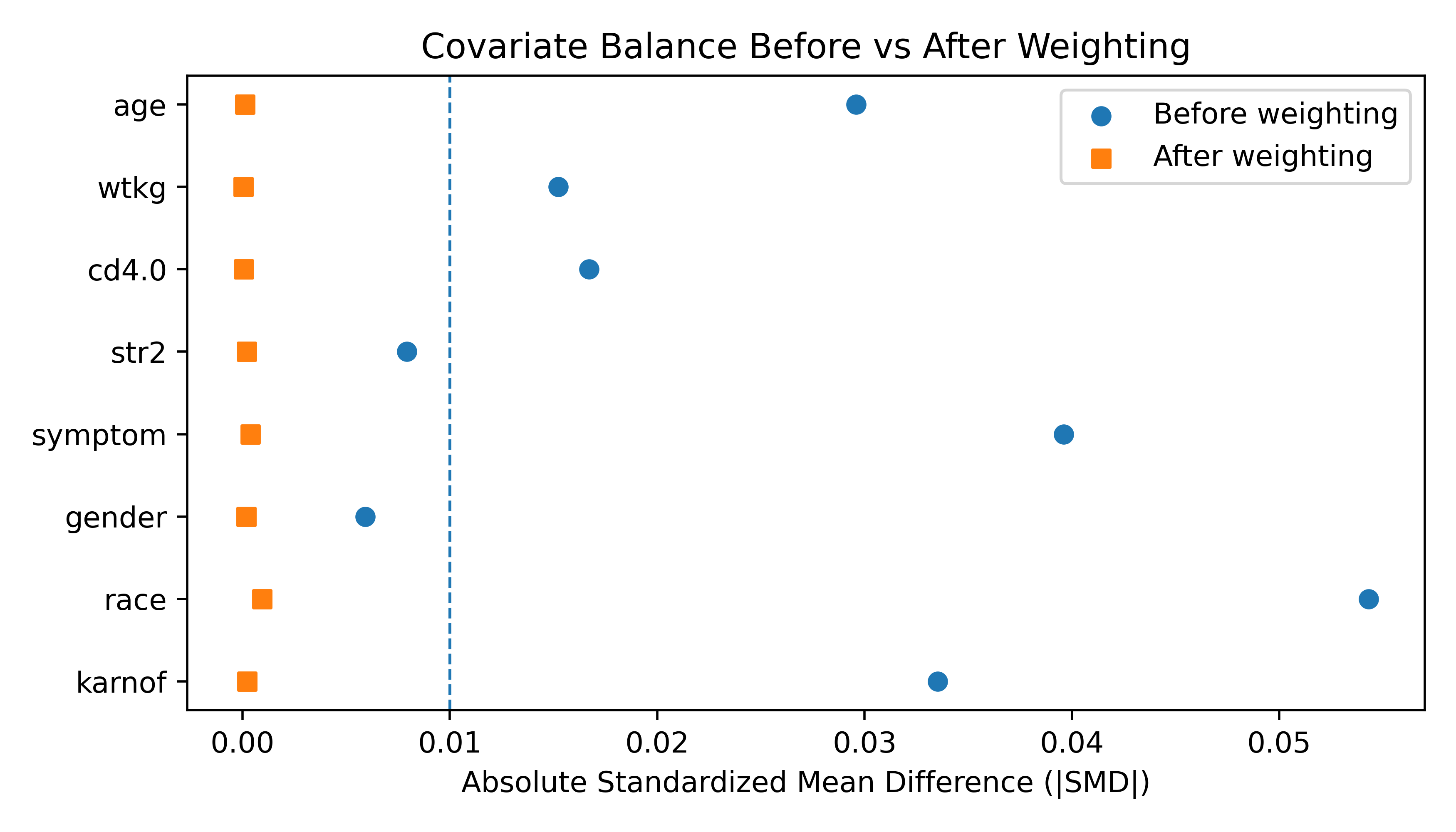}
        \scriptsize (b) Standardized mean differences
    \end{minipage}

    \vspace{0.8em}
    \caption{Propensity score overlapping and standardized mean differences}
    \label{fig:app_PS}
\end{figure}

Figure~\ref{fig:app_PS} provides a visualization of propensity score distribution by treatments and a summary of the standardized mean differences (SMDs) for key baseline covariates, before and after applying propensity score weights. The PS distribution shows satisfactory overlap and weighting reduces all SMDs to values near zero.
\FloatBarrier

\newpage
\subsection{Estimation Surfaces}
\begin{figure}[htbp]
    \centering
    \includegraphics[width=1\linewidth]{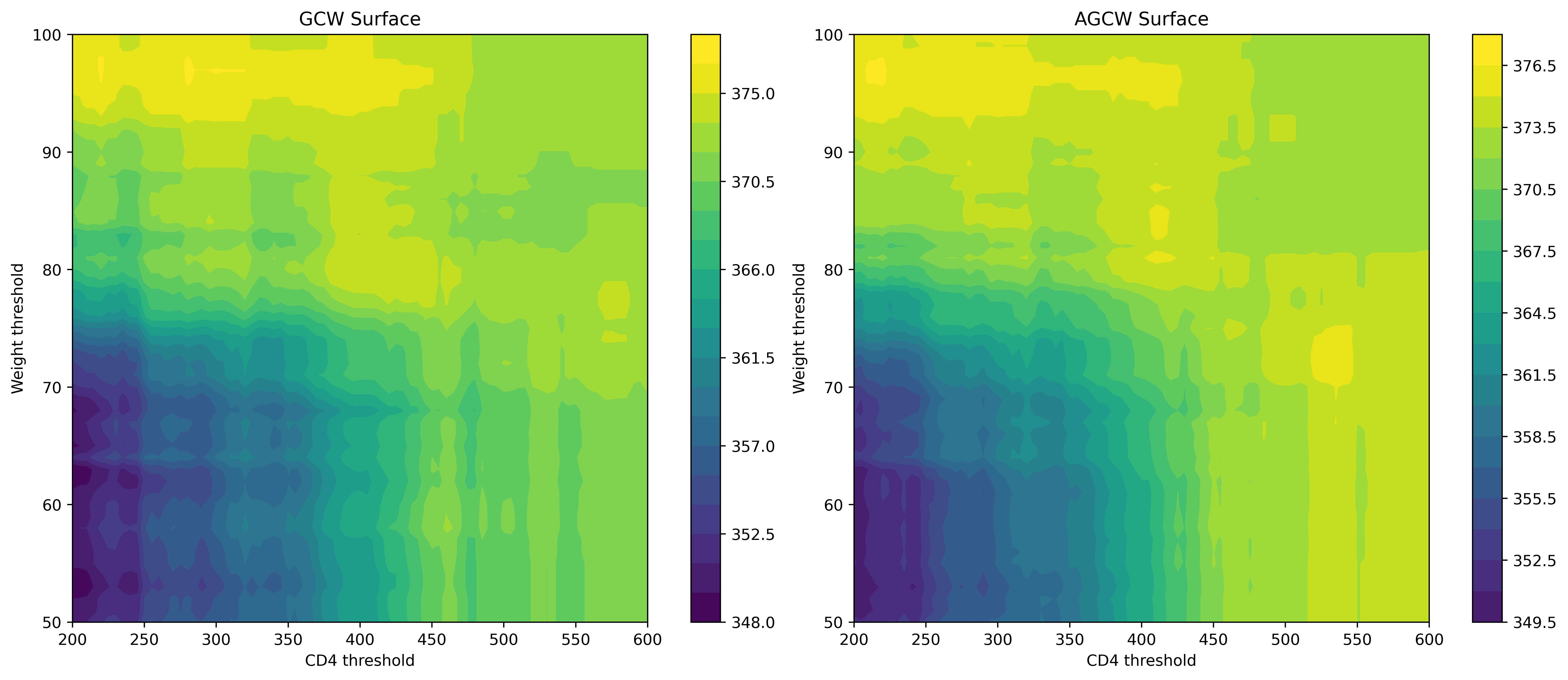}
    \caption{nGCW and nAGCW estimation surfaces}
    \label{fig:app_GCW}
\end{figure}

\begin{figure}[H]
    \centering
    \includegraphics[width=1\linewidth]{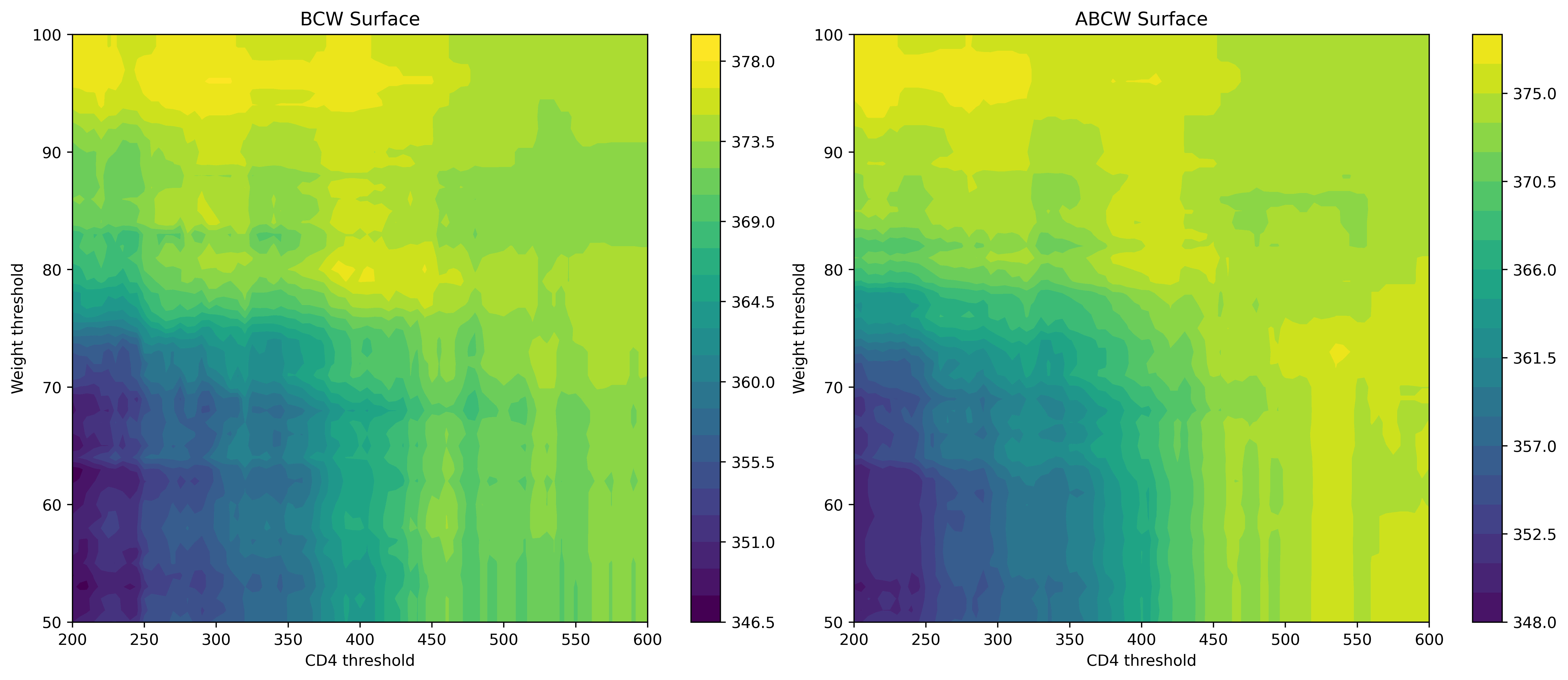}
    \caption{nBCW and nABCW estimation surfaces}
    \label{fig:app_BCW}
\end{figure}

Figures~\ref{fig:app_GCW} and ~\ref{fig:app_BCW} display the estimated value surfaces obtained using the n(A)GCW and n(A)BCW estimators. The overall shapes are similar to those obtained from nIPW and nAIPW. Consistent with the simulation results, augmentation smooths the surface and reduces local irregularities, especially in regions with limited effective sample size.
\FloatBarrier

\subsection{Weight Spread Distributions}
We additionally provide the $95^\text{th} - 5^\text{th}$ percentile spread and the max $-$ min spread for both non-augmented and augmented estimators.
\FloatBarrier
\begin{figure}[htbp]
    \centering
    \begin{subfigure}{0.48\linewidth}
        \centering
        \includegraphics[width=\linewidth]{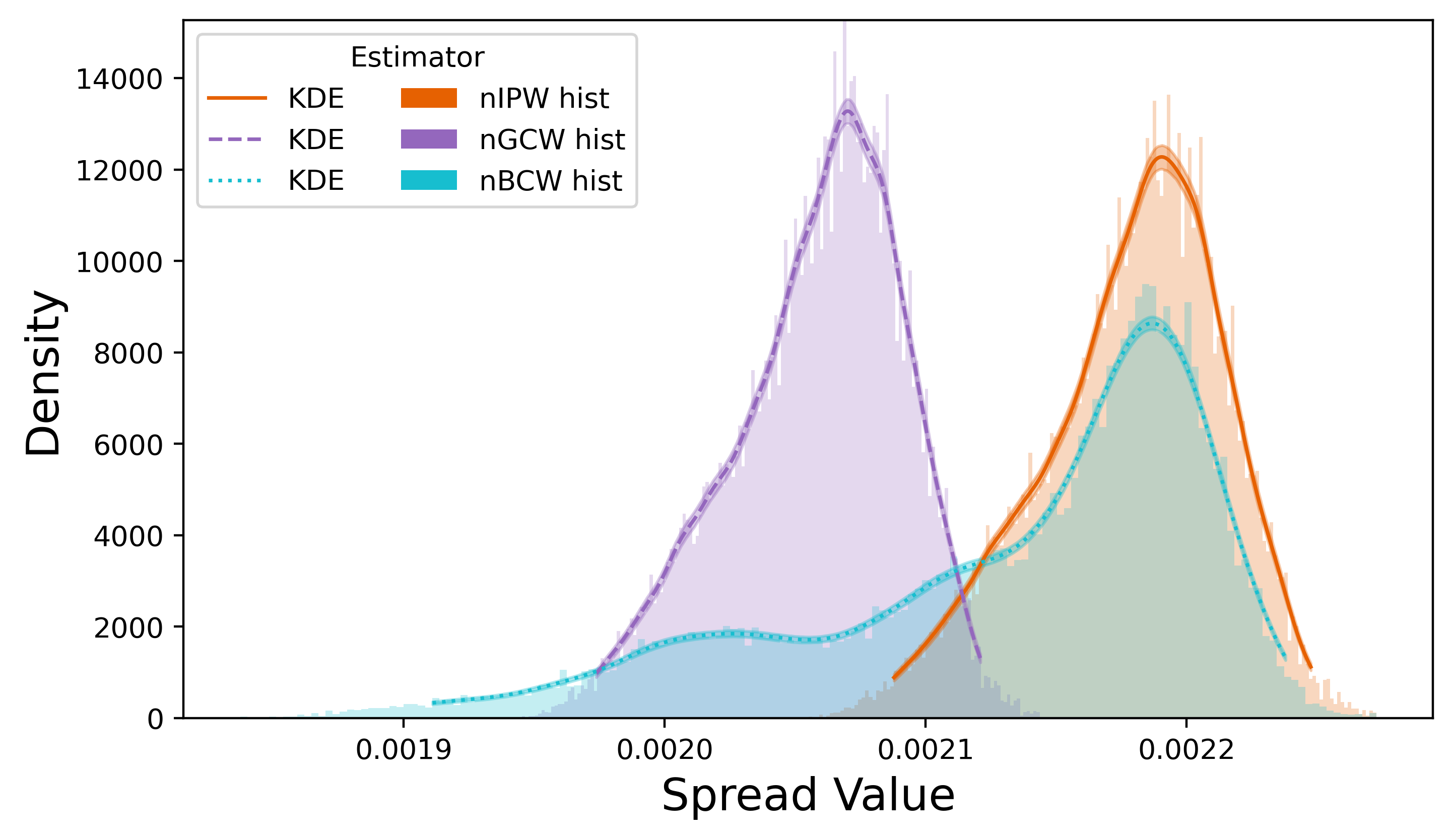}
        \caption{Non-augmented estimators}
    \end{subfigure}
    \hfill
    \begin{subfigure}{0.48\linewidth}
        \centering
        \includegraphics[width=\linewidth]{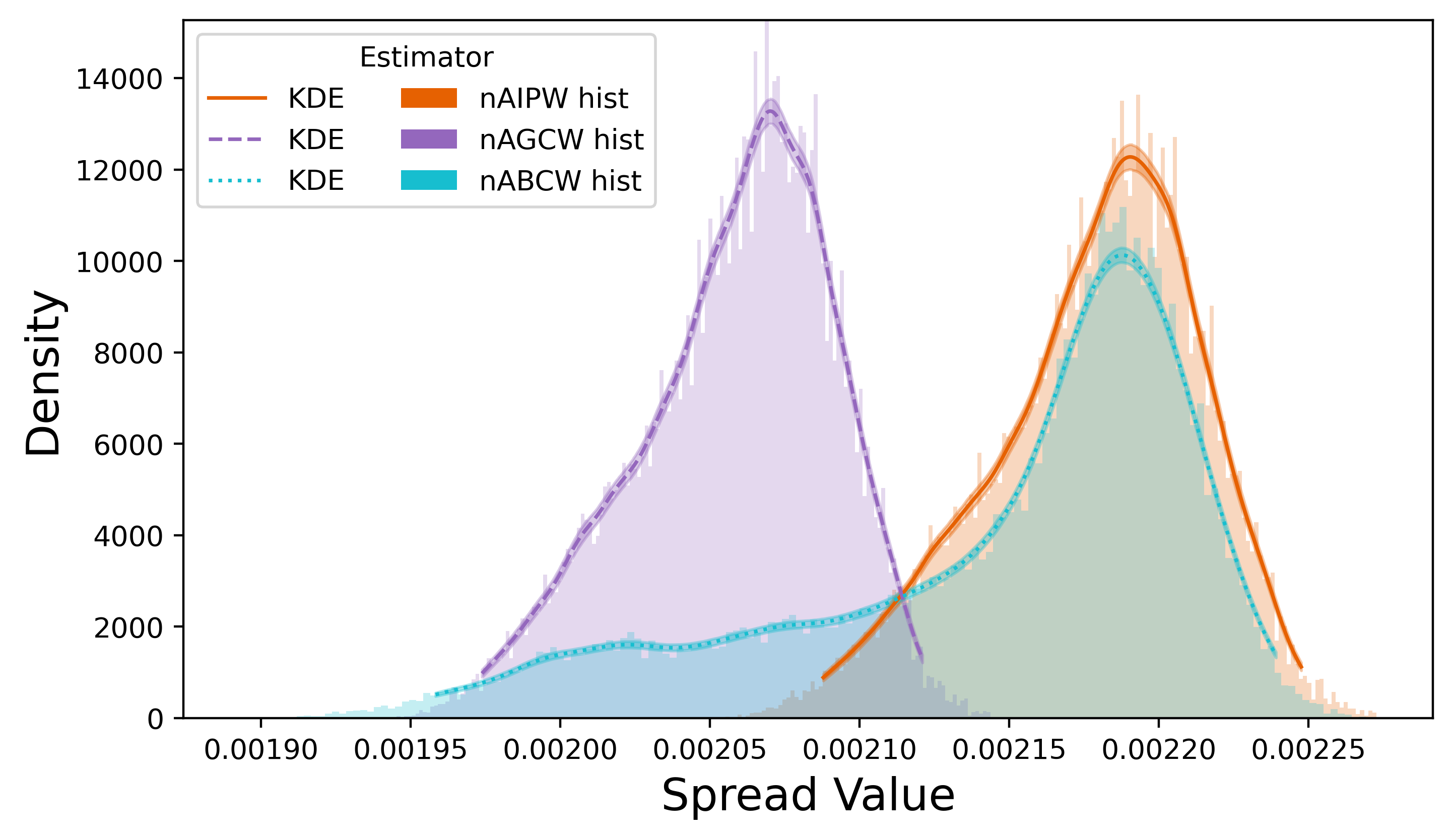}
        \caption{Augmented estimators}
    \end{subfigure}
    \caption{Comparison of weight spread $95^{\text{th}} - 5^{\text{th}}$ percentile.}
    \label{fig:two_weight_distributions_95_5}
\end{figure}

\FloatBarrier
\begin{figure}[htbp]
    \centering
    \begin{subfigure}{0.48\linewidth}
        \centering
        \includegraphics[width=\linewidth]{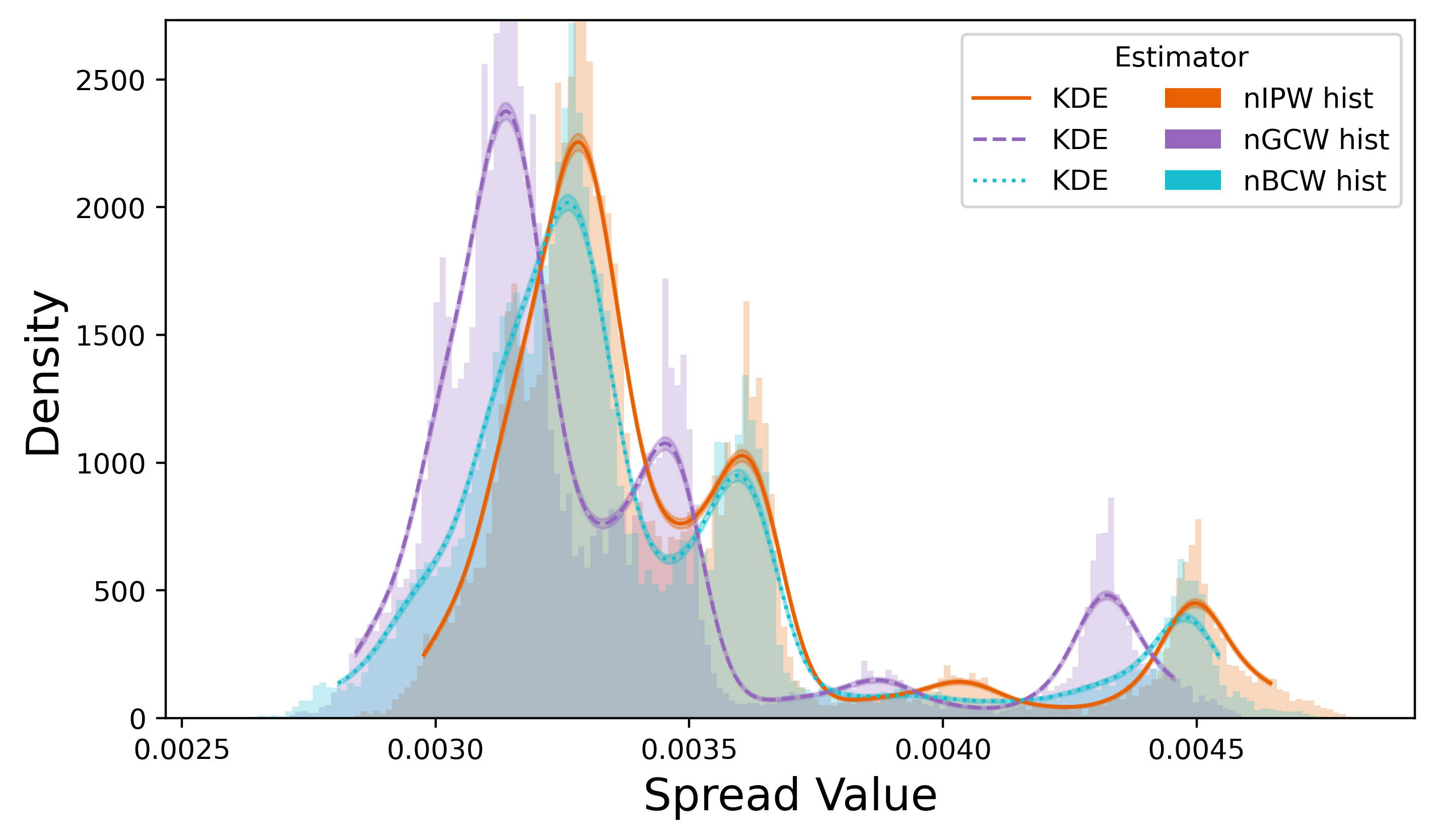}
        \caption{Non-augmented estimators}
    \end{subfigure}
    \hfill
    \begin{subfigure}{0.48\linewidth}
        \centering
        \includegraphics[width=\linewidth]{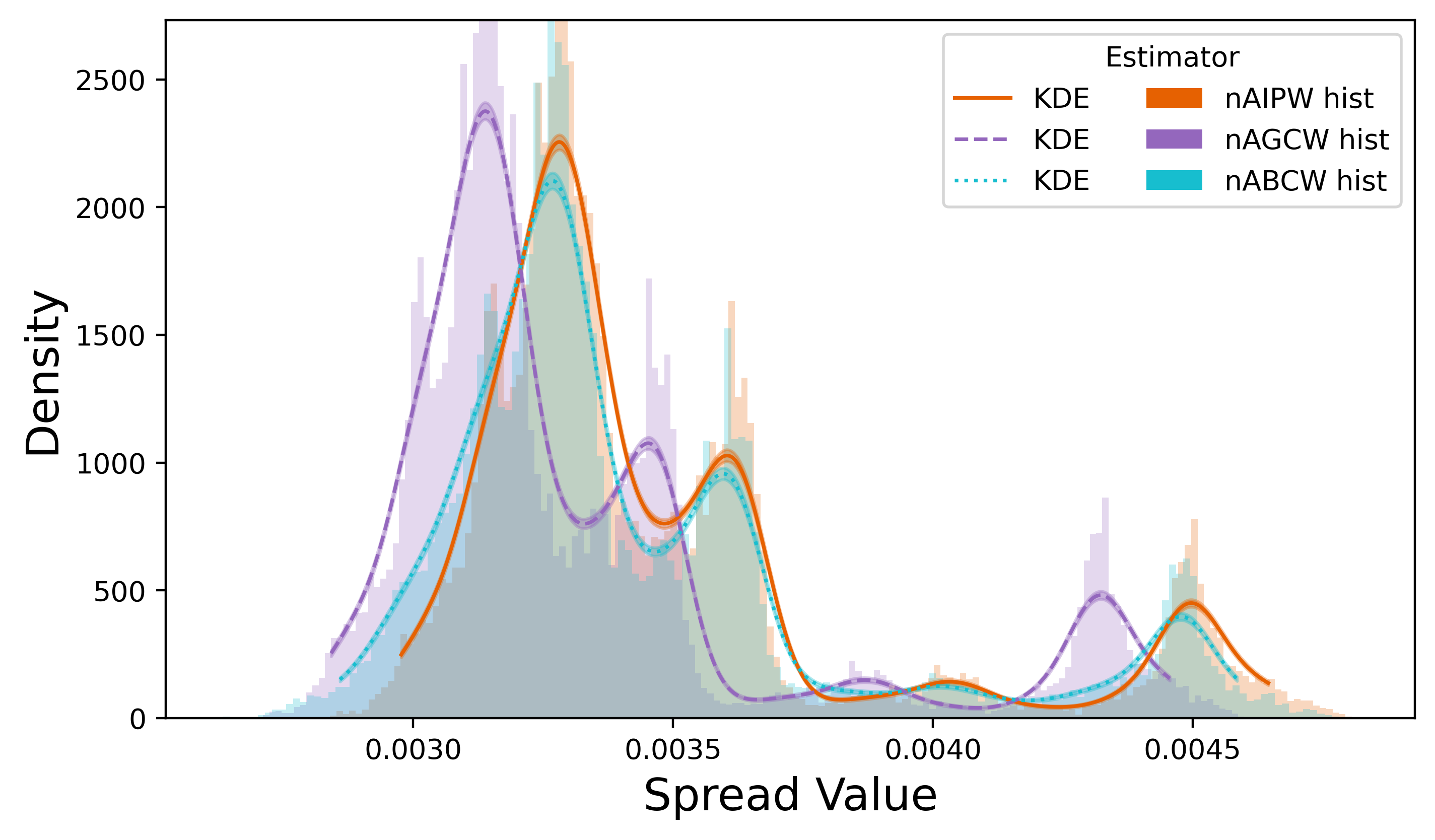}
        \caption{Augmented estimators}
    \end{subfigure}

    \caption{Comparison of weight spread max $-$ min.}
    \label{fig:two_weight_distributions_max_min}
\end{figure}

\end{document}